\documentclass[preprint]{aastex}

\pdfoutput=1

\usepackage{epsfig}
\usepackage{natbib}     
\usepackage{verbatim}   
\usepackage{graphicx}   
\usepackage{amsmath,amsthm,amsfonts,amsopn,amssymb} 

\usepackage{longtable,pdflscape}   
\usepackage{lscape}
\usepackage{longtable}
\usepackage{afterpage}
\usepackage{rotating}	
\usepackage{ulem}

\newcommand{\totaltargets}{360~}           

\newcommand{\totalebs}{207~}              

\newcommand{\totalebsnew}{97~}           

\newcommand{\totalebsknowneclipsing}{110~}   

\newcommand{\rearth}{R$_{\oplus}$}        
\newcommand{\rjup}{R$_{\rm Jup}$}		
\newcommand{\teff}{$T_{\rm eff}$}		

\shorttitle{New eclipsing binaries from K2}
\shortauthors{LaCourse et al.}

\begin{document}


\title{Kepler Eclipsing Binary Stars. VI. Identification of Eclipsing Binaries in the K2 Campaign 0 Data-set}


\author{
Daryll M. LaCourse\altaffilmark{1},
Kian J. Jek\altaffilmark{1}, 
Thomas L. Jacobs\altaffilmark{1},
Troy Winarski\altaffilmark{1},
Tabetha S. Boyajian\altaffilmark{2},
Saul A. Rappaport\altaffilmark{3},
Roberto Sanchis-Ojeda\altaffilmark{3},
Kyle E. Conroy\altaffilmark{4},
Lorne Nelson\altaffilmark{5},
Tom Barclay\altaffilmark{6}, 
Debra A. Fischer\altaffilmark{2},
Joseph R. Schmitt\altaffilmark{2},
Ji Wang\altaffilmark{2},
Keivan G. Stassun\altaffilmark{4,7},
Joshua Pepper\altaffilmark{8},
Jeffrey L. Coughlin\altaffilmark{9},
Avi Shporer\altaffilmark{10,11},
Andrej Pr\v{s}a\altaffilmark{12}
}

\altaffiltext{1}{Amateur Astronomer}
\altaffiltext{2}{Department of Astronomy, Yale University, New Haven, CT 06511, USA}
\altaffiltext{3}{Department of Physics, and Kavli Institute for Astrophysics and Space Research, Massachusetts Institute of Technology, Cambridge, MA 02139, USA}
\altaffiltext{4}{Department of Physics and Astronomy, Vanderbilt University, VU Station B 1807, Nashville, TN 37235}
\altaffiltext{5}{Department of Physics, Bishop's University, 2600 College St., Sherbrooke, QC J1M 1Z7}
\altaffiltext{6}{NASA Ames Research Center, Moffett Field, CA 94035}
\altaffiltext{7}{Department of Physics, Fisk University, Nashville, TN 37208, USA}
\altaffiltext{8}{Department of Physics, Lehigh University, Bethlehem, PA 18015, USA}
\altaffiltext{9}{SETI Institute/NASA Ames Research Center, Moffett Field, CA 94035, USA}
\altaffiltext{10}{Jet Propulsion Laboratory, California Institute of Technology, Pasadena, CA 91109}
\altaffiltext{11}{NASA Sagan Fellow}
\altaffiltext{12}{Department of Astrophysics and Planetary Sciences, Villanova University, 800 E Lancaster Ave, Villanova, PA 19085}


\begin{abstract}

The original {\it Kepler} mission observed and characterized over 2400 eclipsing binaries in addition to its prolific exoplanet detections. Despite the mechanical malfunction and subsequent non-recovery of two reaction wheels used to stabilize the instrument, the {\it Kepler} satellite continues collecting data in its repurposed {\it K2} mission surveying a series of fields along the ecliptic plane. Here we present an analysis of the first full baseline {\it K2} data release: the Campaign 0 data-set. In the 7761 light curves, we have identified a total of \totalebs eclipsing binaries. Of these, \totalebsnew are new discoveries that were not previously identified. Our pixel-level analysis of these objects has also resulted in identification of several false positives (observed targets contaminated by neighboring eclipsing binaries), as well as the serendipitous discovery of two short period exoplanet candidates. We provide catalog cross-matched source identifications, orbital periods, morphologies and ephemerides for these eclipsing systems. We also describe the incorporation of the K2 sample into the Kepler Eclipsing Binary Catalog\footnote{\url{keplerebs.villanova.edu/k2}}, present spectroscopic follow-up observations for a limited selection of nine systems, and discuss prospects for upcoming {\it K2} campaigns.

\end{abstract}



\section{Introduction}
\label{sec:introduction}

Between 2009 and 2013, the {\it Kepler} space telescope recorded high-precision photometry almost continuously for more than 150,000 stars near the constellation Cygnus and Lyra \citep{Koch2010,bor11a,bor11b,bat13,bur14}. This unprecedented data-set was used to detect more than 4000 exoplanet candidates (of which 1031 {\it Kepler} planets have been confirmed to-date) and revolutionized our understanding of the statistical occurrence of exoplanets and stellar astrophysics. The light curves of more than 2400 eclipsing binary stars were also measured \citep{prs11,sla11}, resulting in the detection of over 70 triple systems (\citealt{rap13}; \citealt{con14a}; \citealt{bork15}) and new astrophysical phenomena such as tidally excited Heartbeat binaries \citep{tho12}, self-lensing binaries \citep{kru14}, and Doppler boosting in ordinary stars \citep{ker10}.

Most of these discoveries were made by the {\it Kepler} science teams, using an automated pipeline \citep{jenkins2002, jenkins2010} to analyze the photometric data. In addition, the Planet Hunters Citizen Science project\footnote{\url{www.planethunters.org}} was established to engage the public in helping to analyze the wealth of light curves \citep{fis12}. This turned out to be a complementary scientific asset: to date, the visual inspection of light curves from Planet Hunters has resulted in the recovery and characterization of many of the transiting systems detected by the {\it Kepler} team and more than 50 unique exoplanet candidates. This includes dozens of giant planet candidates residing in the putative habitable zones of their host stars, and a circumbinary exoplanet orbiting one of a pair of binaries in a widely separated quadruple system \citep{Schwamb2013,wan13,sch14}.

{\it Kepler} was about to begin an extended mission when the loss of the second of four reaction wheels ended the primary mission.  At this juncture, a new plan of operation was devised to point the spacecraft along the ecliptic plane. This orbit minimizes the perturbations from solar illumination and pointing is maintained with the remaining two reaction wheels \citep{Howell2014}. This repurposed mission, known as {\it K2}, will survey ten new fields along the ecliptic plane over the next two to three years. Each of these ecliptic pointings constitutes one campaign, and the duration of observations for each field is limited to about 75 days because of solar radiation pressure constraints. All data are immediately non-proprietary, and will be released from the Mikulski Archive for Space Telescope (MAST) archive about three months after observations to engage the community in rapid follow-up. The first {\it K2} observations began in 2014 March to assess the quality of the photometric precision. This Campaign 0 (C0) data was uploaded to MAST in 2014 September. 
 
The {\it K2} C0 data set consists of observations from an $\sim 80$~day long engineering run, designed to assess whether a long term two-wheeled, ecliptic orientated mission was viable. Significant differences exist between the {\it Kepler} and the {\it K2} data sets in both quantity and quality. While the {\it Kepler} targets were carefully vetted before the launch of the mission and nominally selected based on their magnitude and luminosity class, the {\it K2} C0 targets were requested by the community and selected from 89 diverse Guest Observer proposals that aimed to carry out a broad range of science goals. This method of {\it K2} target selection has introduced different selection biases than for the stars observed in the {\it Kepler} field. The {\it K2} target information for each campaign is stored in the Ecliptic Plane Input Catalog\footnote{\url{\emph http://archive.stsci.edu/k2/epic/search.php}} (EPIC).

Unlike the {\it Kepler} mission, the quality of the {\it K2} C0 photometry is compromised by quasi-periodic, six hour thruster firings, a more rapidly changing thermal environment, and the emergent failure of detector Module~7. These effects have introduced significant and non-uniform systematics into the data. Targets that are farther from the bore sight sweep out larger arcs on the detectors than targets that are in the center of the field. Because of complications extracting and cleaning light curves, the {\it K2} mission is currently only providing MAST with a calibrated Target Pixel File (TPF) for each target. 

Despite these new challenges, the precision photometry offered by the {\it K2} mission will continue to make significant contributions to the field of eclipsing binaries (EBs). This has already been evidenced by the \textquoteleft Two-Wheel Concept Engineering Test\textquoteright~data set, a nine-day target limited observation performed in 2014 February that resulted in the detection of 31 EBs \citep{con14b}. Upon release of the C0 TPFs, we developed a custom light curve extraction pipeline in order to reduce and review the data. We present results from our pilot study to identify light curves that exhibit periodic variations, in particular, eclipsing binaries in the C0 data set. These results join EBs detected from {\it Kepler} and the {\it K2}: Two-Wheel Concept Engineering Test within the Kepler Eclipsing Binary Catalog\footnote{\url{\emph http://keplerebs.villanova.edu/}}.

\section{Campaign 0 Light Curves}
\label{sec:C0_data}

We initially extracted 7761 light curves in an automated fashion and independently de-correlated each of them against the spacecraft's known pointing motions, using the same technique described in \citet{vand14}. In addition, the aperture extraction mask size was adjusted as a function of the EPIC catalog listed source magnitude, with minimum mask dimensions of $3 \times 3$ pixels designated for the faintest observed C0 targets ($K_p$ of 19 or less) ranging up to $15 \times 15$ for the brightest ($K_p \sim 9$).  Unfortunately, a systemic spacecraft pointing error resulted in degradation for the first 40 days of observations for all targets (about half of the full C0 observation time baseline). While this early data range was not entirely compromised, we only used the second half of C0 observations in our bulk analysis. However, in our final analysis we were able to recover some of this early data to aid in characterizing several long period eclipsing binaries.

Visual inspection of the extracted C0 light curves produced an initial list of 258 EPIC targets that were flagged to be investigated for transient or periodic variations suggestive of an EB. A Fast Fourier Transform (FFT) routine subsequently recovered an additional 53 candidate signals of varying periods.

A 384,000 pixel \textquoteleft superstamp\textquoteright~was used in C0 to capture flux from the open clusters NGC~2168 (M35) and NGC~2158. The superstamp is coverage oriented, and no explicit EPIC target identifications are associated with each of the 153 uniform dimension TPFs contained within the superstamp. We assessed each of the superstamp TPFs for crowding. In this selection, we excluded dim background sources comprised of fewer than four pixels and any brighter source truncated by the borders of the TPF mask. We tailored our photometric extraction routine to identify source centroids and process light curves, again adjusting aperture masks as a function of source magnitude. These superstamp extracted light curves were then visually reviewed and vetted at the pixel-level in the same manner as the regular C0 target sample.  Although there is increased incidence of source confusion for superstamp TPFs covering the nominal center of each cluster, in many cases the exterior areas of the superstamp yielded high quality light curves at an average of 23 extractions per TPF. Examination of the output from this superstamp pipeline resulted in 49 additional periodic variables which were added to our initial investigation pool, bringing the total number of flagged targets to \totaltargets.

Because of concern about the quality of the {\it K2} C0 data, we employed open source Guest Observer software \textsc{PyKE} \citep{sti12}\footnote{\url{\emph http://keplerscience.arc.nasa.gov/PyKE.shtml}} to perform a second series of extractions for the flagged point sources, reducing the data into two new light curves using two different reduction techniques; Self Field Flattening (SFF) \citep{vand14} and Principal Component Analysis (PCA). These two reduction routines are included in the \textsc{PyKE} software package as \textsc{KepSFF} and \textsc{KepPCA} and have been updated for use with {\it K2}. The SFF method corrects for the {\it K2} aperture photometry artifacts by isolating short term variability and calculating a position-flux relation for each detrended iteration of the extracted light curve. PCA accomplishes a similar isolation of systematics by assessing the time series variability of each individual pixel in the chosen extraction mask and then segregating these trends into component groups that are correlated or anti-correlated to instrument movement \citep{har12}.

Aperture photometry pixel masks common to this second series of reductions and pixel level analysis were manually defined on a case by case basis in order to minimize flux losses from the intended target and to exclude neighboring sources wherever possible. We also extracted and inspected light curves from bright neighboring sources that were potentially contaminating our target of interest.

The light curves extracted with \textsc{KepPCA} and \textsc{KepSSF} were compared to assess the degree each routine had muted or obfuscated the intrinsic astrophysical signal. Testing this across a variety of point sources showed that neither reduction technique offers a clear advantage over the other in all situations. As such, we chose to employ light curves derived from both \textsc{KepSFF} and \textsc{KepPCA} for each of the targets reviewed in the course of our dedicated pixel-level analysis.

\section{Culling}
\label{sec:Culling}

Due to {\it Kepler}'s sensitivity, pixel size, and observing fields, the likeliness for source confusion common to any given target is high. The most common forms of contamination occur as a result of {\it Kepler}'s pixel response function (PRF), which is quantified by the spacecraft's pointing precision, pixel resolution and point spread function \citep{brys10}. When the PRF of two or more point sources directly overlap and blend, or when the PRF wings of sufficiently bright stars encroach into the photometric apertures of separated targets, the variable signal from one point source can be introduced into another point source as a diluted alias \citep{van09, cald10}. For this reason we take a conservative approach in the vetting of EBs for our final C0 sample. If the source of the eclipsing signal for a particular EB could not confidently be determined during a pixel level review, it was excluded from our final sample. Of the 360 light curves we flagged for eclipsing or ellipsoidal activity, 153 were ultimately rejected as spurious detections or blended binaries that could not be confidently associated with a a specific point source. The itemized breakdown of these rejected EB candidates is detailed as follows:

1. Fifteen flagged EPIC targets were rejected as suspected cases of direct PRF contamination. Examination of the TPF in conjunction with review of optical all-sky databases (e.g., \textsc{2MASS}, \citealt{skr06}; \textsc{DSS}\footnote{\url{https://archive.stsci.edu/dss/}}; \textsc{WISE}, \citealt{wri10}) indicated that multiple point sources were blended or crowded within our initially employed photometric aperture. Subsequent efforts to re-extract and isolate the eclipse signal failed to tease apart the originating pixel location and as such these low signal-to-noise (SNR) candidates were culled from our sample.

2. Twenty-nine additional cases where a point source in the halo of an EPIC's TPF was found to contaminate the intended EPIC target's light curve.  Aliases were identified where it was evident that the halo source displayed the same ephemerides but at a much greater amplitude. We have cross matched the 2MASS identifications of these non-EPIC EBs and appropriately noted the victimized EPICs as false positives in the \textsc{Kepler Eclipsing Binary Catalog}. In eight cases where the suspected source of contamination was truncated by the edge of the TPF definition or itself appeared to be a blend, the target was removed from our final sample.

3. Nine targets were culled after an investigation for periods and epochs that matched or were closely similar to another EB candidate in our sample. Ephemerides matching has previously proven an effective tool to root out false positives {\it Kepler} data (e.g., \citealt{bat13, cou14}). With the small size of our vetting sample it was possible to perform this search manually with large allowances for period error. No maximum arc second search radius was imposed in order to safeguard against potential instrument related forms of contamination that allow variable signals to afflict victim sources positioned on entirely different area of the detector, such as CCD Crosstalk and Antipodal Reflection \citep{cald10, cou14}. Although no clear cases of such exotic contamination were clearly identified, we identified nine cases of EPICs that were contaminating other observed EPICs.

4. Fourteen pairs of EPICs were noted during ephemerides matching and other vetting processes as apparent duplicate observations: the TPFs for each EPIC center on the same apparent point source, but with minutely differing target catalog coordinates listed. After reviewing the data to ensure that a dim field companion star was not actually the source of either EPIC proposal, we retained the EPIC that appeared best centered on the intended target and removed the other seven EPICs from our sample.

5. Twenty-eight targets were rejected as cases of mistaken identity with variable stars displaying periodic or near-periodic pulsations or rotational modulation.

6. Sixty-five flagged EPICs were rejected without conclusive final dispositions and are suspected to be largely spurious detections from two general categories: faint background EBs unresolved in the TPF pixel data and available all-sky data or artifact escapements common to our reduction processes. 

The 207 survivors of these culling reviews were selected for further characterization and calculation of ephemerides. A serendipitous by-product of these analyses were the identification of two exoplanet candidates.

\section{Ephemerides, Periods, Phase Curves, Cross Matched Source Identifications, and Meta Data}
\label{sec:pixel_level_analysis}

The periodicity for each target displaying three or more eclipses was initially determined based on the phase dispersion minimization (PDM) method \citep{ste78}. The PDM variant method described by \citealt{pla08} was also used to find periods. The PDM methods essentially fold the light curve at different periods and assign a \textquoteleft dispersion score\textquoteright~ for each detected period. These scores are then sorted, the minima associated with the best scores are picked and iteratively phase folded at increasingly narrower period ranges to find the best local solution. Our routine based on the original \citet{ste78} work proved effective at correctly identifying the estimated period with no constraints on initial period range. However, we also utilized the PDM variant by \citep{pla08} as it assigns scores differently: a boxcar average is subtracted from the flux for each test fold and the residuals are squared and summed in order to provide ranking.

We evaluated mid-eclipse times by employing the methods of \textquoteleft KWM\textquoteright\  \citep{kwe56}, Barycenter (e.g., \citealt{osh12}) and Least Squares (e.g., \citealt{smi12}). The output parameters were then vetted by eye to ensure that BJD0 coincides with the deeper eclipse in cases where secondary eclipses are also present in the light curve.  

\begin{figure*}
\begin{center}
\begin{tabular}{cc}
\includegraphics[width=0.38\linewidth]{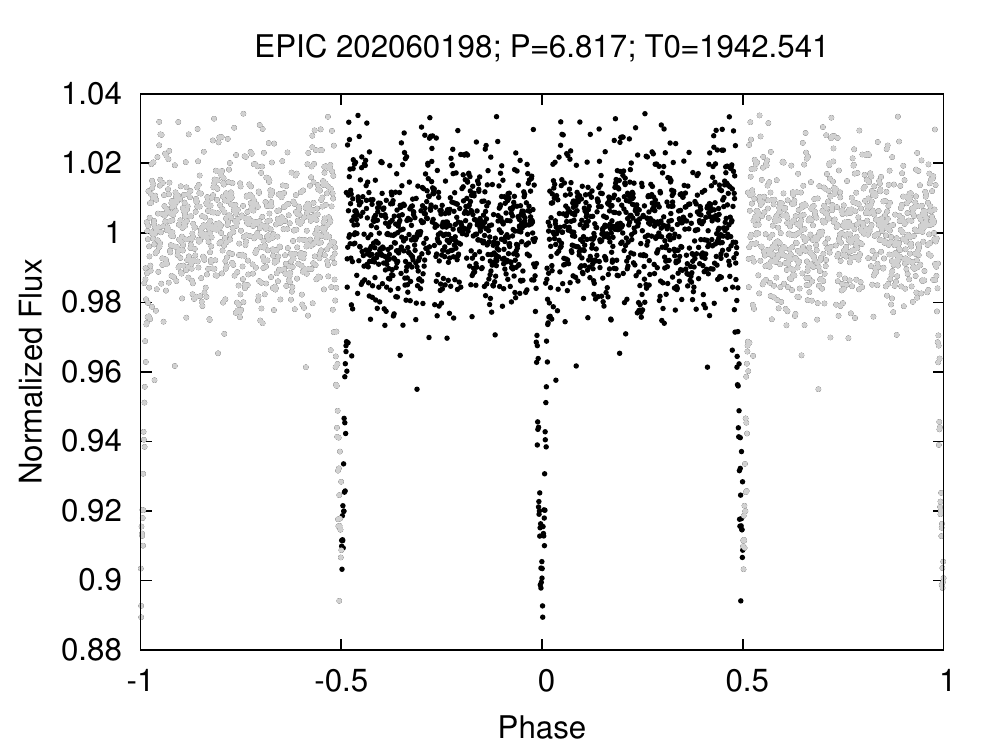} &
\includegraphics[width=0.38\linewidth]{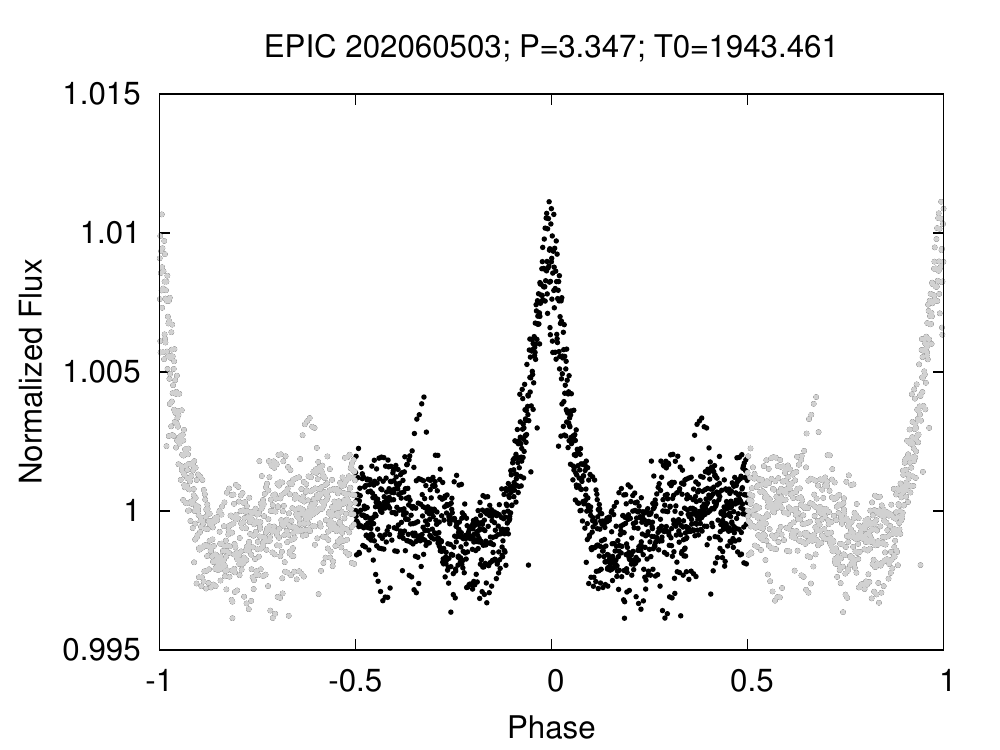} \\
\includegraphics[width=0.38\linewidth]{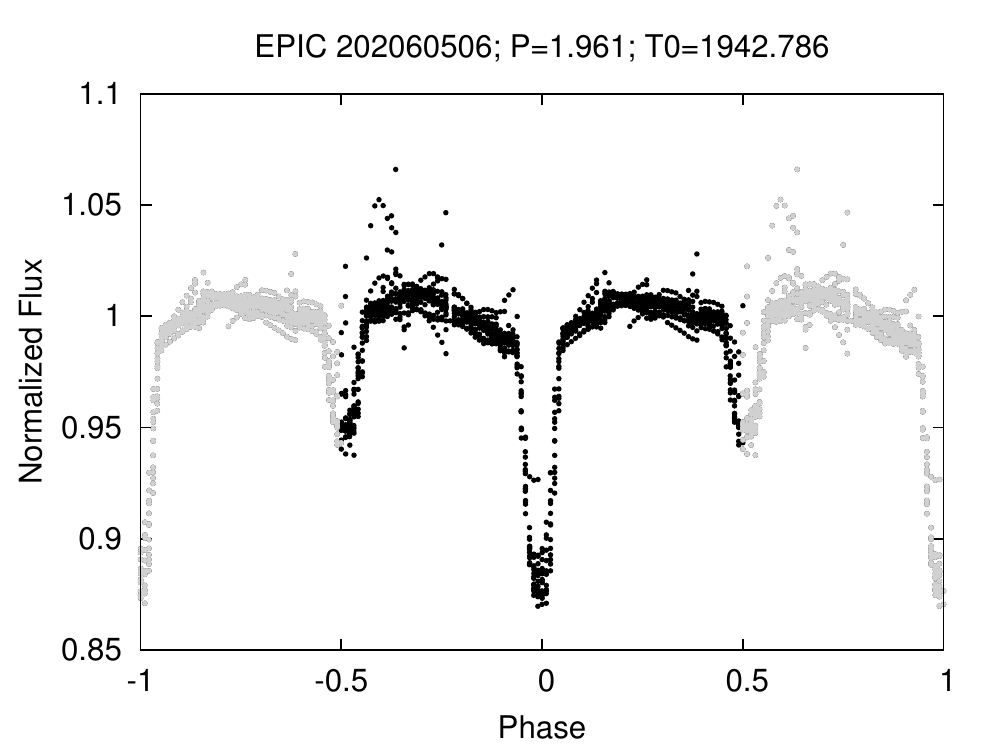} &
\includegraphics[width=0.38\linewidth]{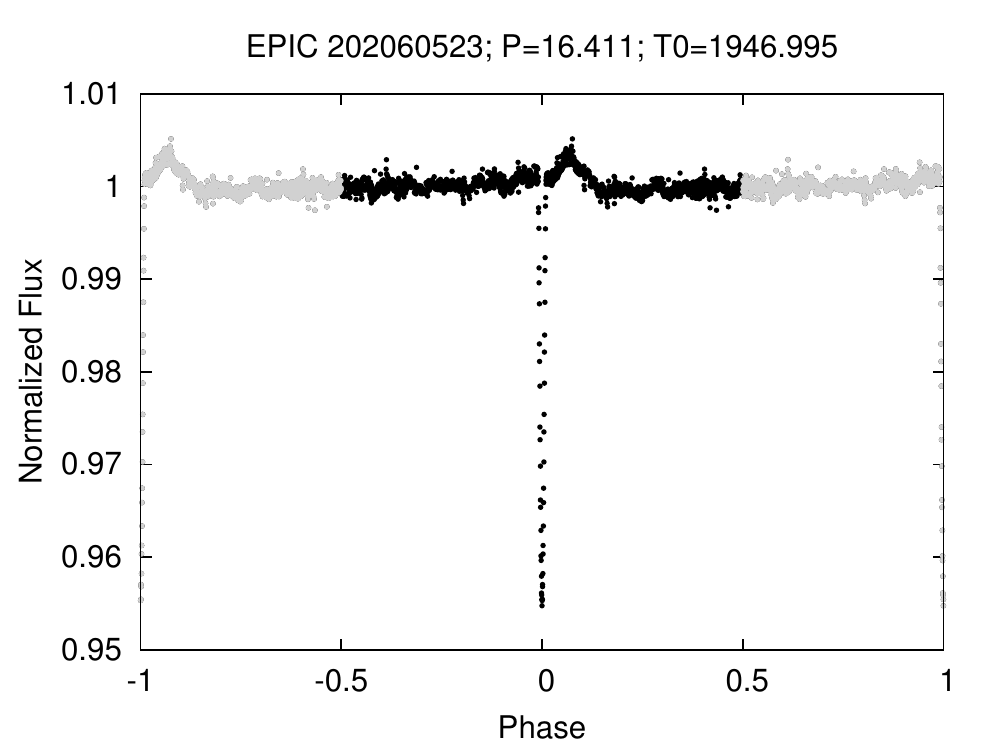} \\
\includegraphics[width=0.38\linewidth]{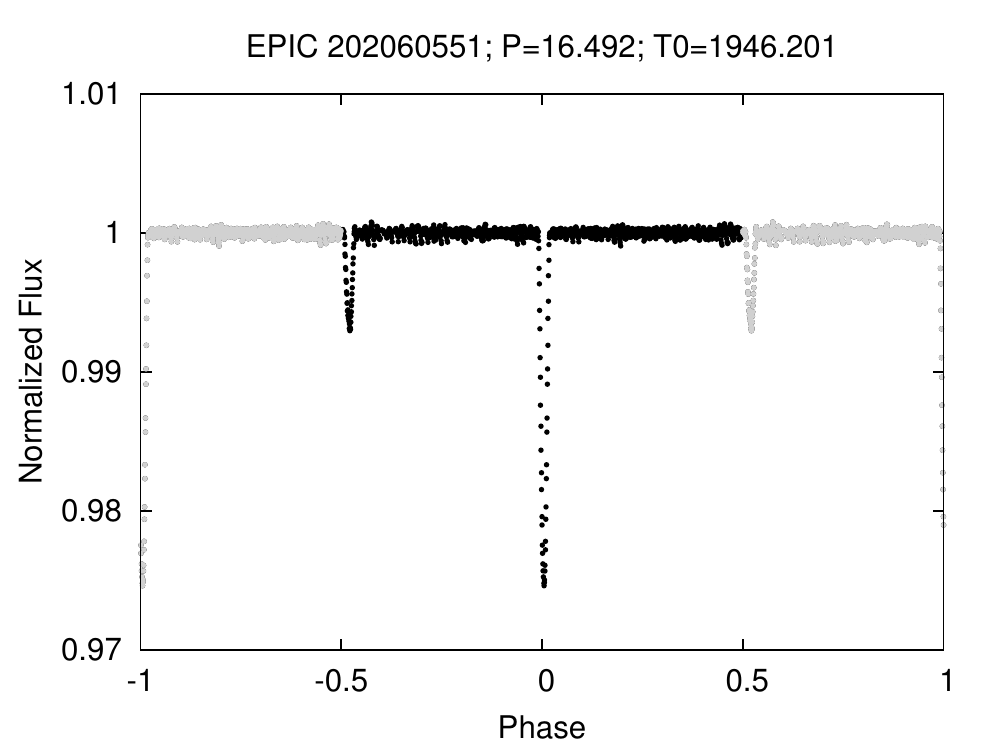} &
\includegraphics[width=0.38\linewidth]{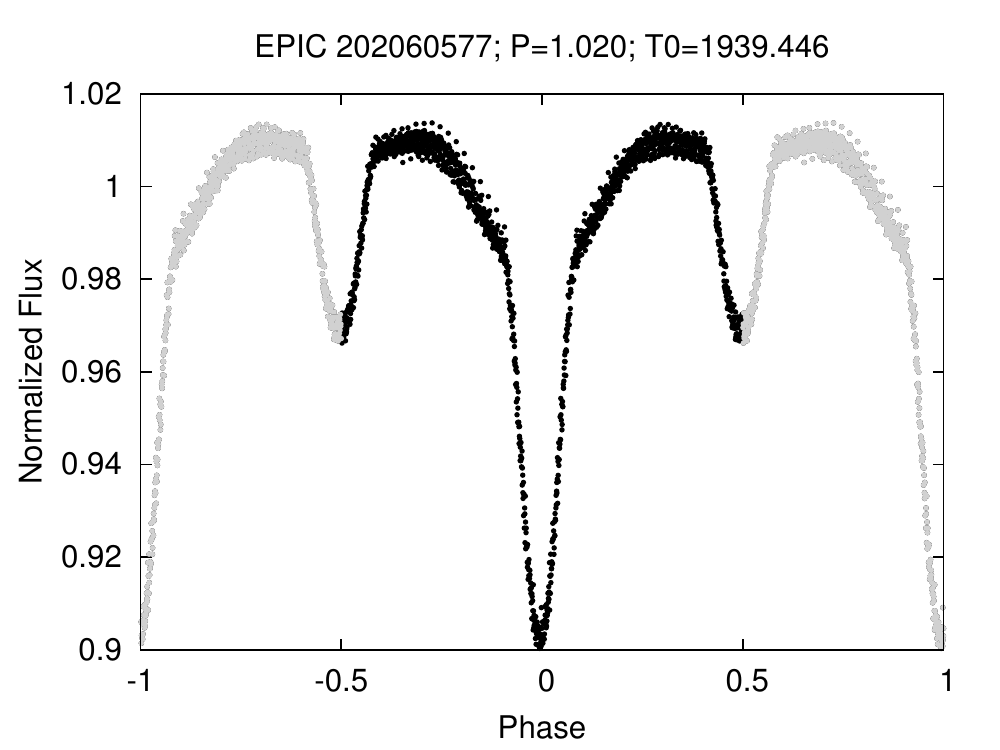} \\
\includegraphics[width=0.38\linewidth]{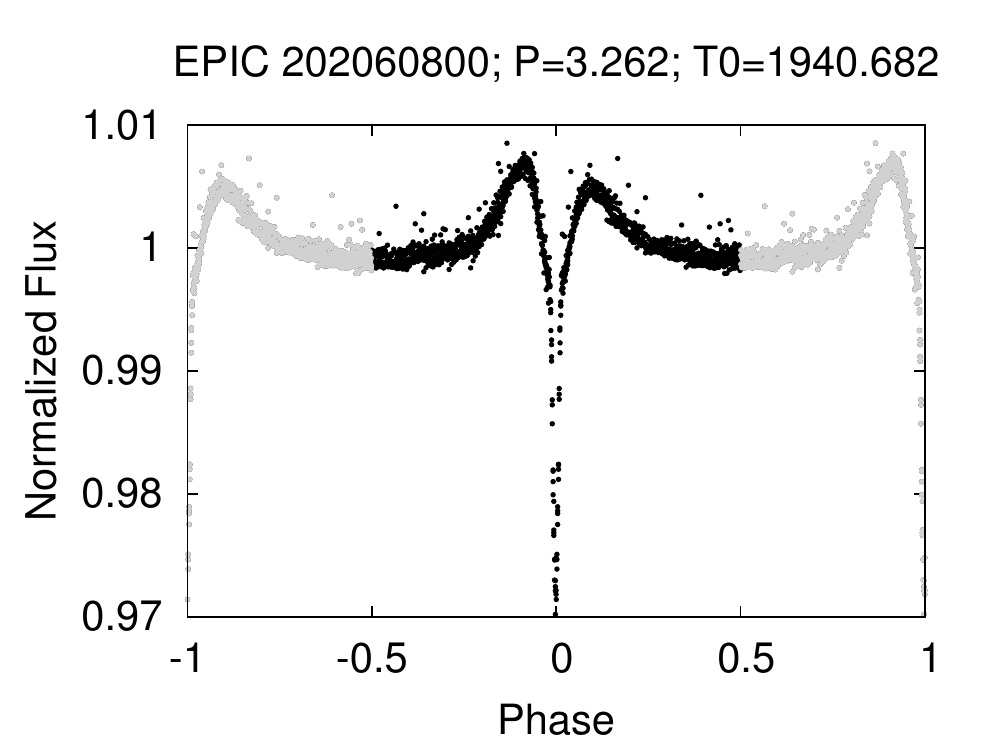} &
\includegraphics[width=0.38\linewidth]{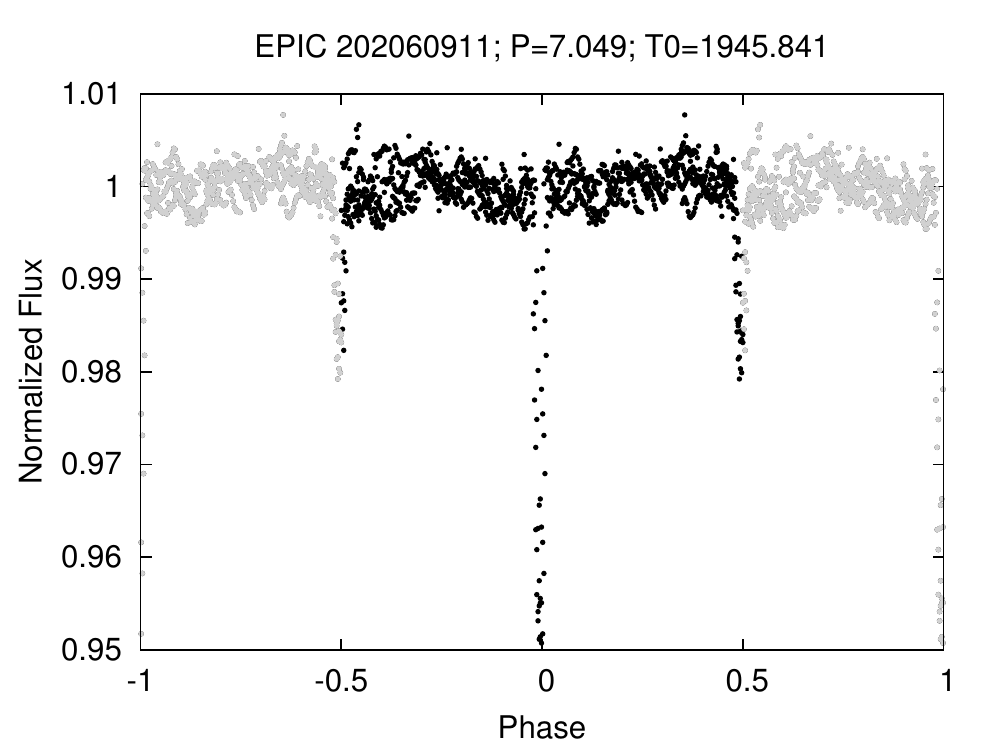} 
\end{tabular}
\end{center}
\caption{Plots of phased light curves for new eclipsing systems. Ephemerides are presented in Tables~\ref{tab:newEBs}, \ref{tab:m35eb}, and \ref{tab:nonEPIC}.  See Section~\ref{sec:pixel_level_analysis} for details.}
\label{fig:phasedplots_1}
\end{figure*}

\begin{figure*}
\begin{center}
\begin{tabular}{cc}
\includegraphics[width=0.38\linewidth]{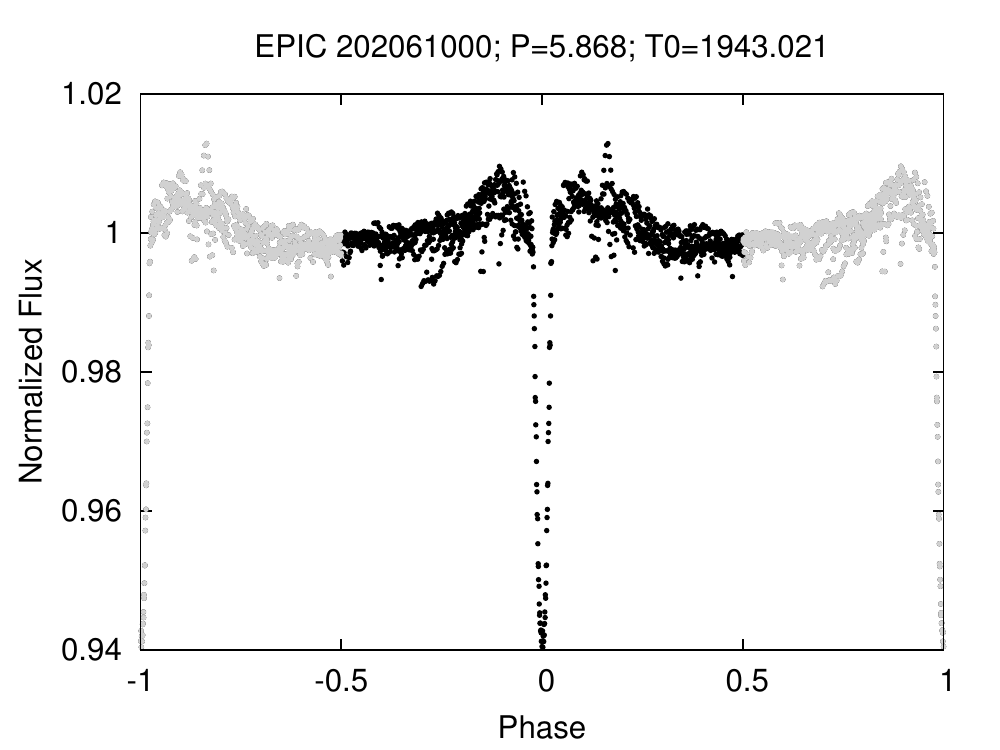} &
\includegraphics[width=0.38\linewidth]{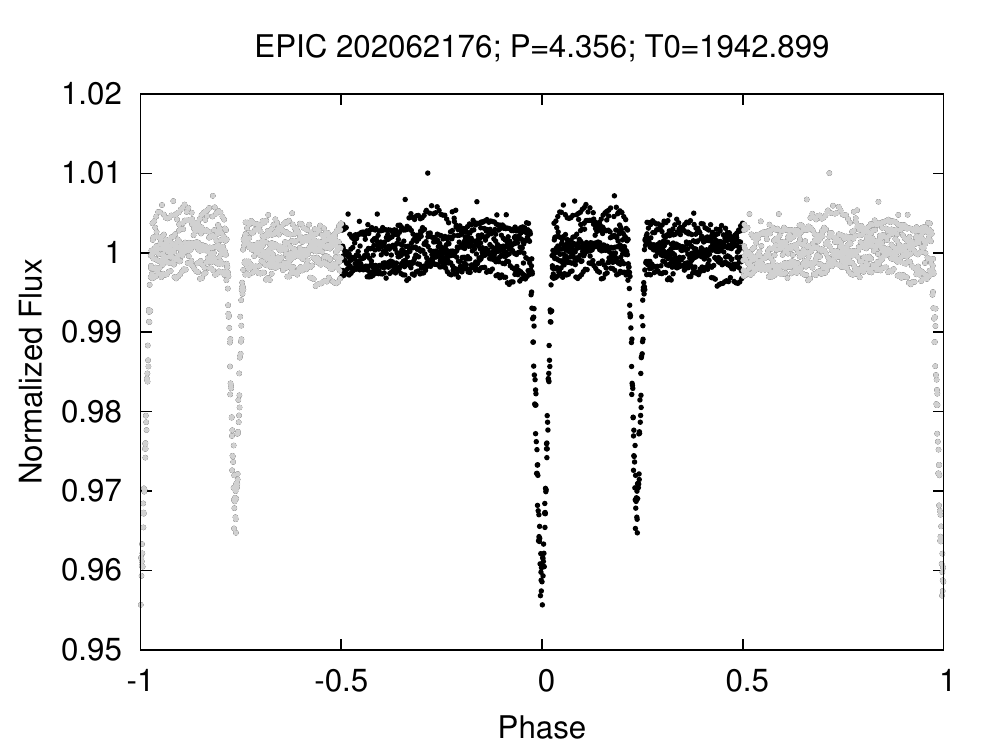} \\
\includegraphics[width=0.38\linewidth]{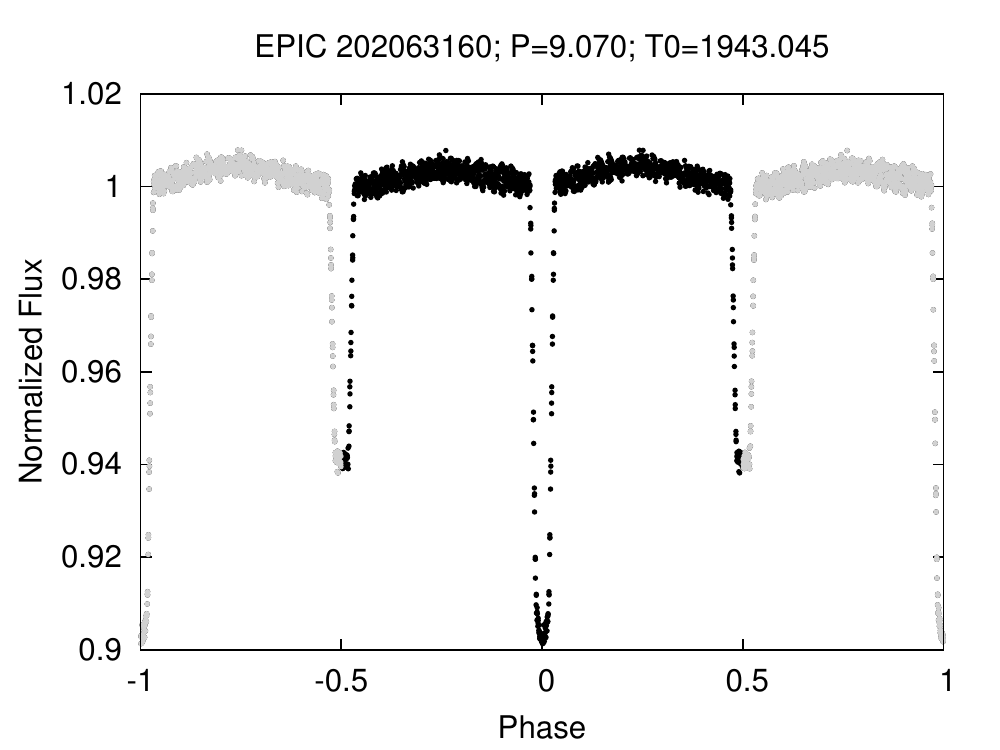} &
\includegraphics[width=0.38\linewidth]{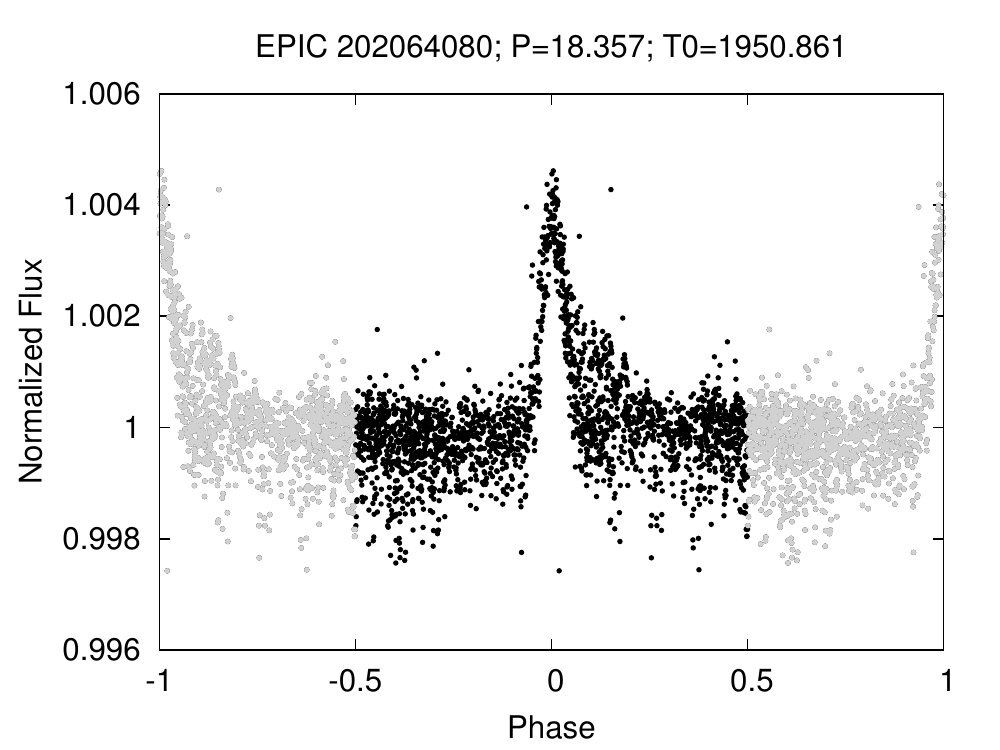} \\
\includegraphics[width=0.38\linewidth]{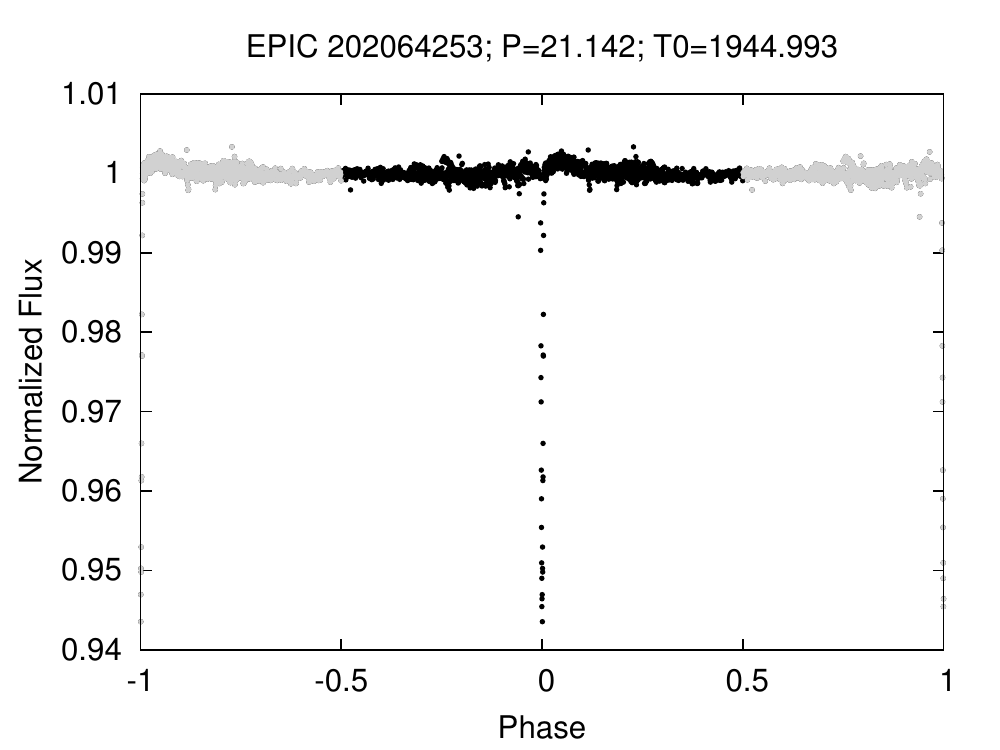} &
\includegraphics[width=0.38\linewidth]{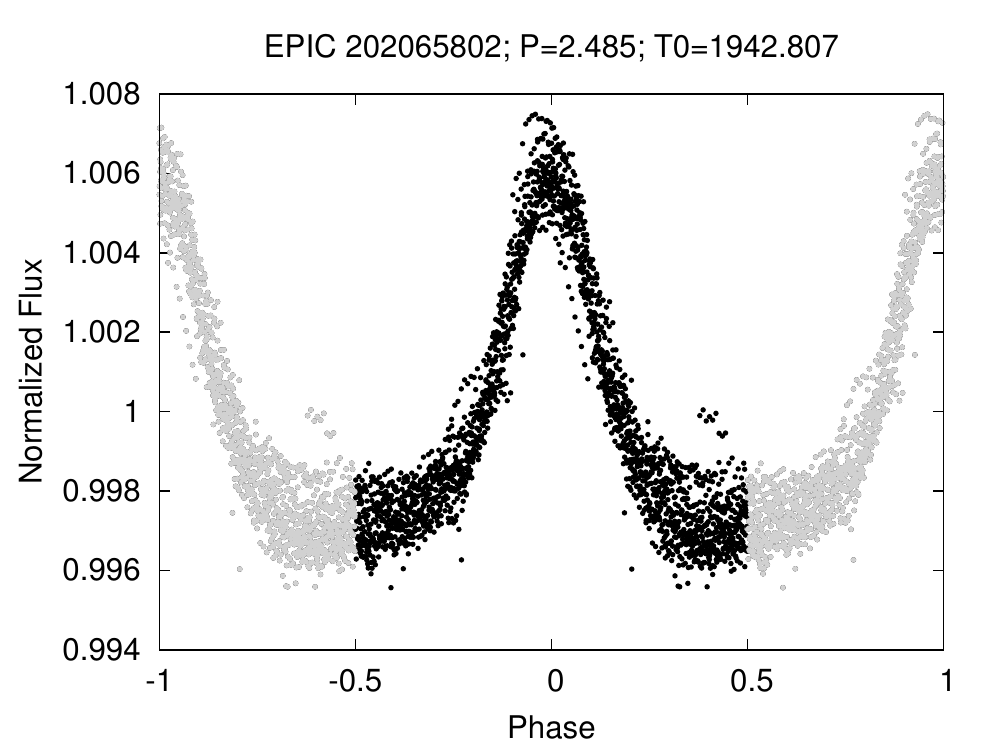} \\
\includegraphics[width=0.38\linewidth]{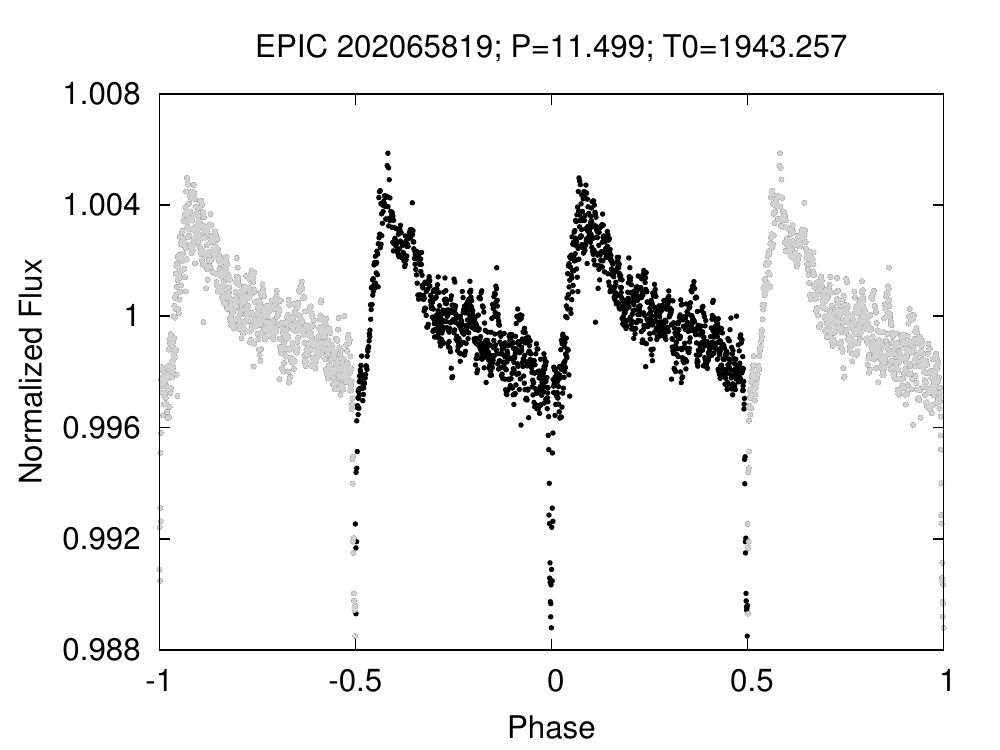} &
\includegraphics[width=0.38\linewidth]{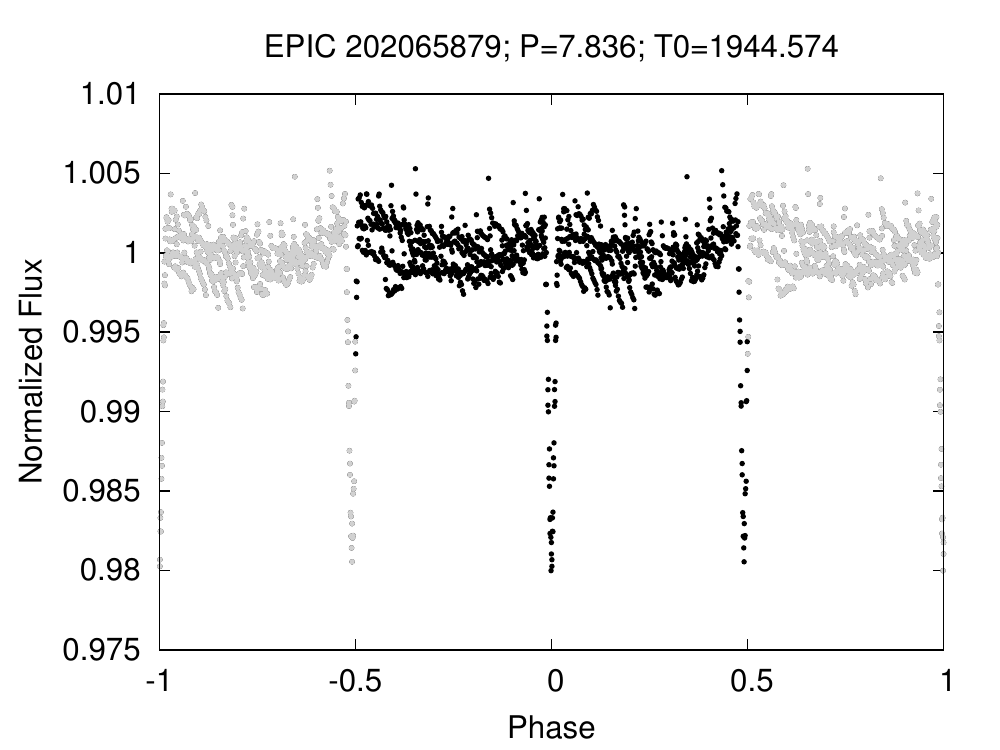} 
\end{tabular}
\end{center}
\caption{Plots of phased light curves for new eclipsing systems. Ephemerides are presented in Tables~\ref{tab:newEBs}, \ref{tab:m35eb}, and \ref{tab:nonEPIC}.  See Section~\ref{sec:pixel_level_analysis} for details.}
\label{fig:phasedplots_2}
\end{figure*}

\begin{figure*}
\begin{center}
\begin{tabular}{cc}
\includegraphics[width=0.38\linewidth]{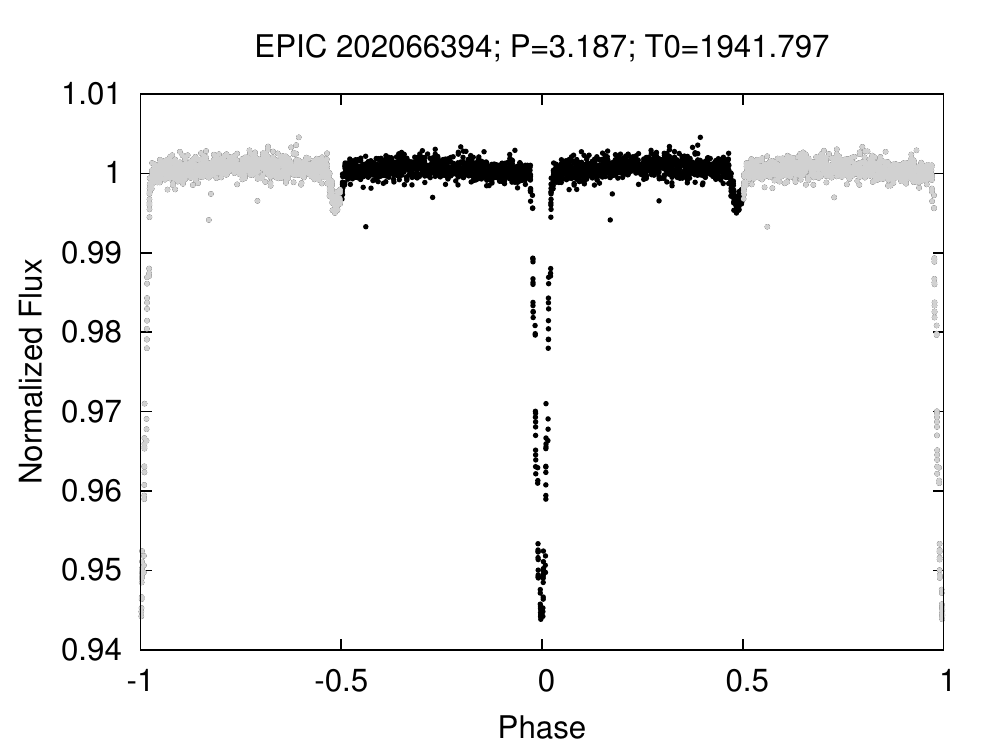} &
\includegraphics[width=0.38\linewidth]{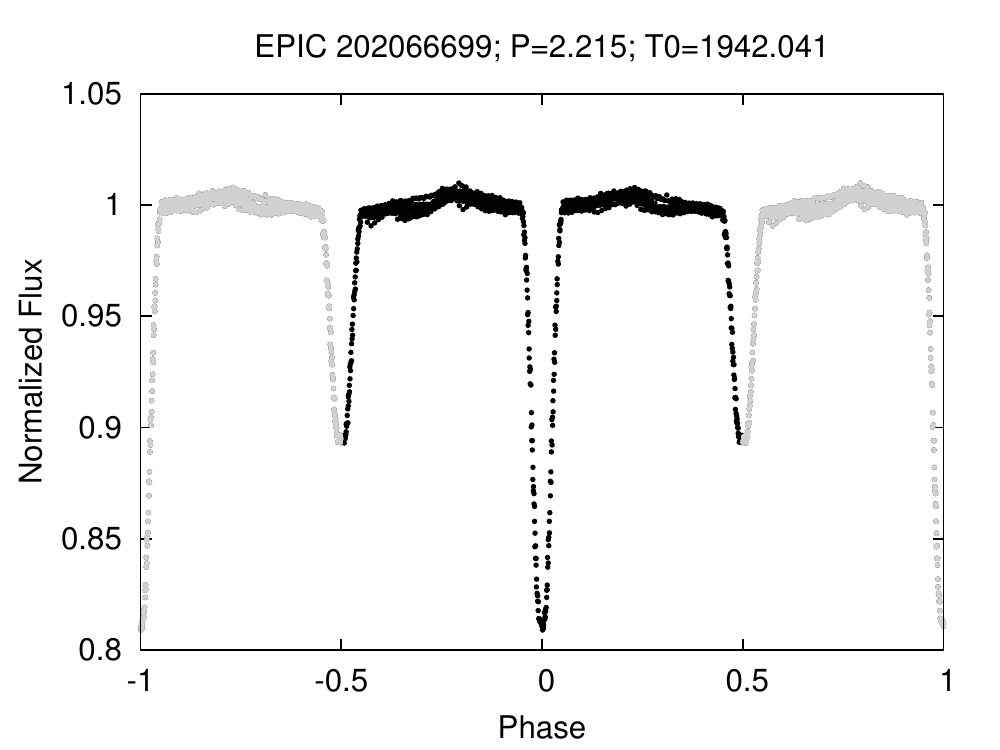} \\
\includegraphics[width=0.38\linewidth]{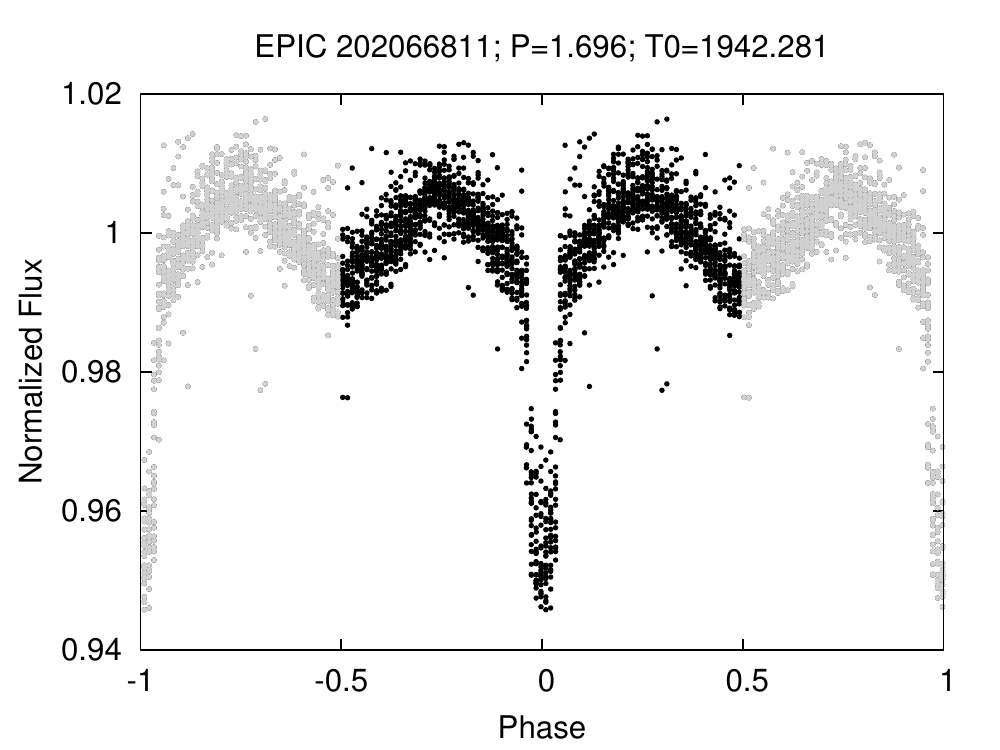} &
\includegraphics[width=0.38\linewidth]{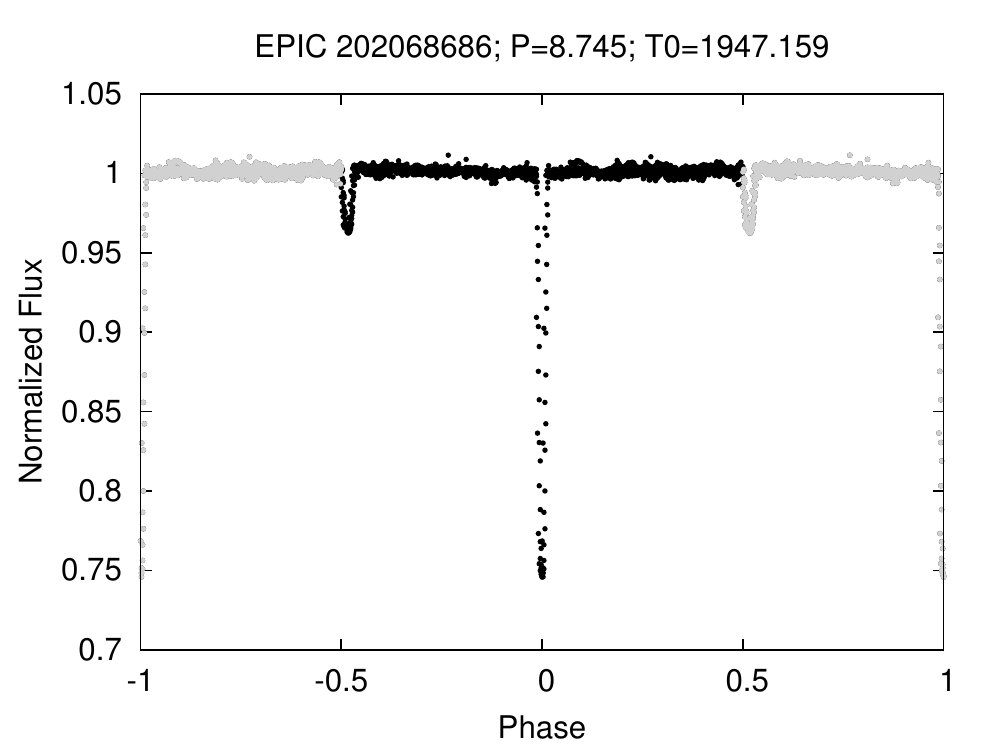} \\
\includegraphics[width=0.38\linewidth]{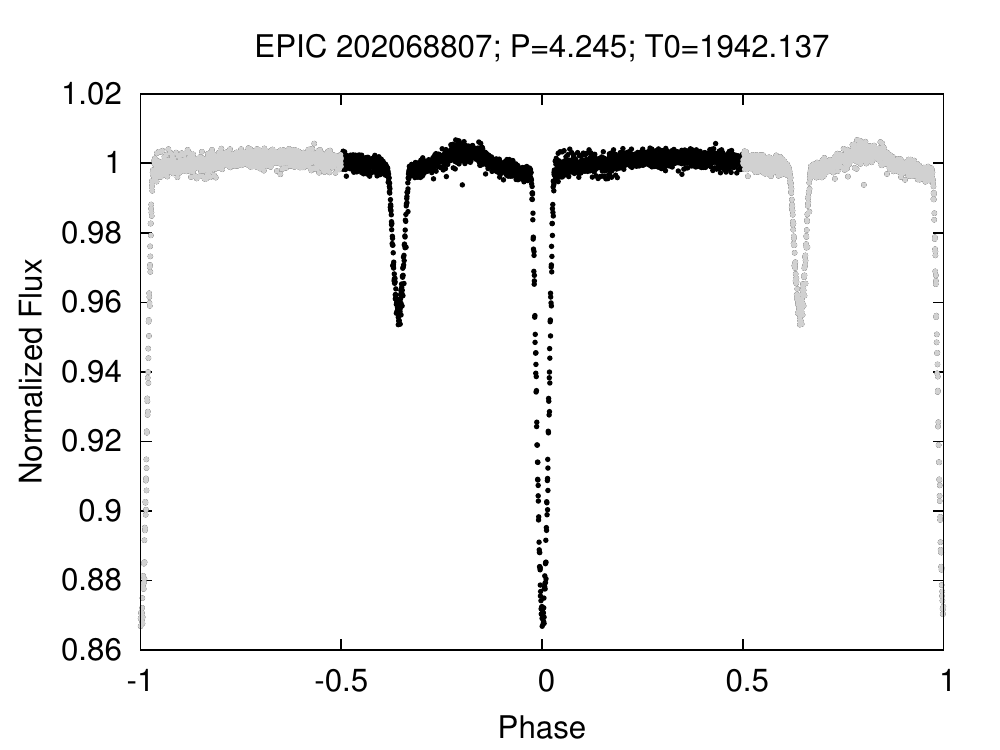} &
\includegraphics[width=0.38\linewidth]{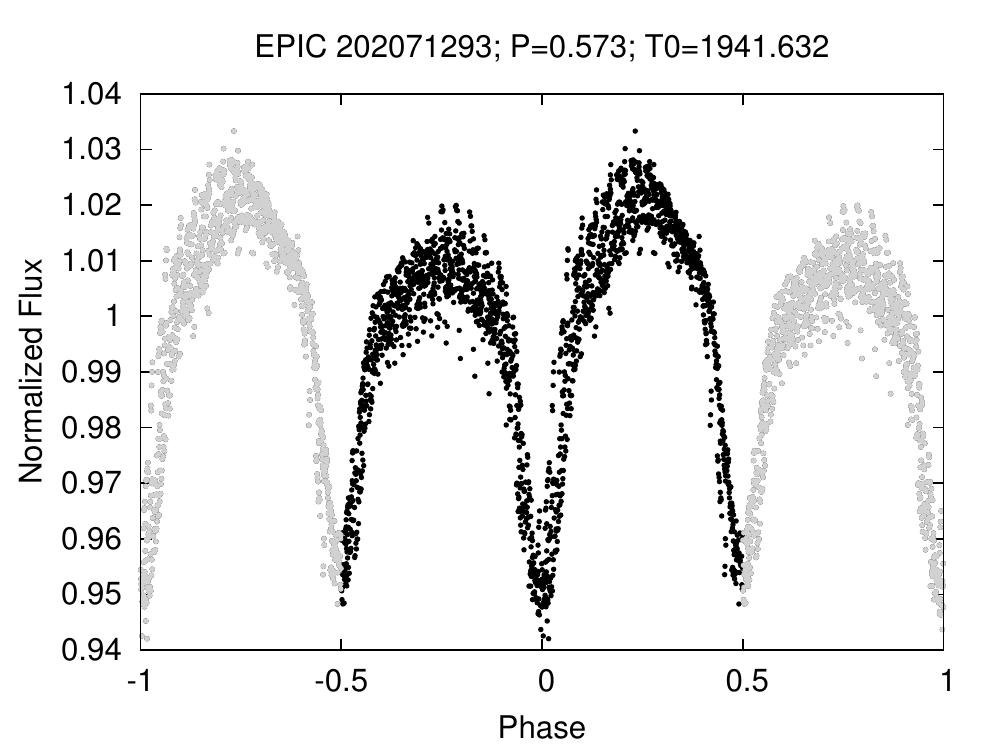} \\
\includegraphics[width=0.38\linewidth]{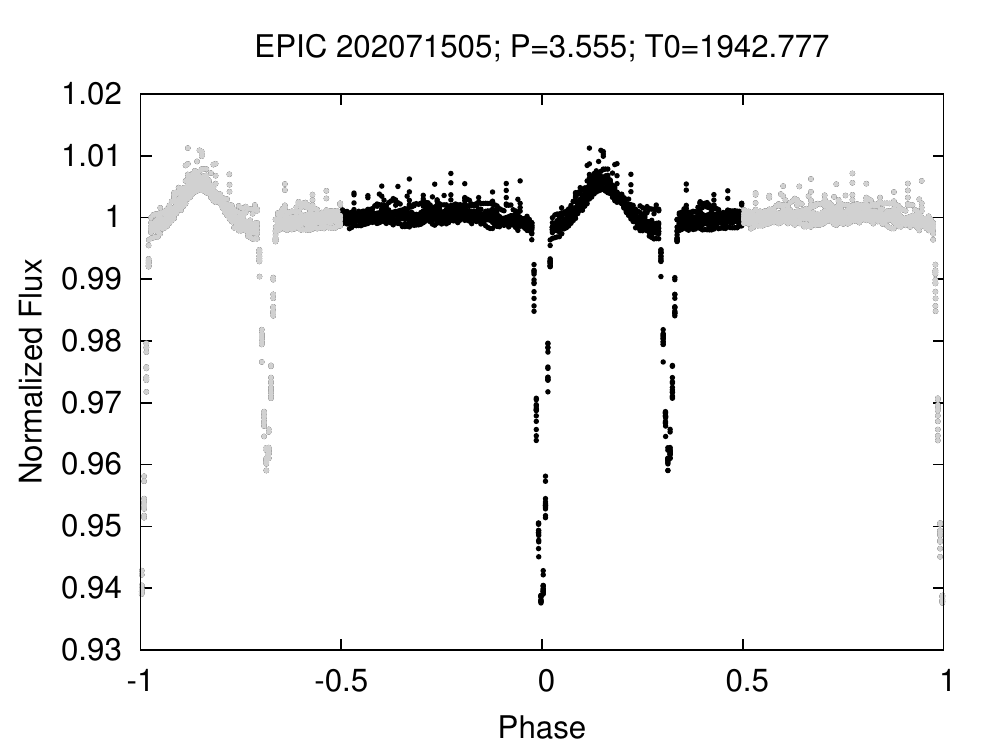} &
\includegraphics[width=0.38\linewidth]{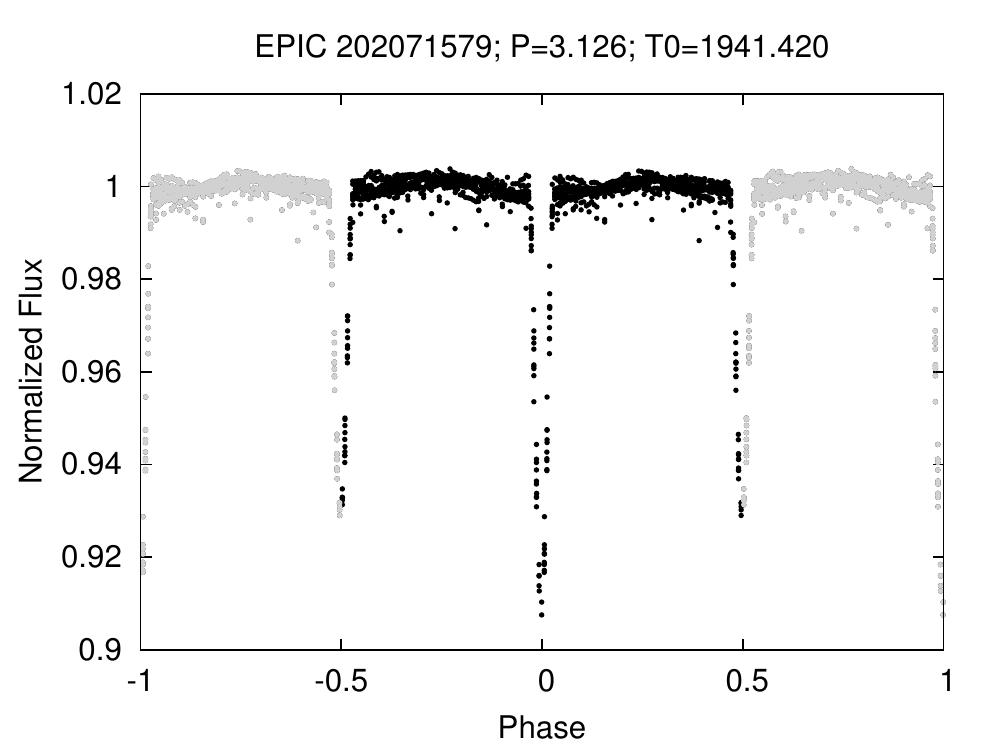} 
\end{tabular}
\end{center}
\caption{Plots of phased light curves for new eclipsing systems. Ephemerides are presented in Tables~\ref{tab:newEBs}, \ref{tab:m35eb}, and \ref{tab:nonEPIC}.  See Section~\ref{sec:pixel_level_analysis} for details.}
\label{fig:phasedplots_3}
\end{figure*}

\begin{figure*}
\begin{center}
\begin{tabular}{cc}
\includegraphics[width=0.38\linewidth]{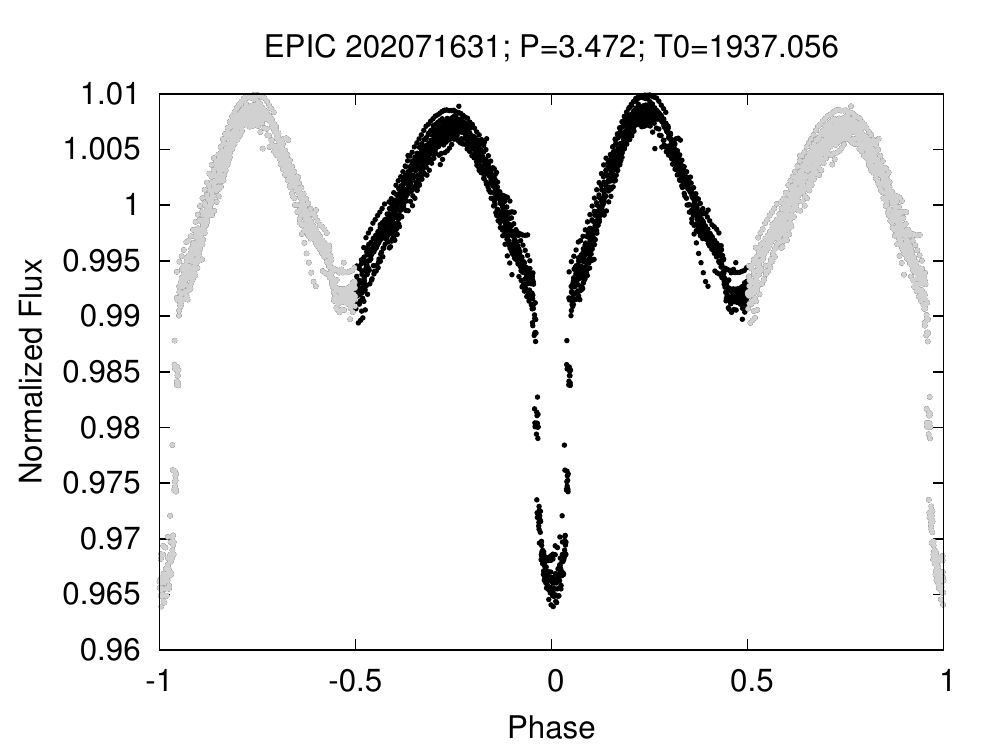} &
\includegraphics[width=0.38\linewidth]{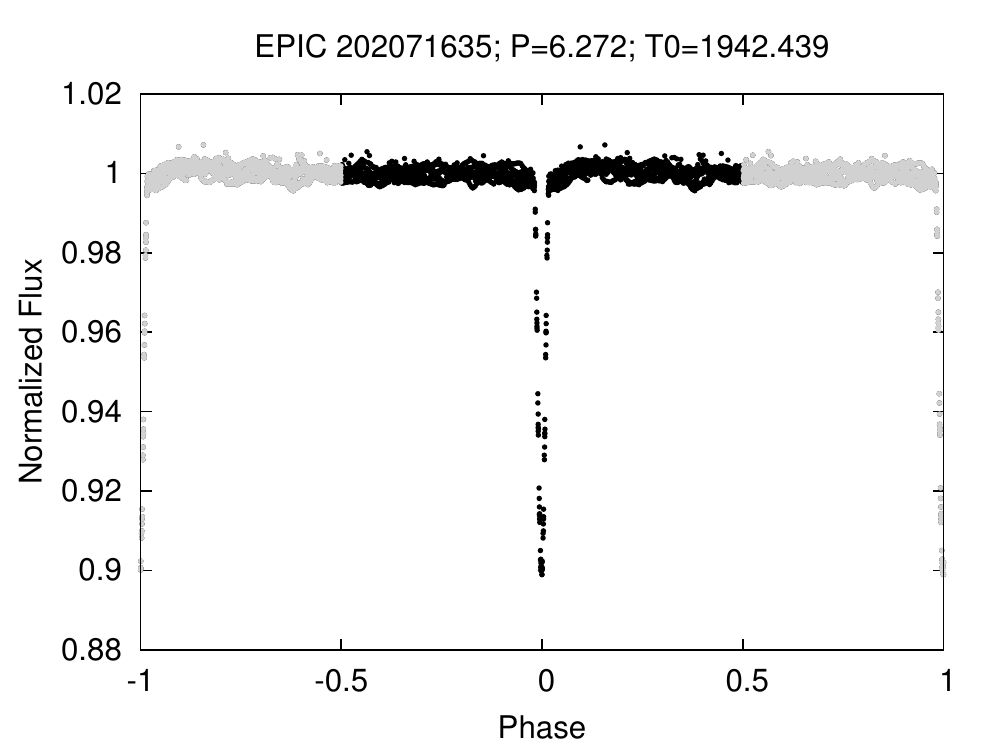} \\
\includegraphics[width=0.38\linewidth]{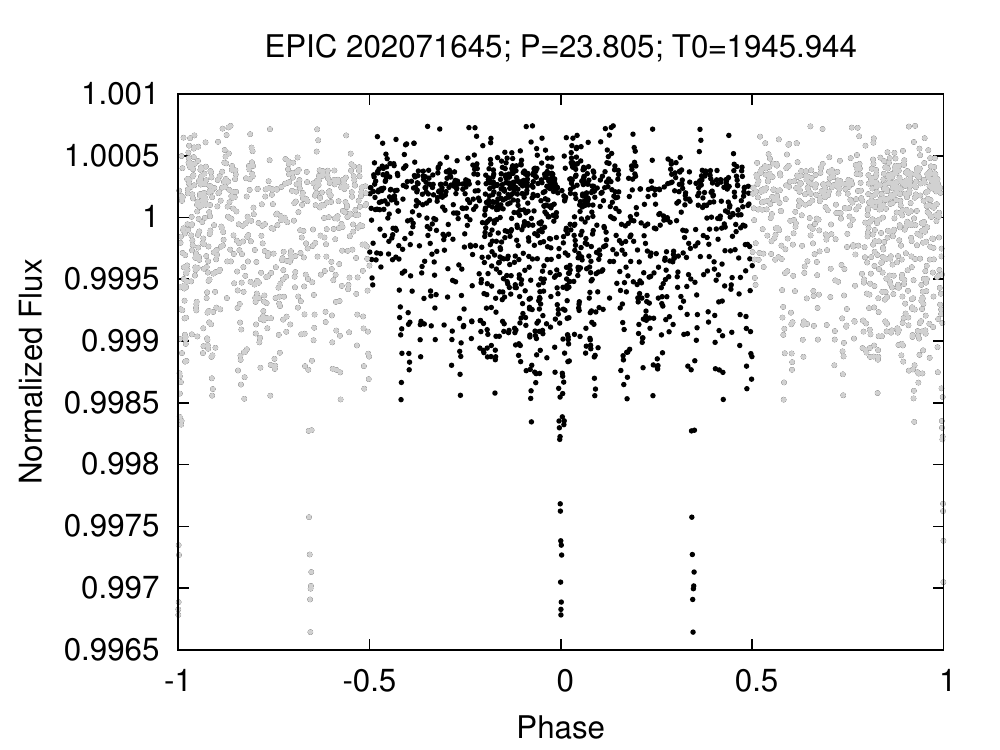} &
\includegraphics[width=0.38\linewidth]{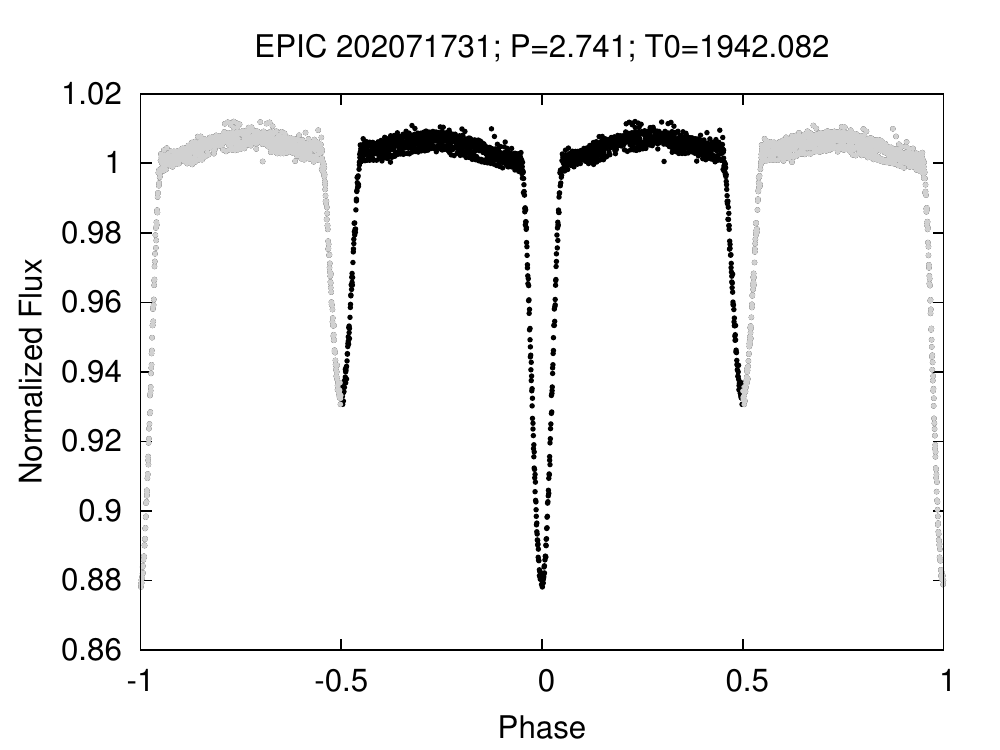} \\
\includegraphics[width=0.38\linewidth]{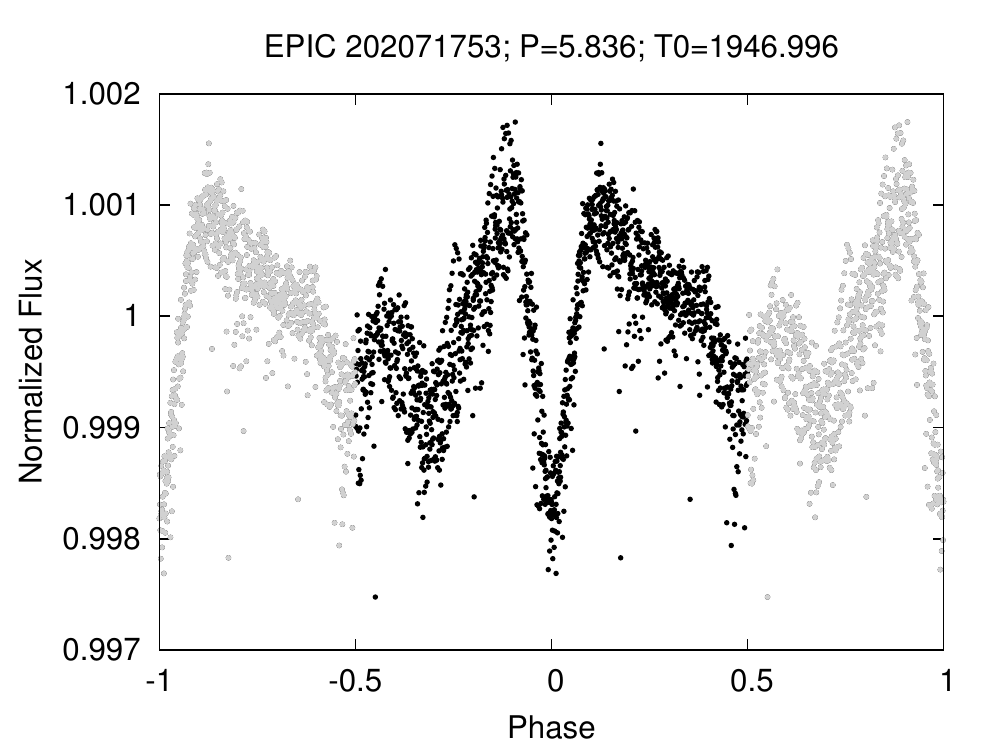} &
\includegraphics[width=0.38\linewidth]{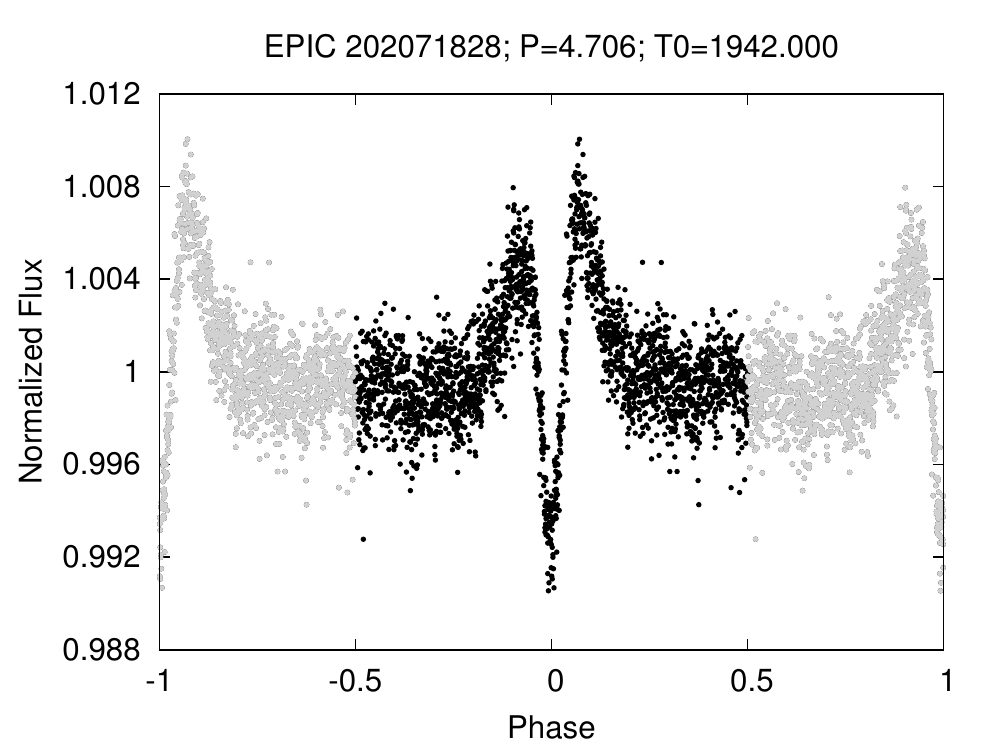} \\
\includegraphics[width=0.38\linewidth]{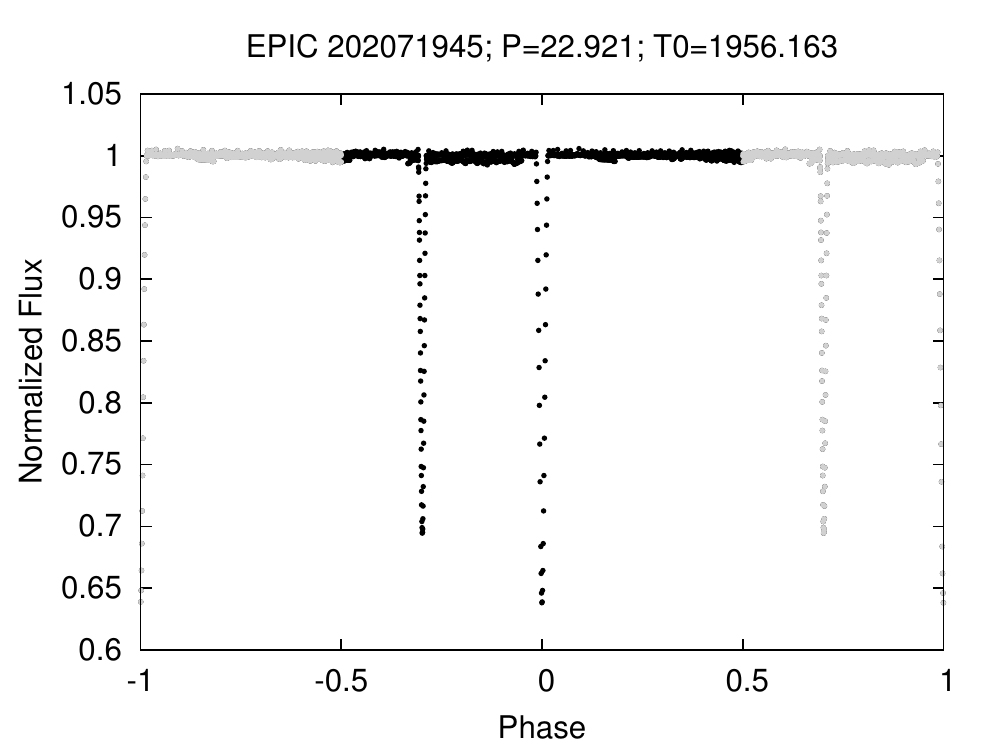} &
\includegraphics[width=0.38\linewidth]{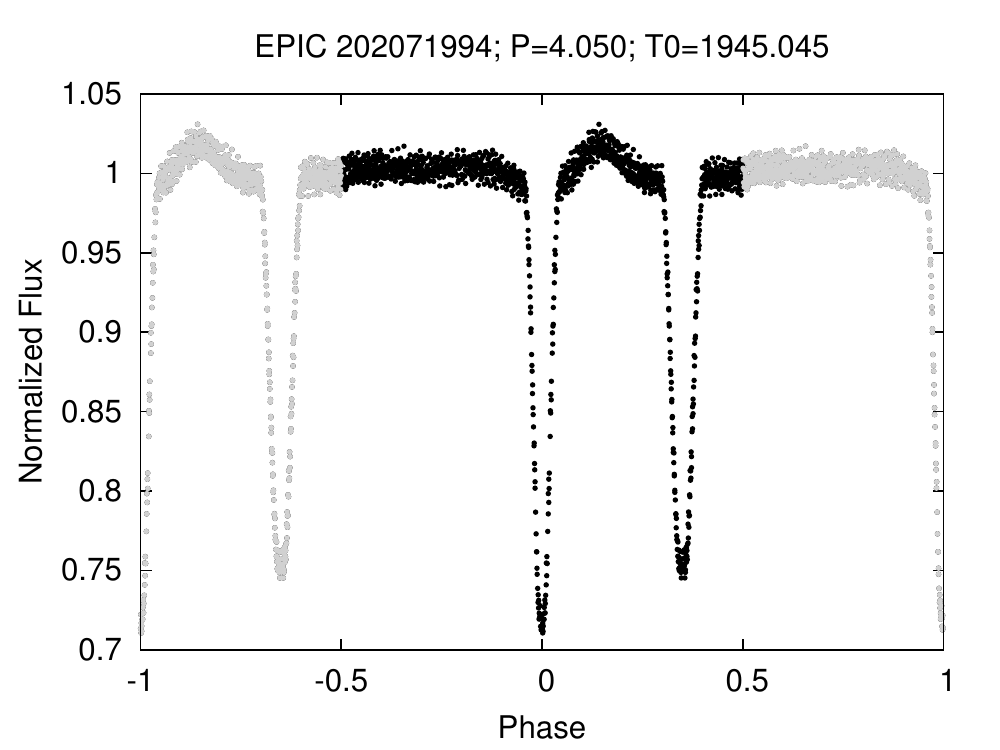} 
\end{tabular}
\end{center}
\caption{Plots of phased light curves for new eclipsing systems. Ephemerides are presented in Tables~\ref{tab:newEBs}, \ref{tab:m35eb}, and \ref{tab:nonEPIC}.  See Section~\ref{sec:pixel_level_analysis} for details.}
\label{fig:phasedplots_4}
\end{figure*}

\begin{figure*}
\begin{center}
\begin{tabular}{cc}
\includegraphics[width=0.38\linewidth]{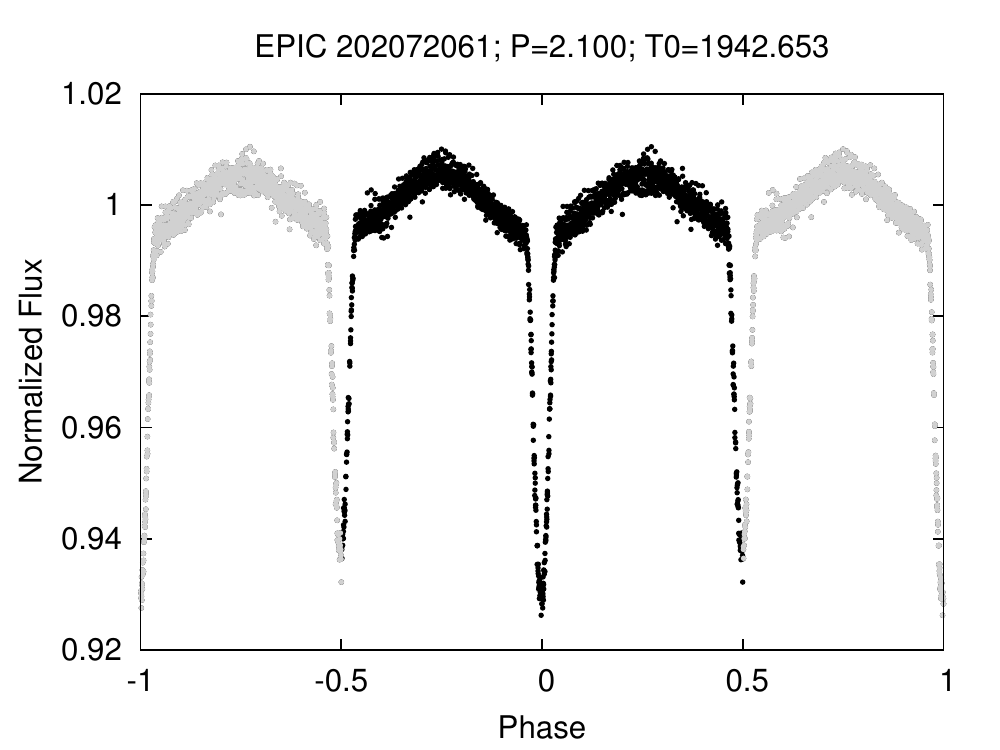} &
\includegraphics[width=0.38\linewidth]{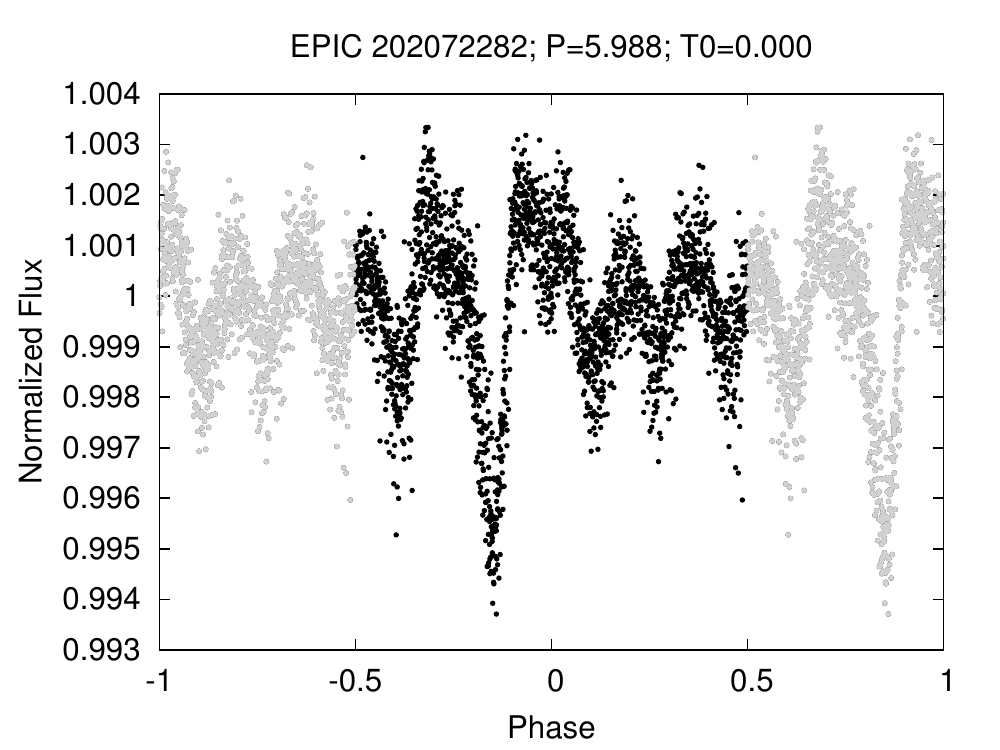} \\
\includegraphics[width=0.38\linewidth]{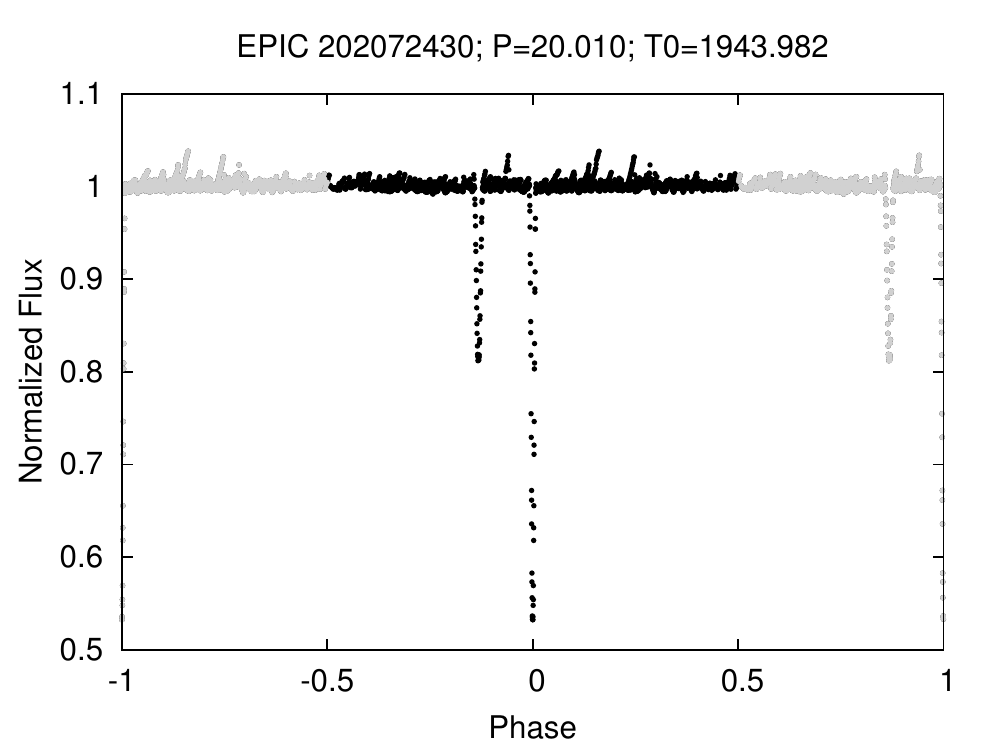} &
\includegraphics[width=0.38\linewidth]{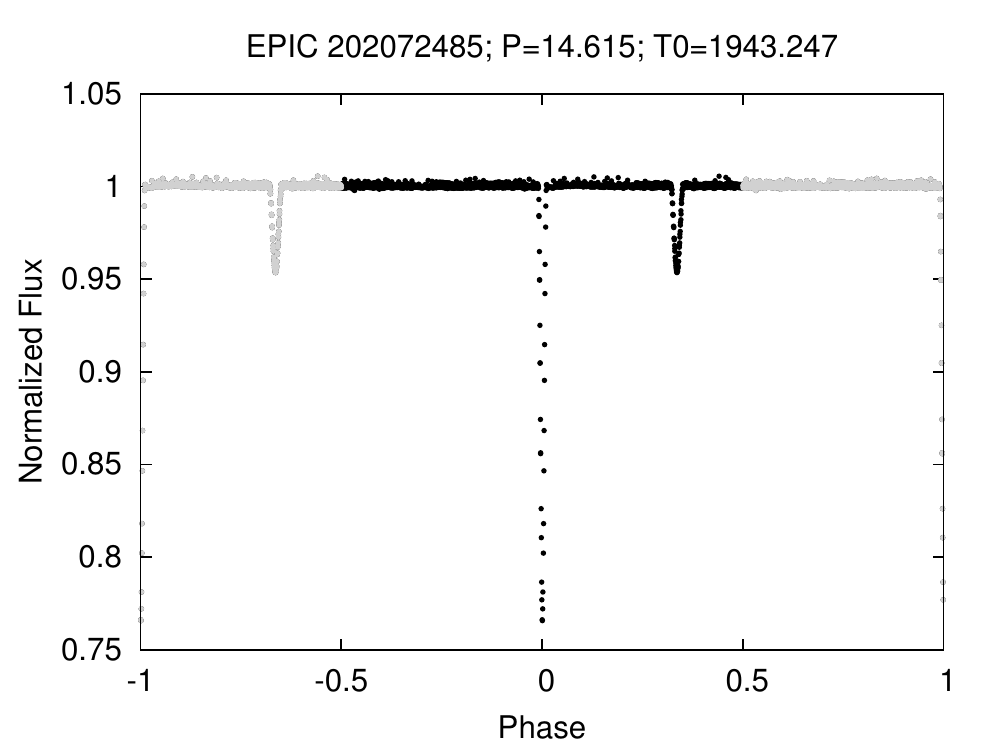} \\
\includegraphics[width=0.38\linewidth]{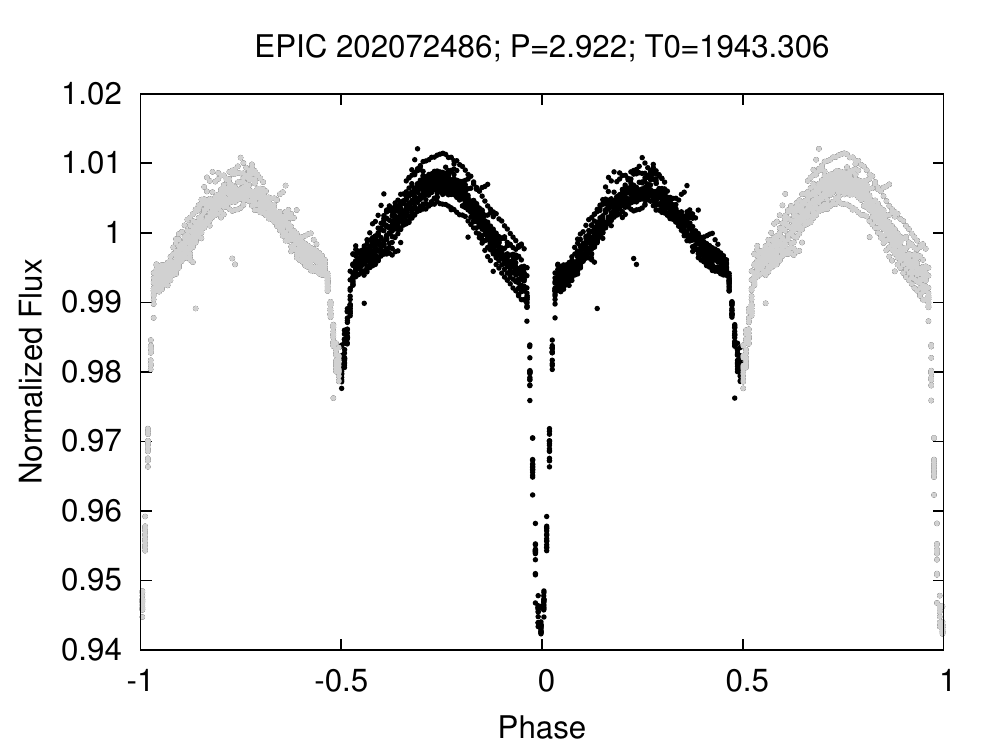} &
\includegraphics[width=0.38\linewidth]{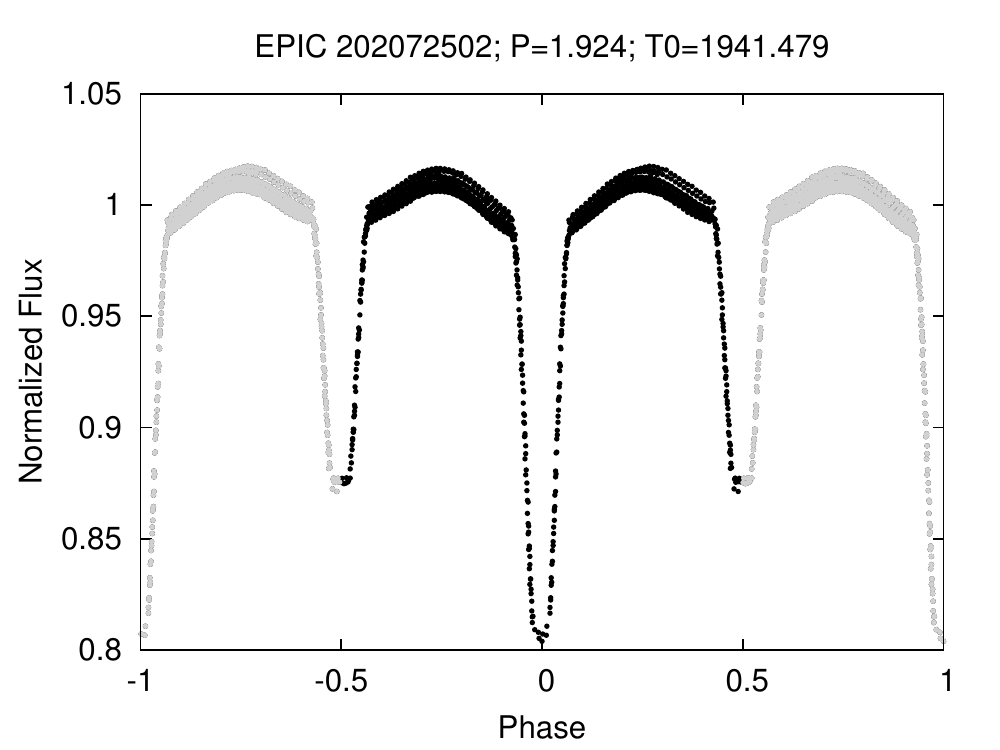} \\
\includegraphics[width=0.38\linewidth]{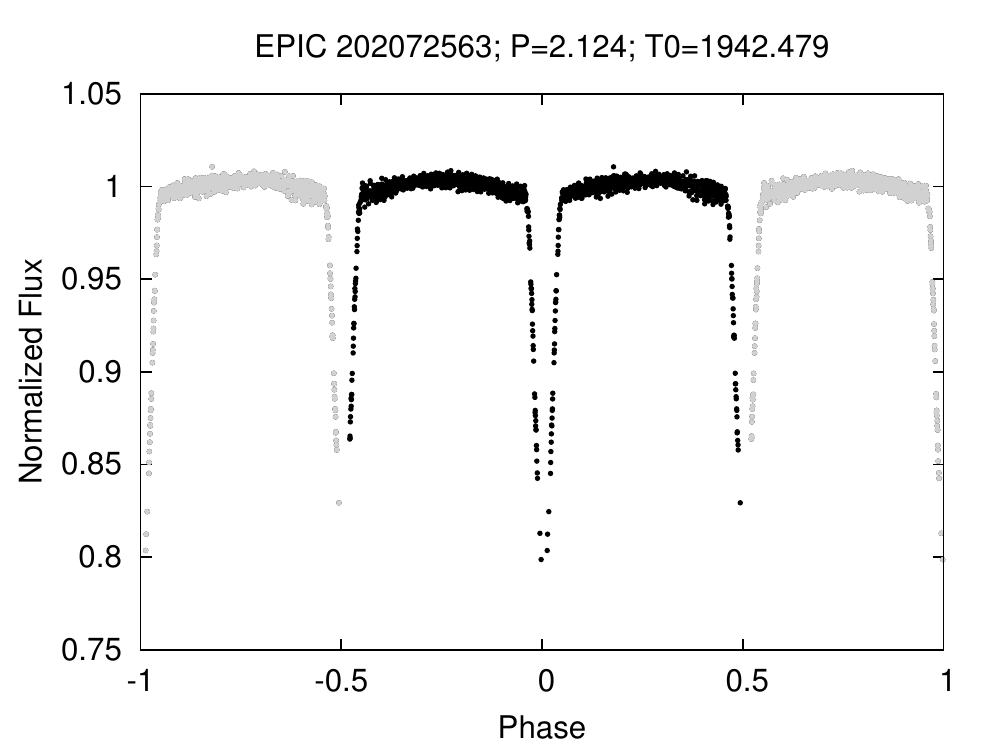} &
\includegraphics[width=0.38\linewidth]{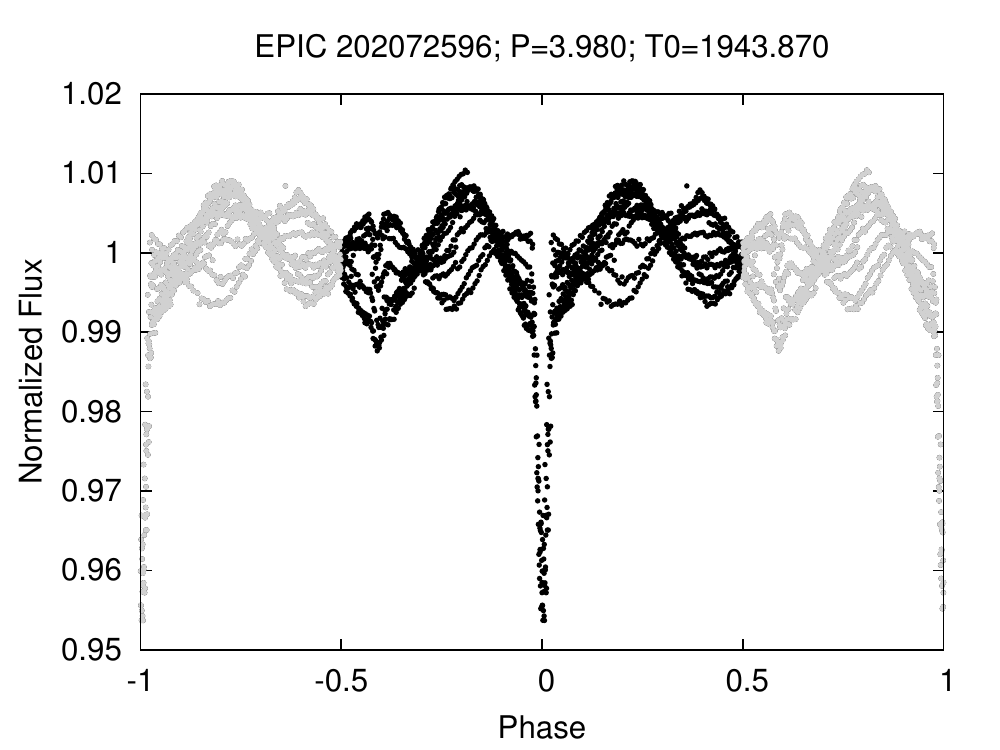} 
\end{tabular}
\end{center}
\caption{Plots of phased light curves for new eclipsing systems. Ephemerides are presented in Tables~\ref{tab:newEBs}, \ref{tab:m35eb}, and \ref{tab:nonEPIC}.  See Section~\ref{sec:pixel_level_analysis} for details.}
\label{fig:phasedplots_5}
\end{figure*}

\begin{figure*}
\begin{center}
\begin{tabular}{cc}
\includegraphics[width=0.38\linewidth]{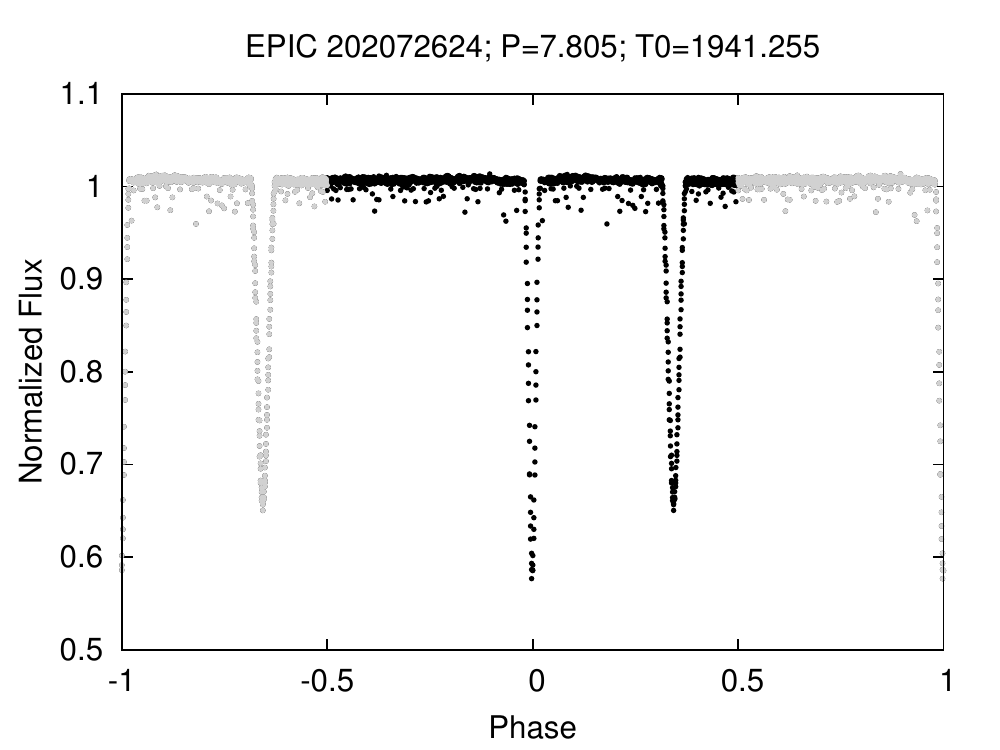} &
\includegraphics[width=0.38\linewidth]{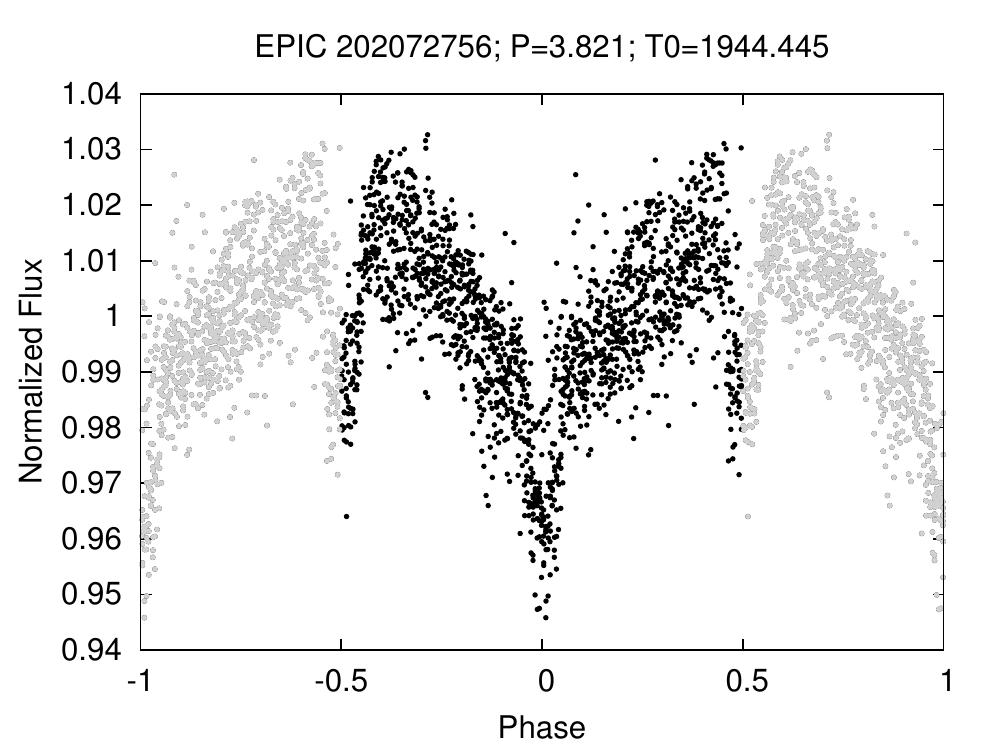} \\
\includegraphics[width=0.38\linewidth]{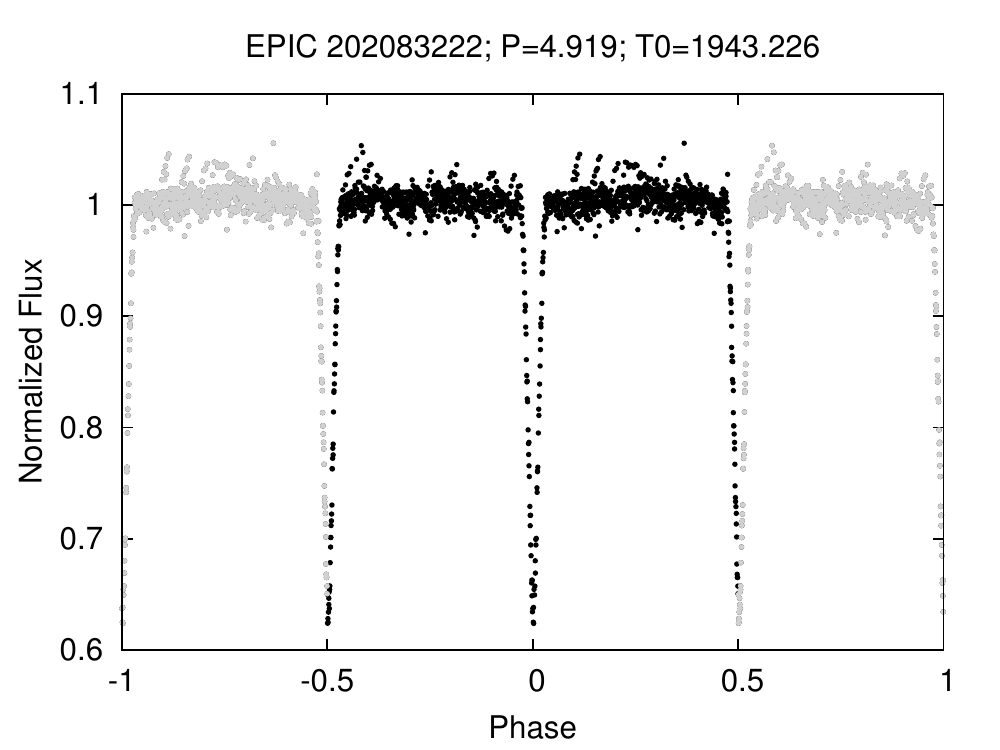} &
\includegraphics[width=0.38\linewidth]{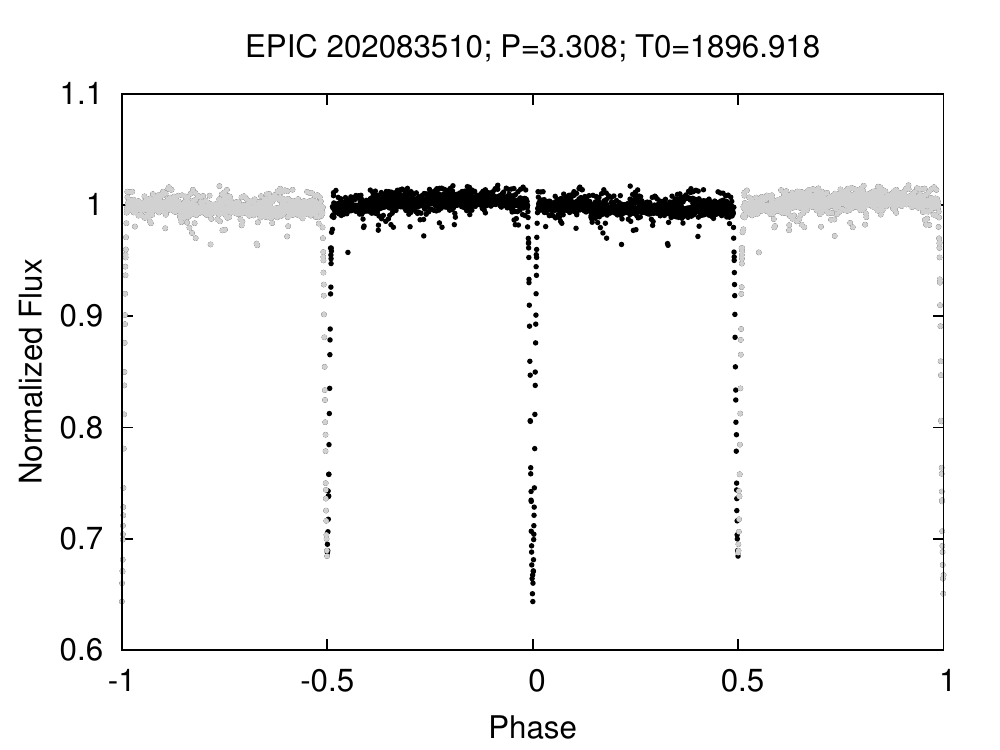} \\
\includegraphics[width=0.38\linewidth]{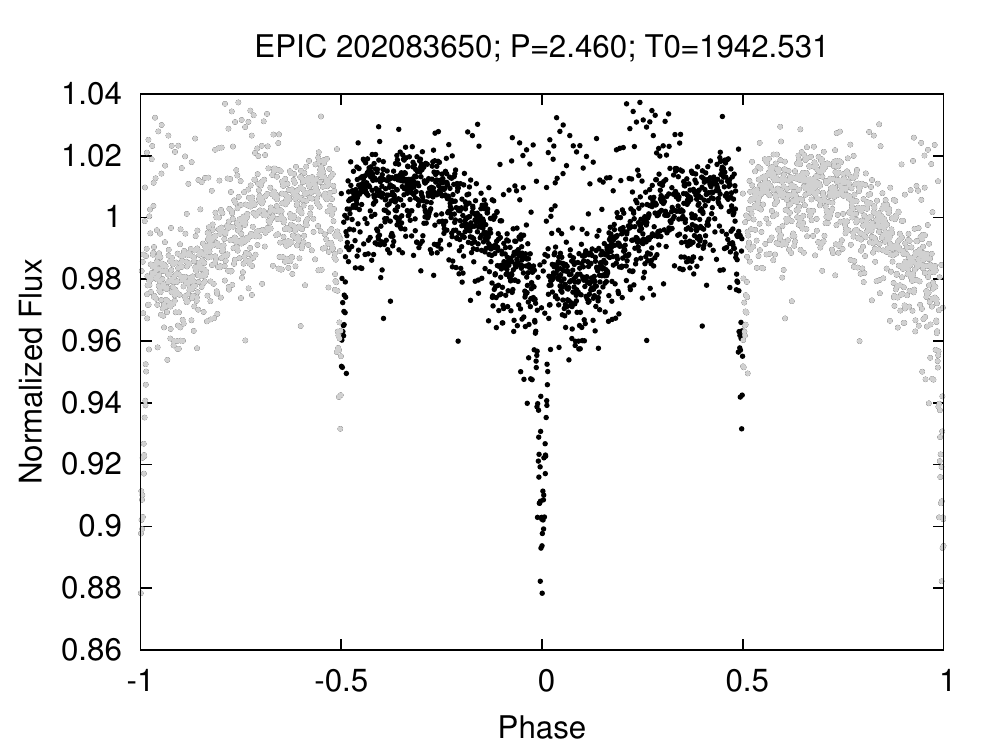} &
\includegraphics[width=0.38\linewidth]{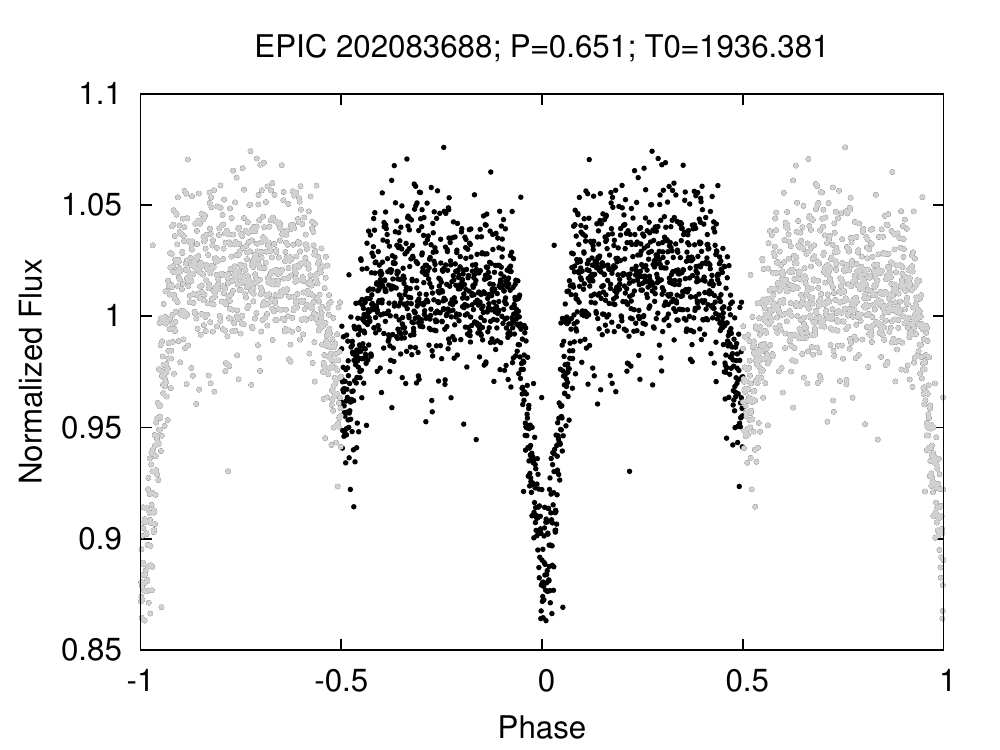} \\
\includegraphics[width=0.38\linewidth]{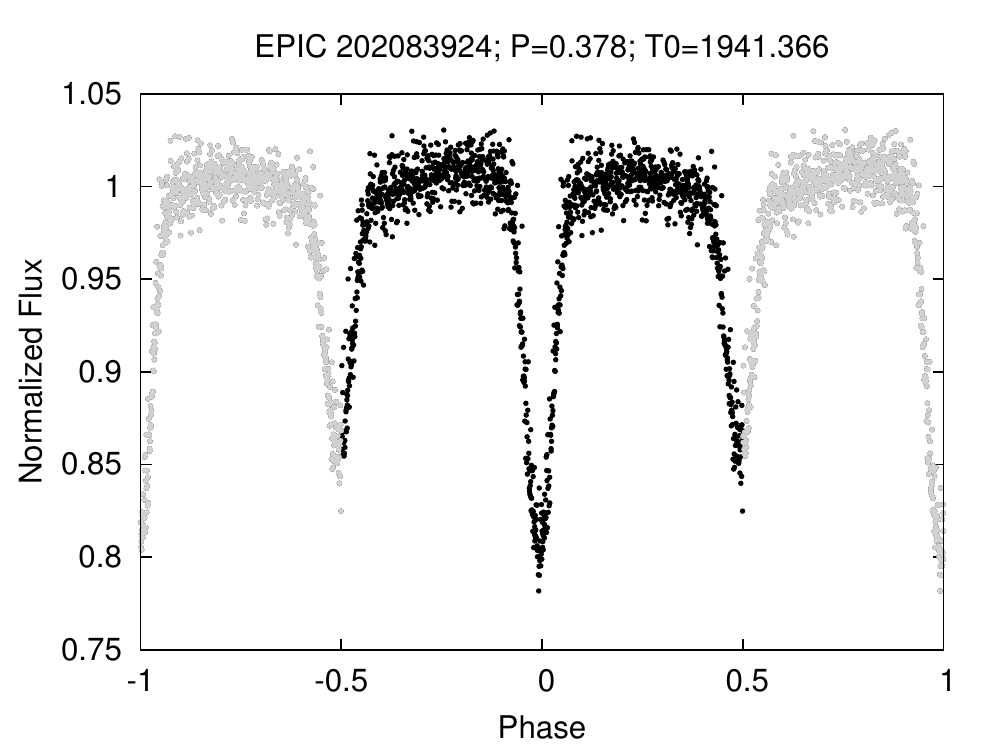} &
\includegraphics[width=0.38\linewidth]{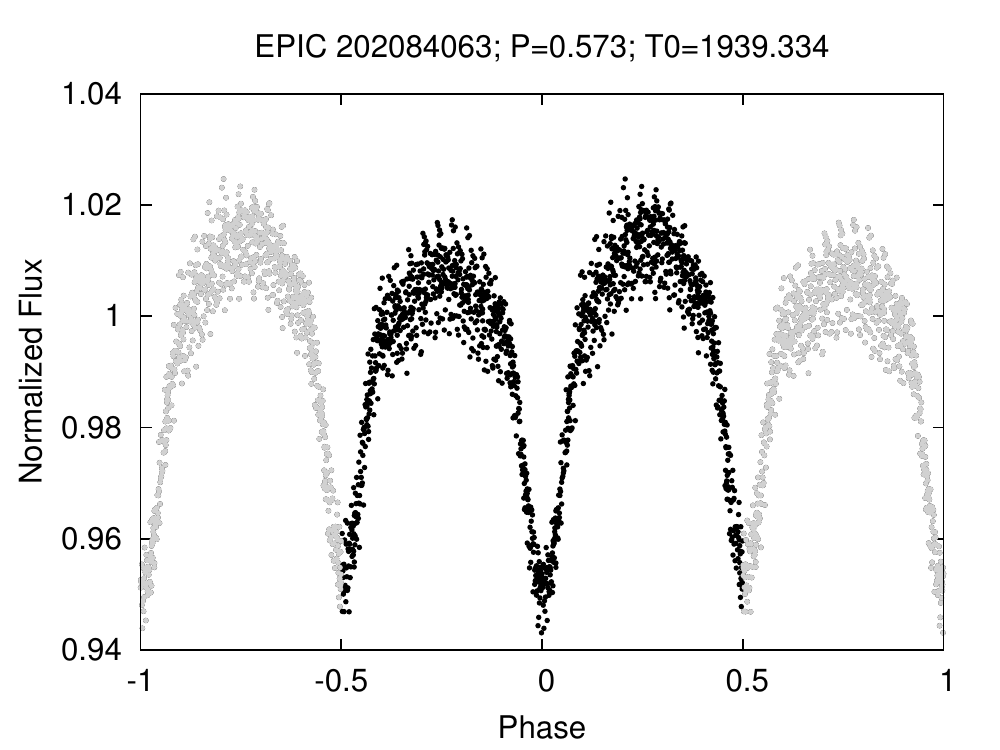} 
\end{tabular}
\end{center}
\caption{Plots of phased light curves for new eclipsing systems. Ephemerides are presented in Tables~\ref{tab:newEBs}, \ref{tab:m35eb}, and \ref{tab:nonEPIC}.  See Section~\ref{sec:pixel_level_analysis} for details.}
\label{fig:phasedplots_6}
\end{figure*}

\begin{figure*}
\begin{center}
\begin{tabular}{cc}
\includegraphics[width=0.38\linewidth]{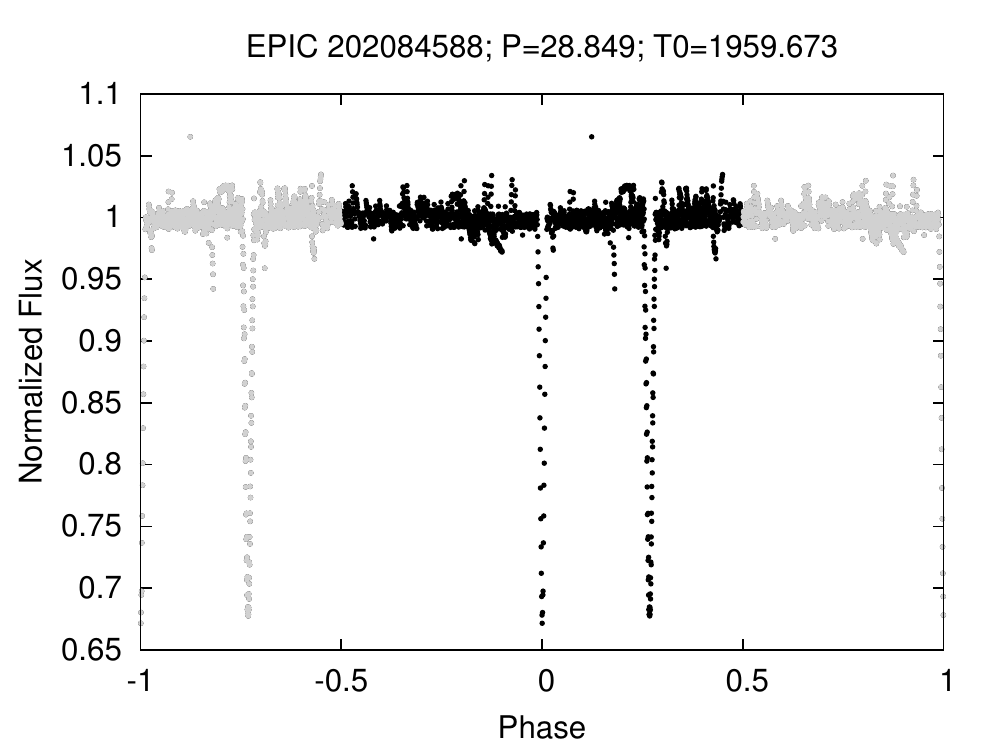} &
\includegraphics[width=0.38\linewidth]{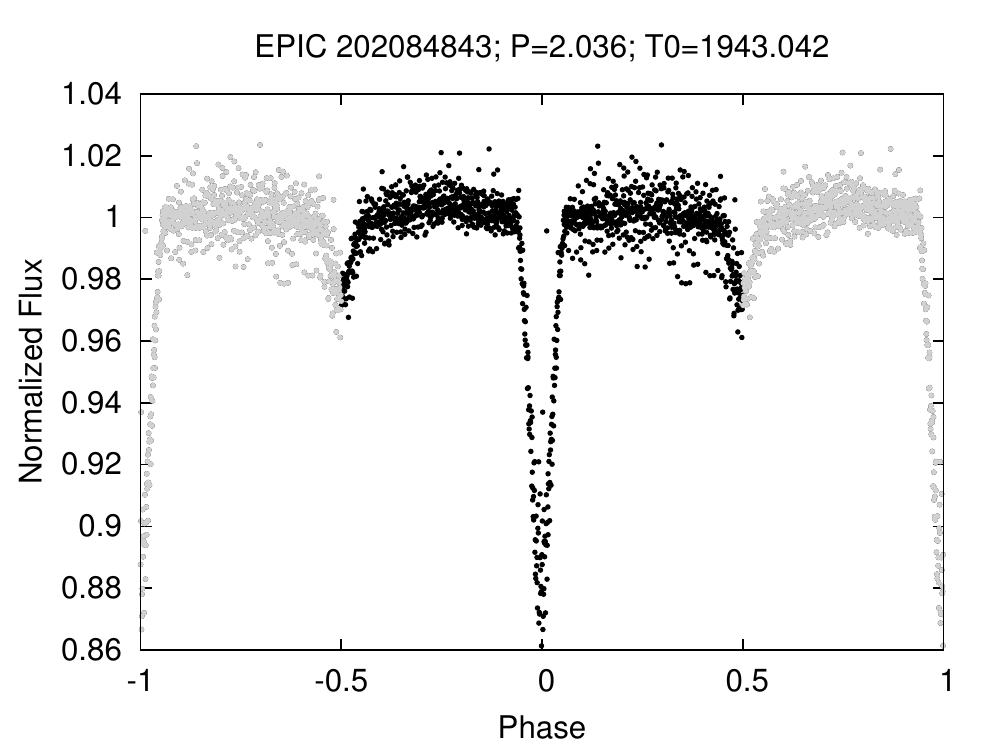} \\
\includegraphics[width=0.38\linewidth]{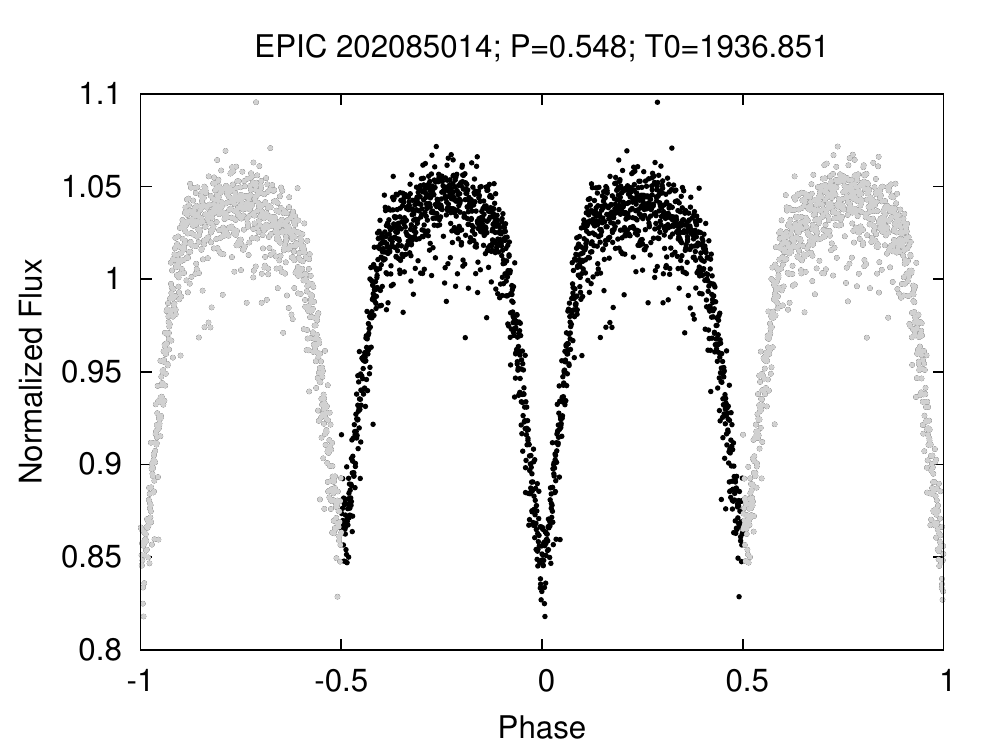} &
\includegraphics[width=0.38\linewidth]{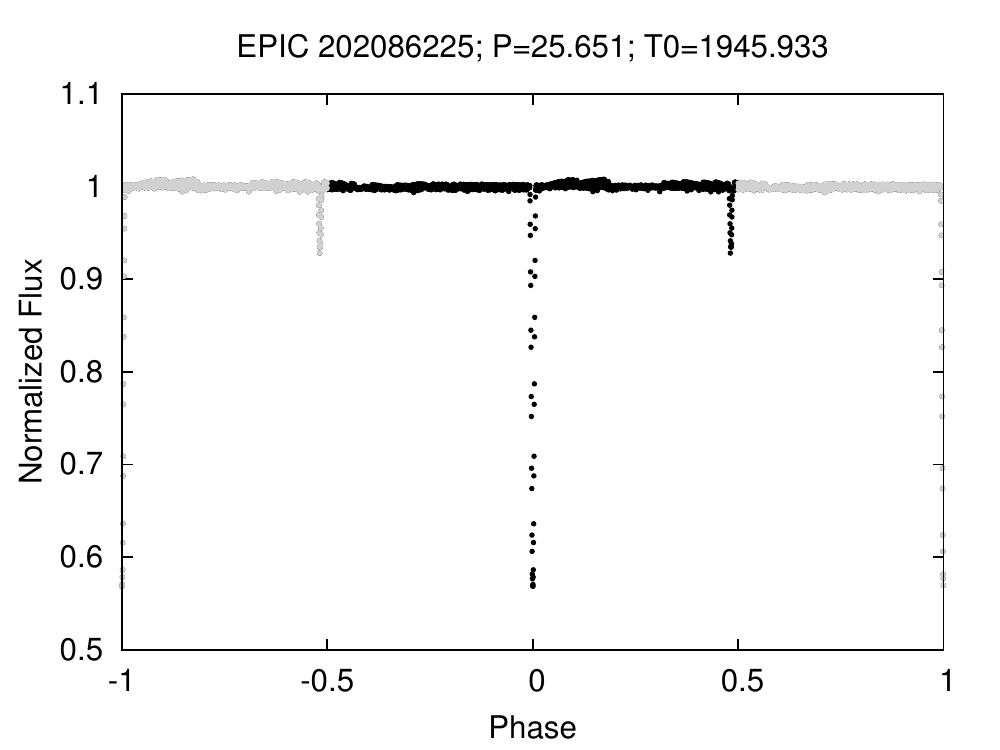} \\
\includegraphics[width=0.38\linewidth]{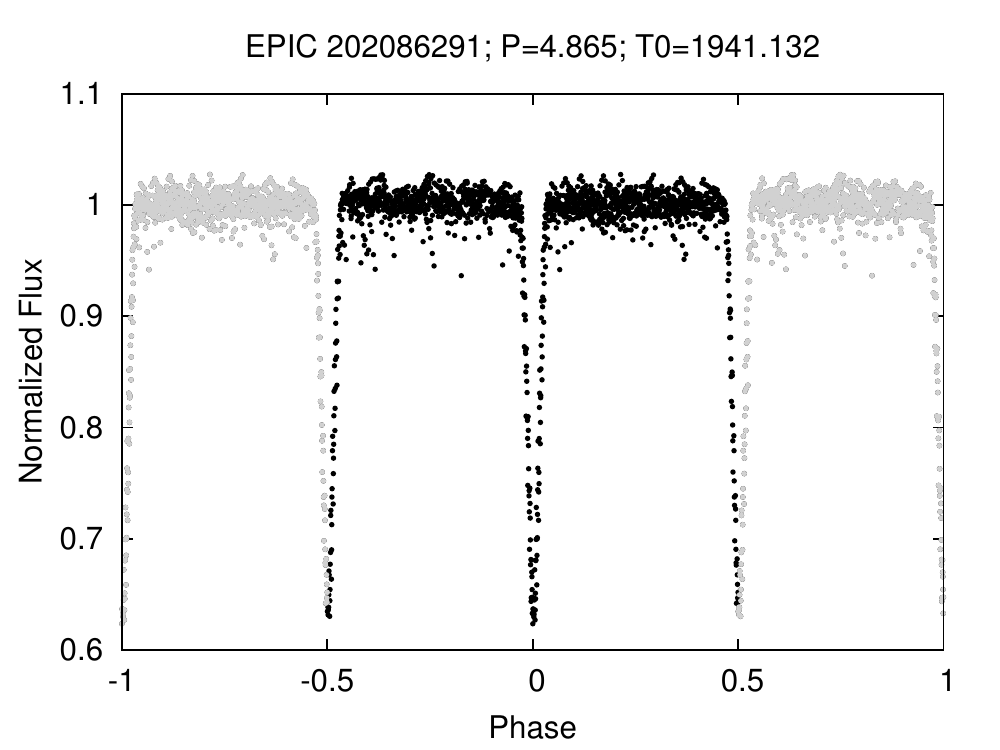} &
\includegraphics[width=0.38\linewidth]{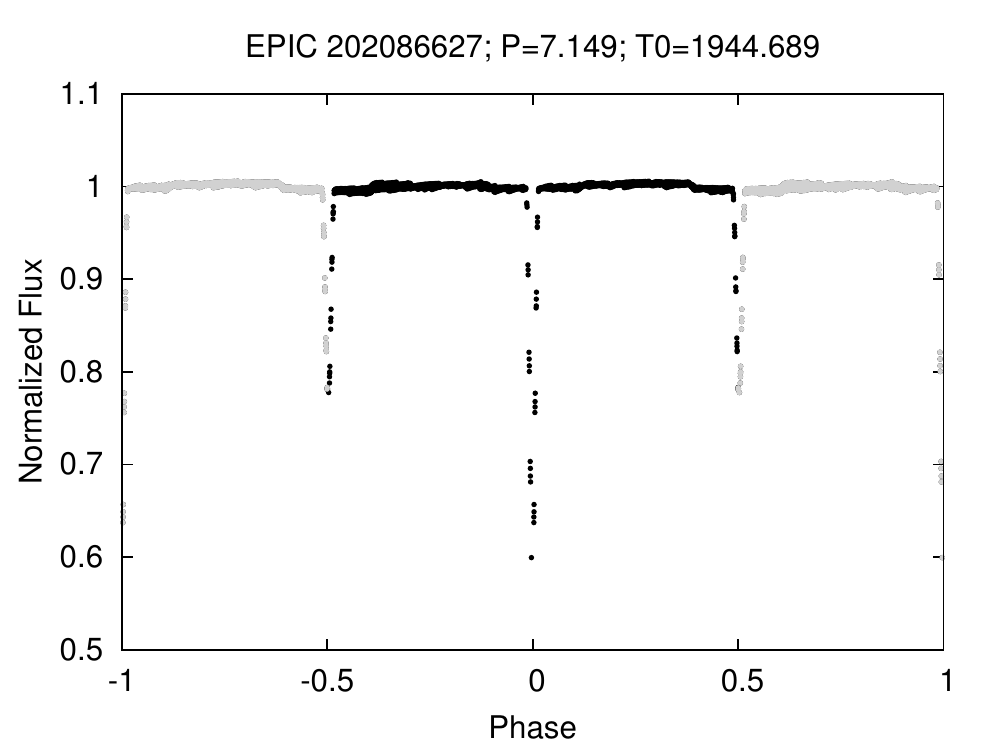} \\
\includegraphics[width=0.38\linewidth]{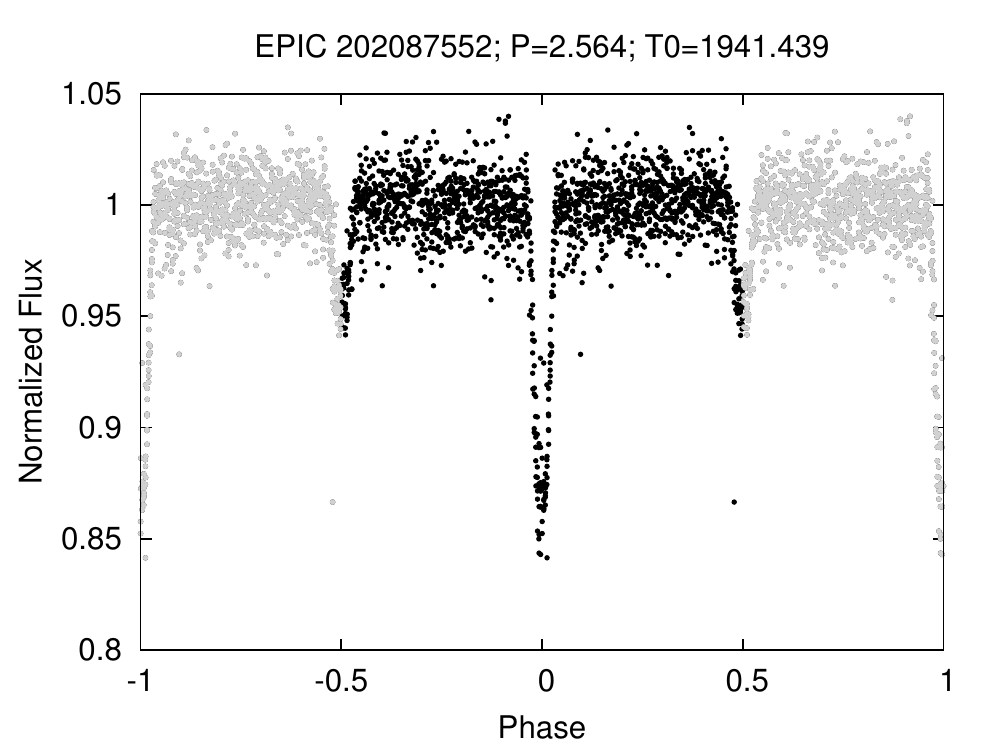} &
\includegraphics[width=0.38\linewidth]{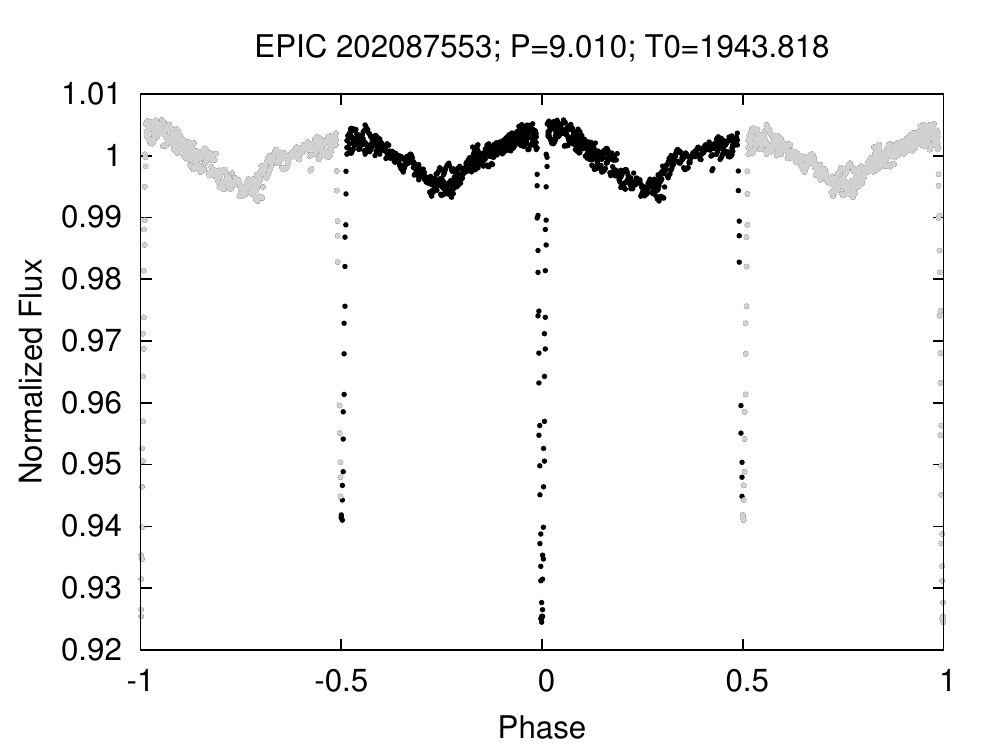} 
\end{tabular}
\end{center}
\caption{Plots of phased light curves for new eclipsing systems. Ephemerides are presented in Tables~\ref{tab:newEBs}, \ref{tab:m35eb}, and \ref{tab:nonEPIC}.  See Section~\ref{sec:pixel_level_analysis} for details.}
\label{fig:phasedplots_7}
\end{figure*}

\begin{figure*}
\begin{center}
\begin{tabular}{cc}
\includegraphics[width=0.38\linewidth]{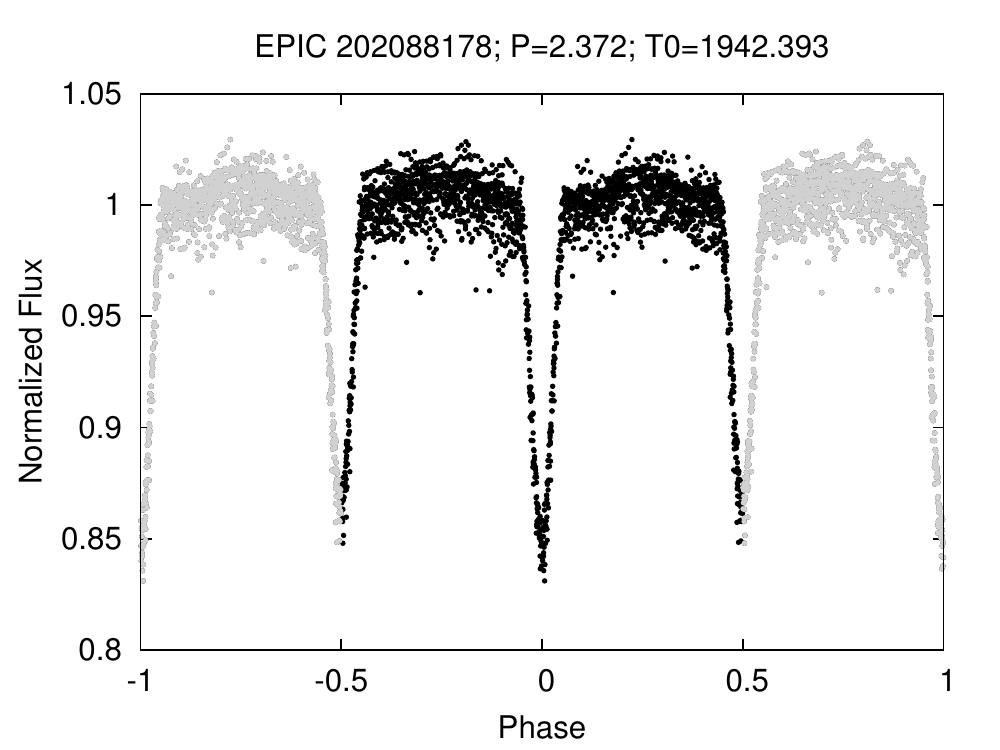} &
\includegraphics[width=0.38\linewidth]{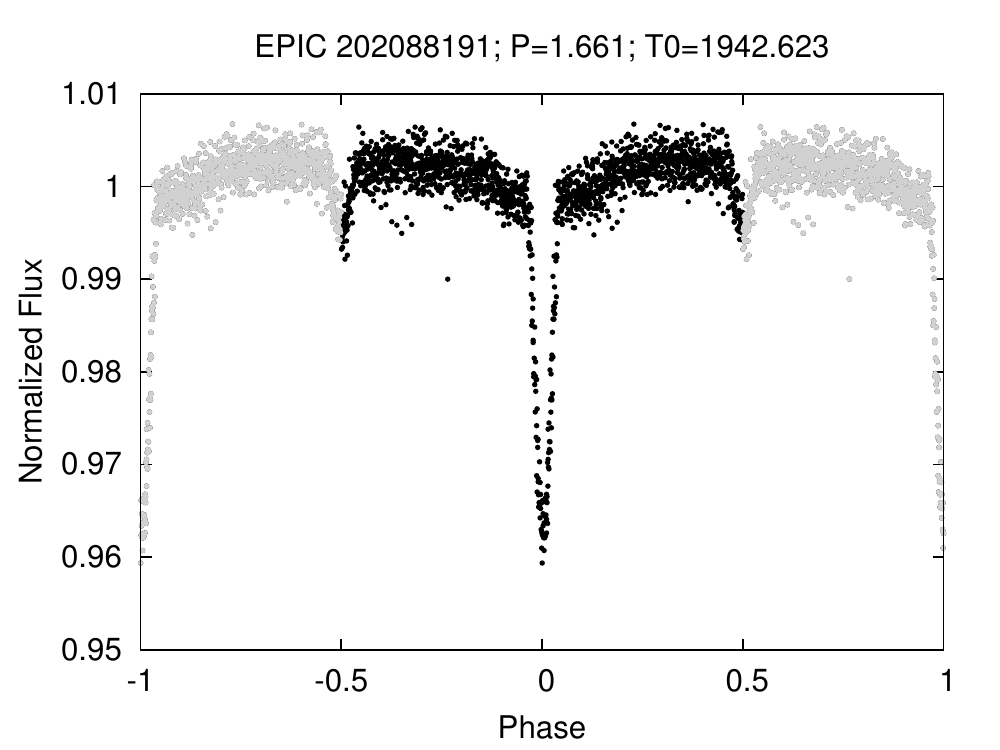} \\
\includegraphics[width=0.38\linewidth]{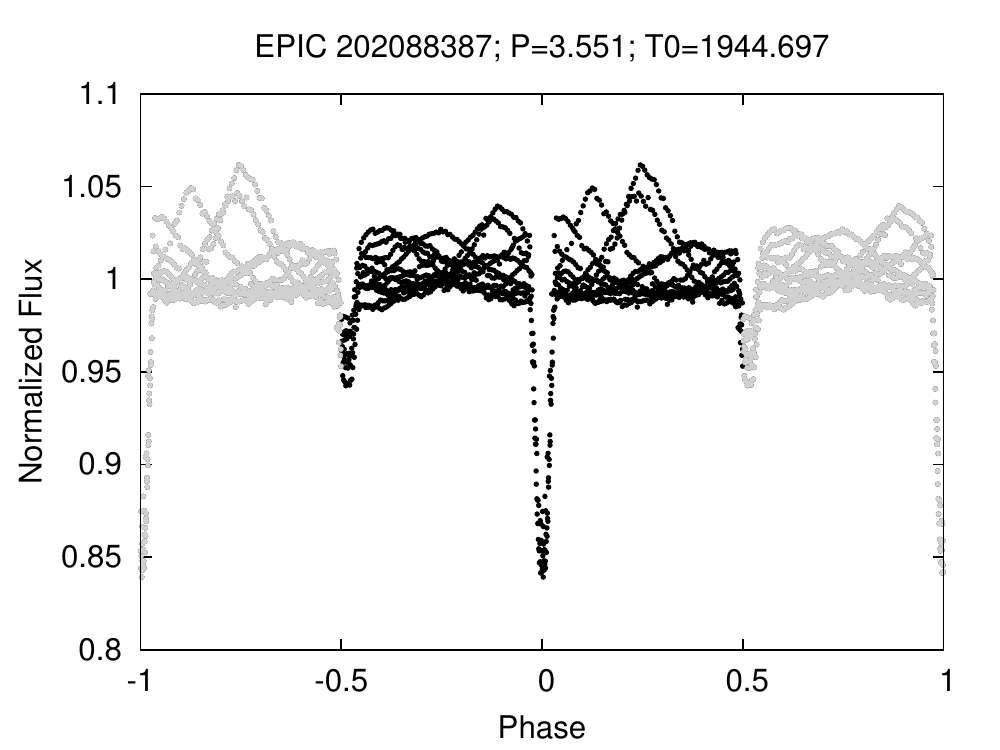} &
\includegraphics[width=0.38\linewidth]{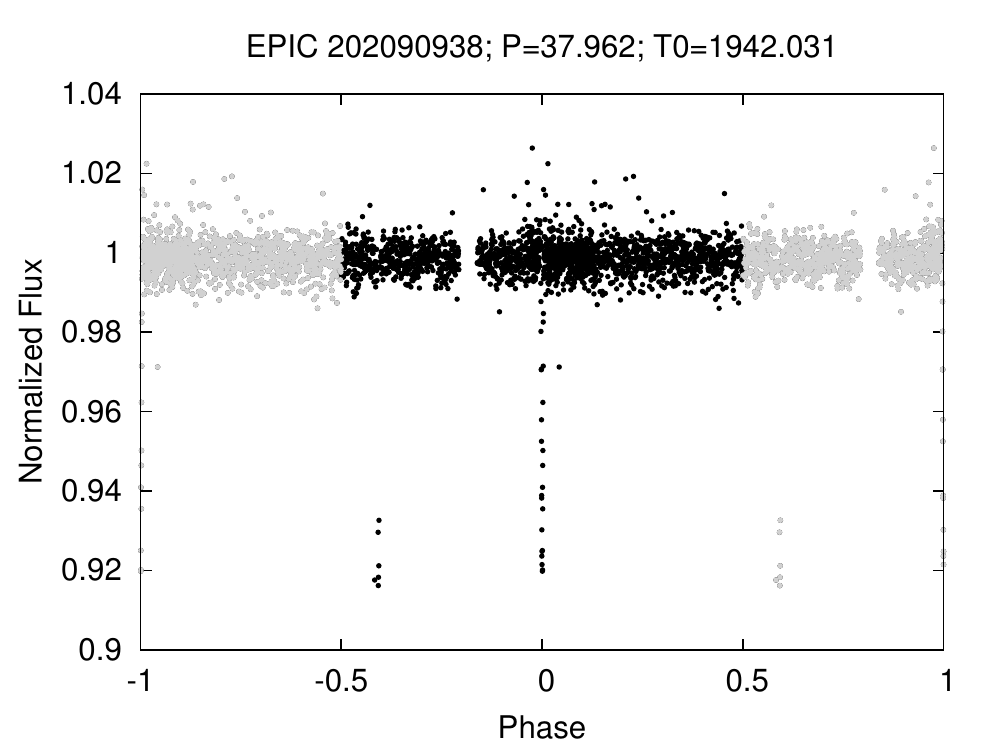} \\
\includegraphics[width=0.38\linewidth]{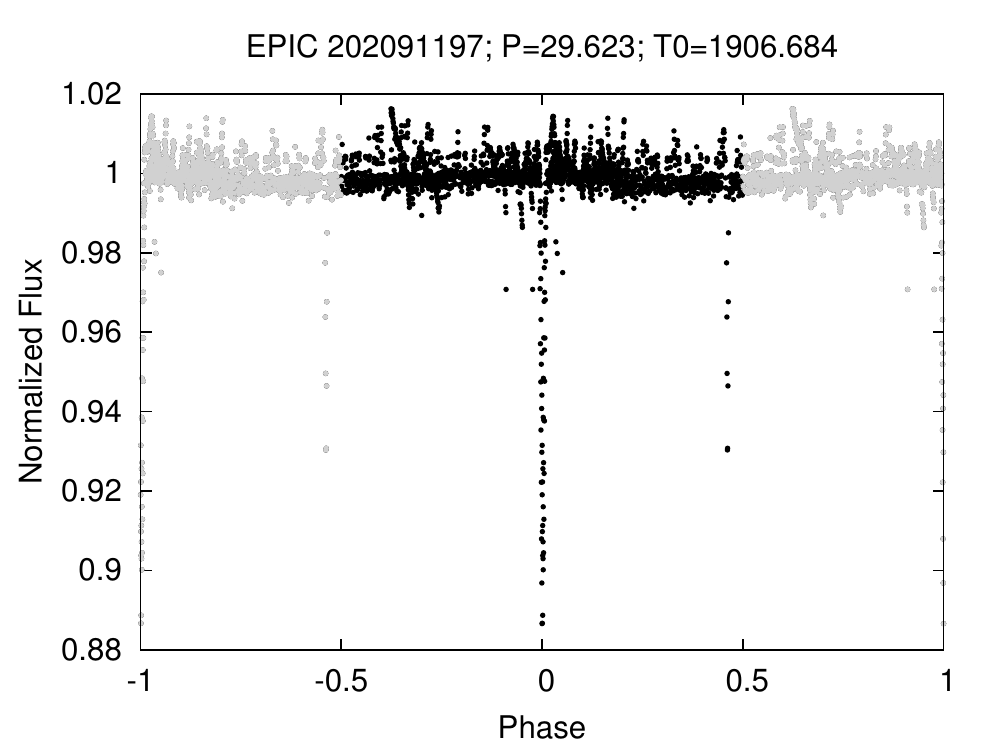} &
\includegraphics[width=0.38\linewidth]{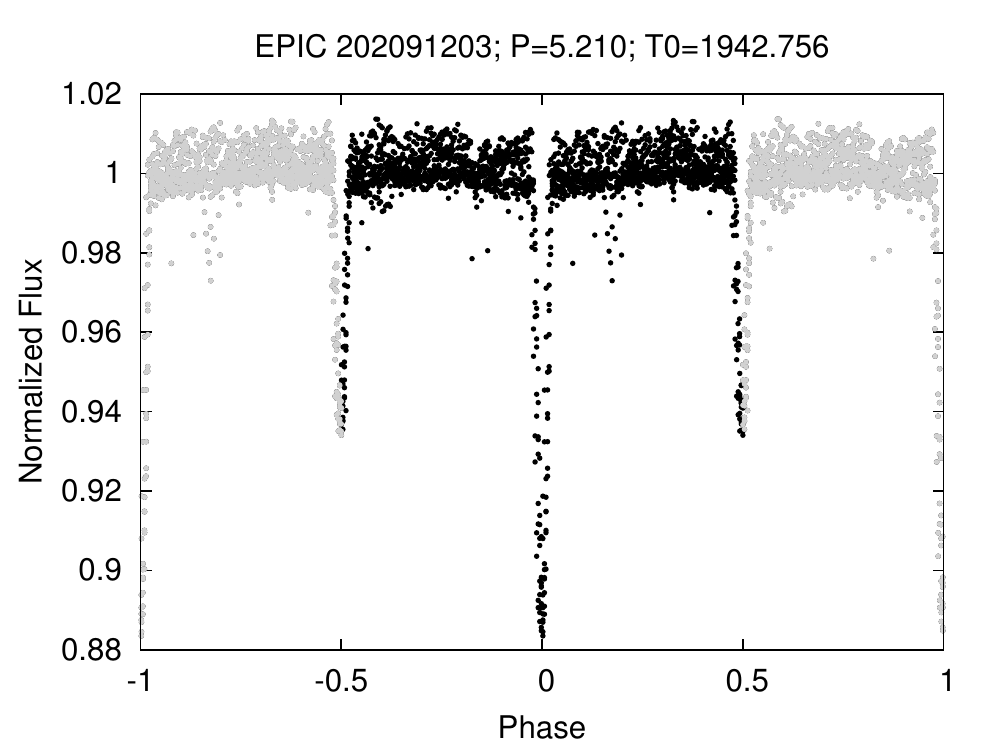} \\
\includegraphics[width=0.38\linewidth]{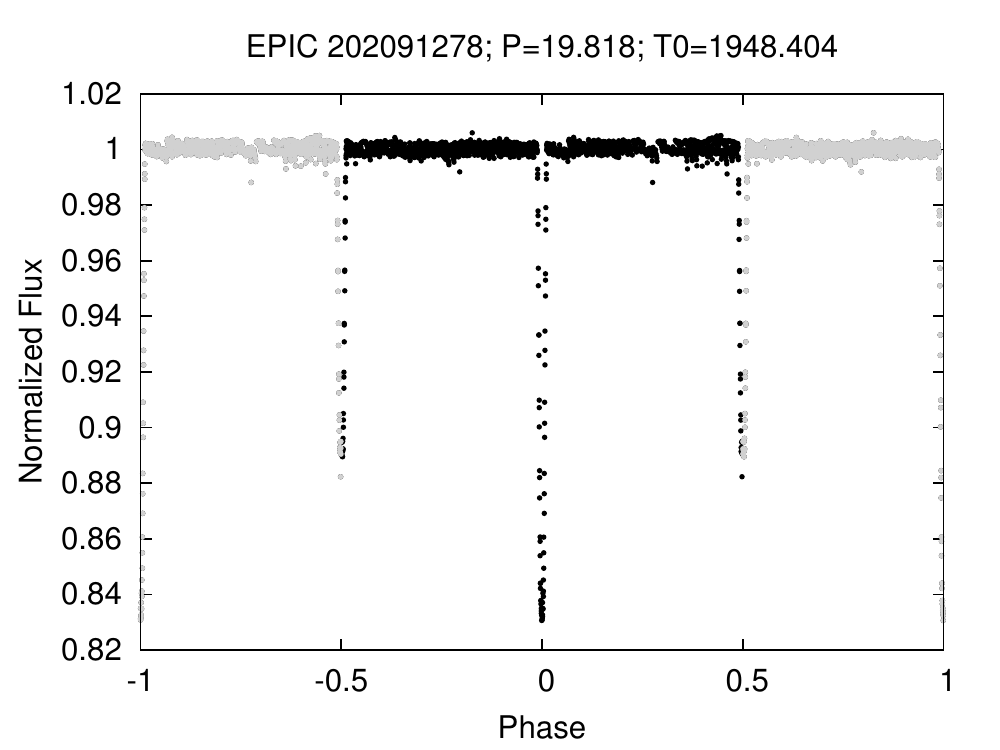} &
\includegraphics[width=0.38\linewidth]{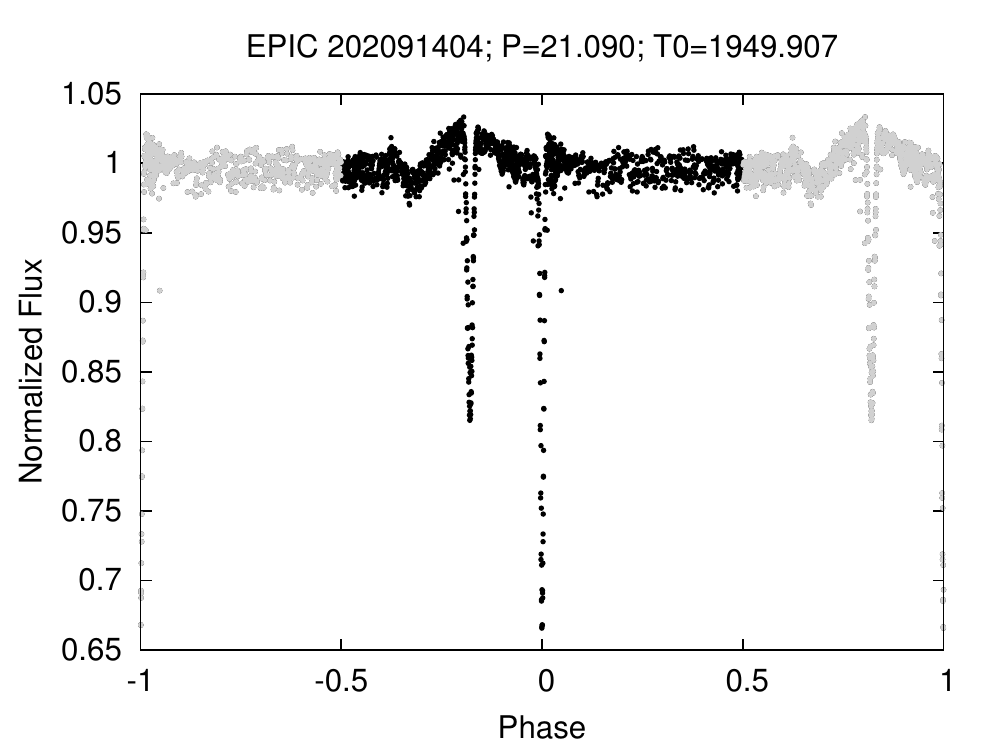} 
\end{tabular}
\end{center}
\caption{Plots of phased light curves for new eclipsing systems. Ephemerides are presented in Tables~\ref{tab:newEBs}, \ref{tab:m35eb}, and \ref{tab:nonEPIC}.  See Section~\ref{sec:pixel_level_analysis} for details.}
\label{fig:phasedplots_8}
\end{figure*}

\begin{figure*}
\begin{center}
\begin{tabular}{cc}
\includegraphics[width=0.38\linewidth]{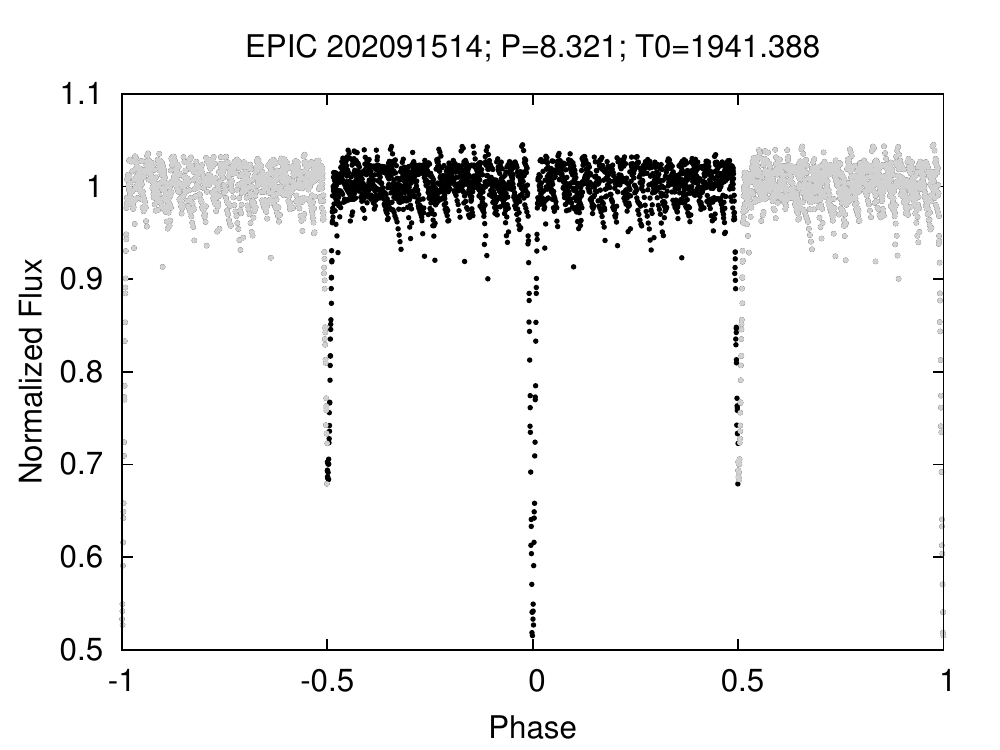} &
\includegraphics[width=0.38\linewidth]{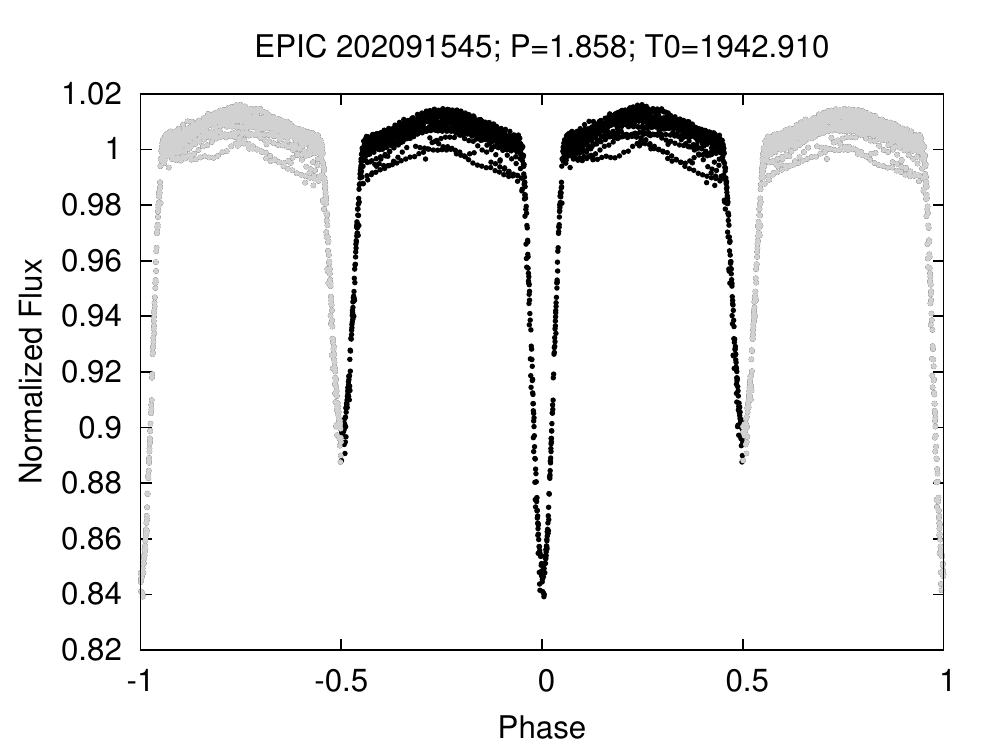} \\
\includegraphics[width=0.38\linewidth]{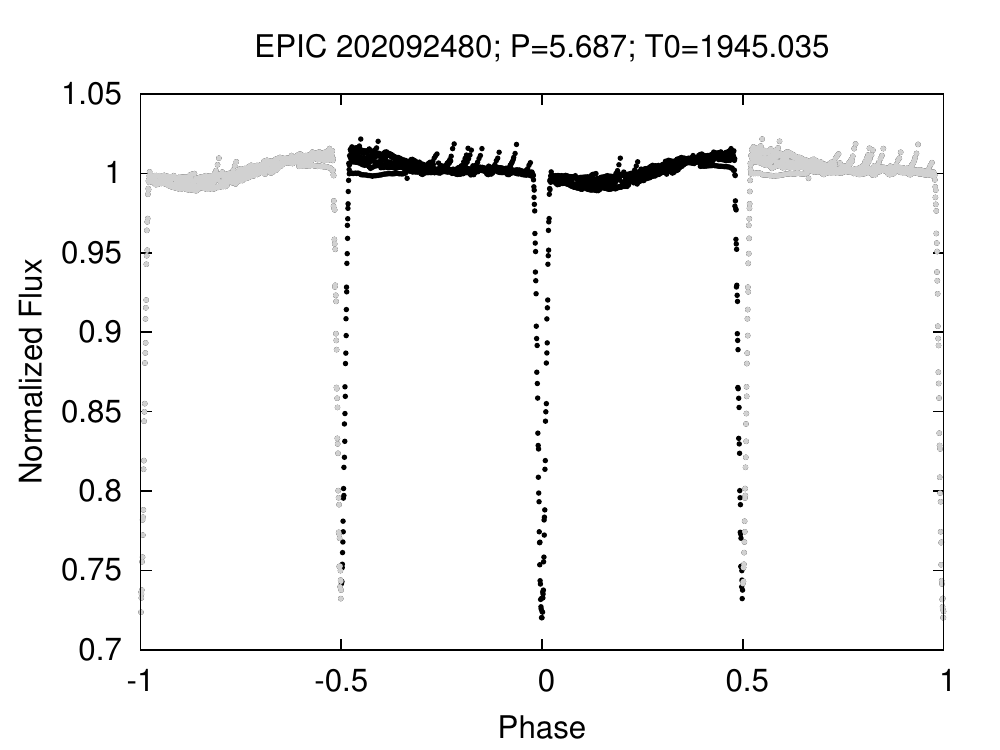} &
\includegraphics[width=0.38\linewidth]{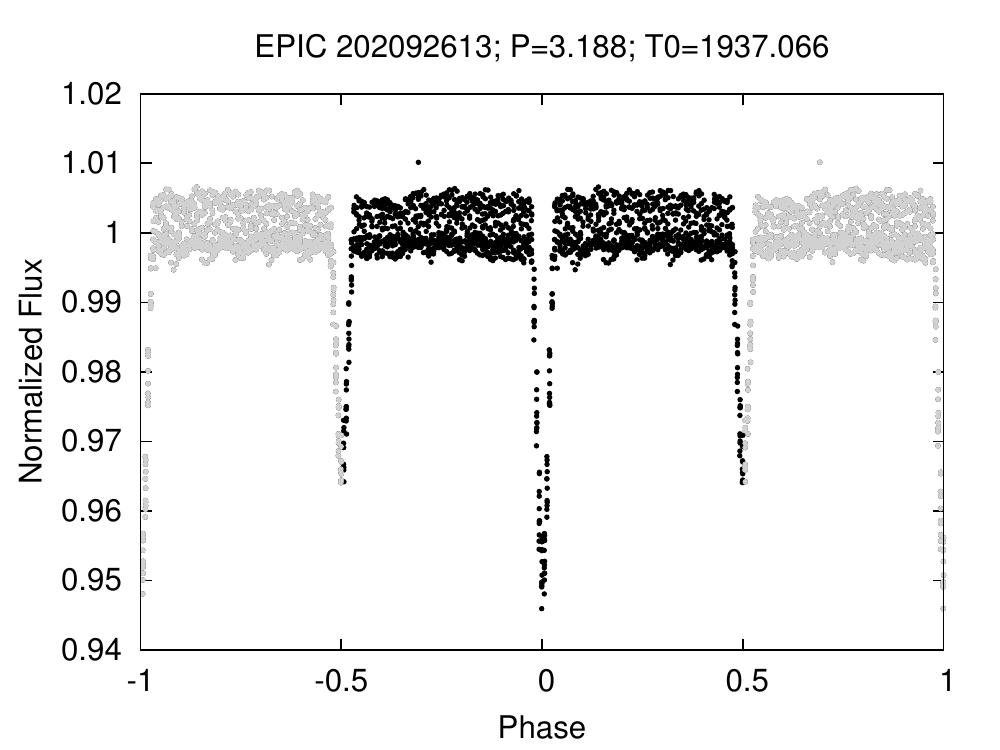} \\
\includegraphics[width=0.38\linewidth]{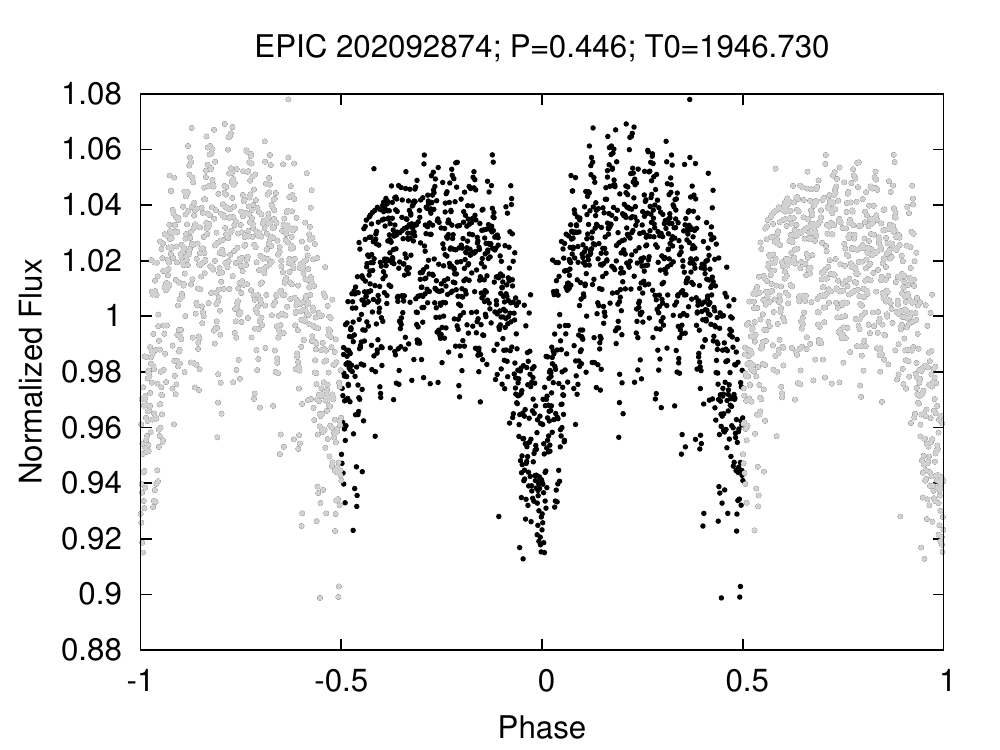} &
\includegraphics[width=0.38\linewidth]{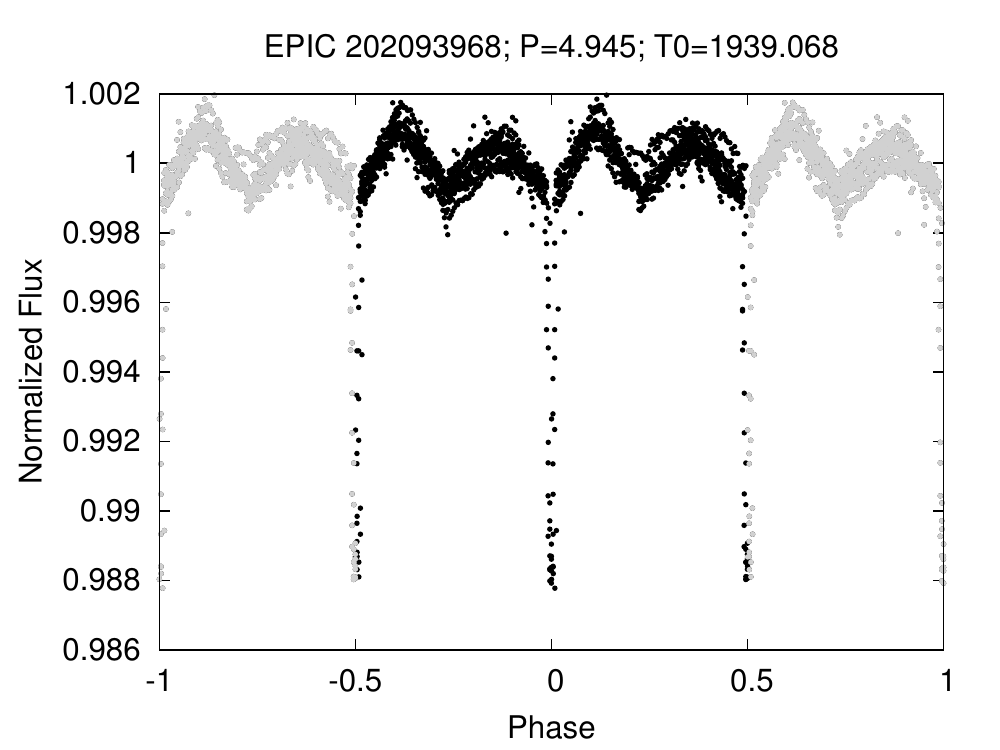} \\
\includegraphics[width=0.38\linewidth]{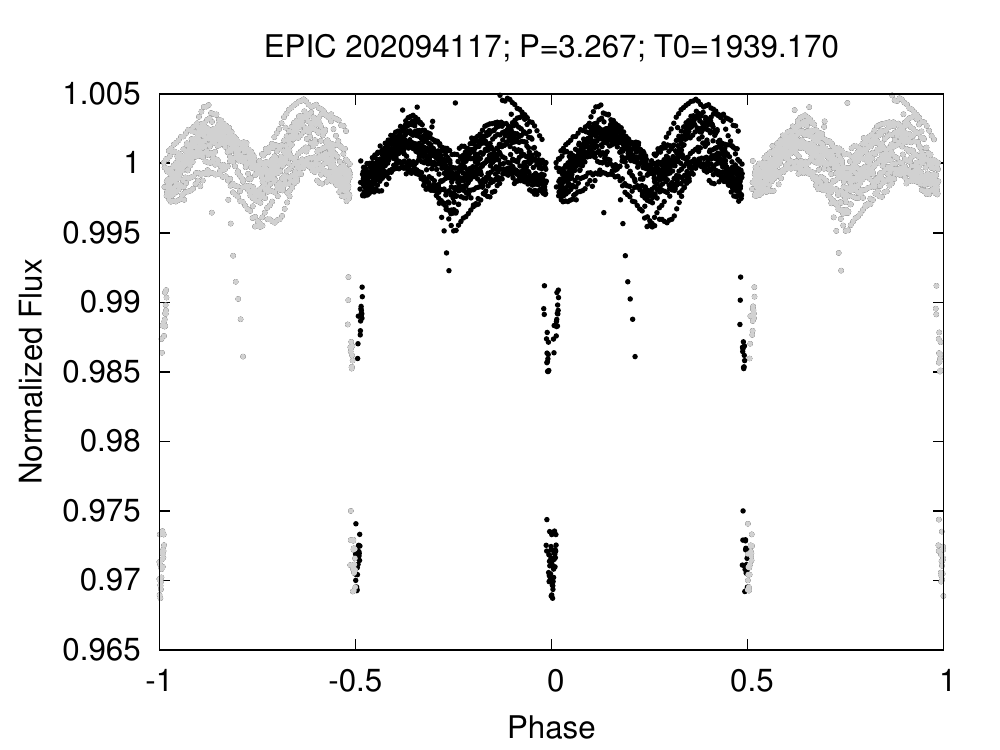} &
\includegraphics[width=0.38\linewidth]{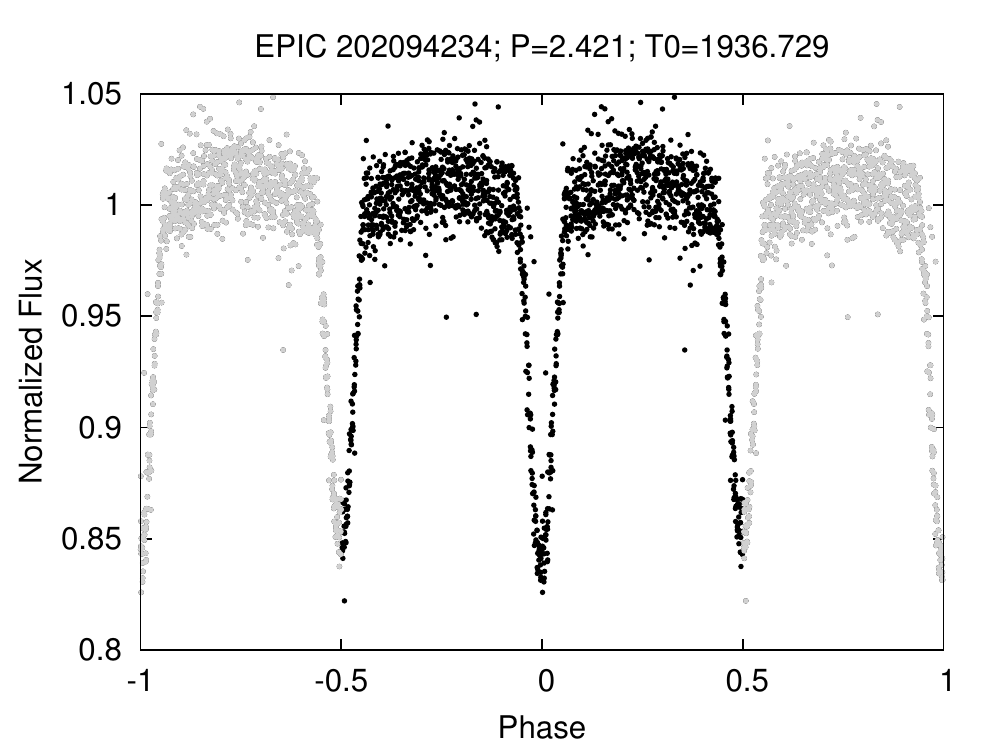} 
\end{tabular}
\end{center}
\caption{Plots of phased light curves for new eclipsing systems. Ephemerides are presented in Tables~\ref{tab:newEBs}, \ref{tab:m35eb}, and \ref{tab:nonEPIC}.  See Section~\ref{sec:pixel_level_analysis} for details.}
\label{fig:phasedplots_9}
\end{figure*}

\begin{figure*}
\begin{center}
\begin{tabular}{cc}
\includegraphics[width=0.38\linewidth]{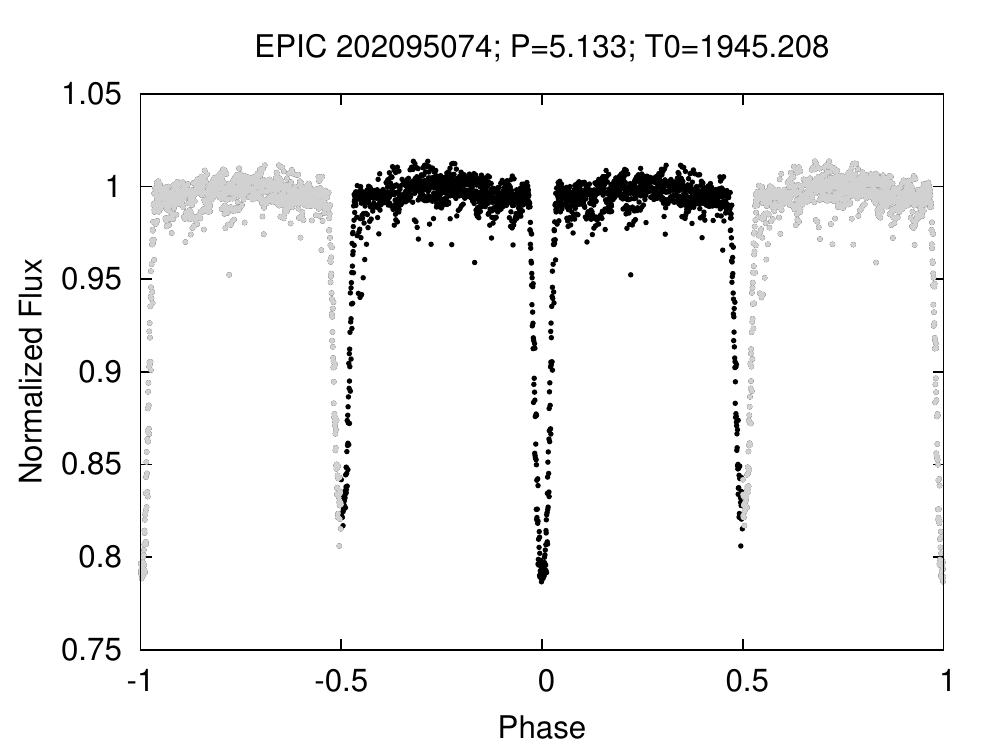} &
\includegraphics[width=0.38\linewidth]{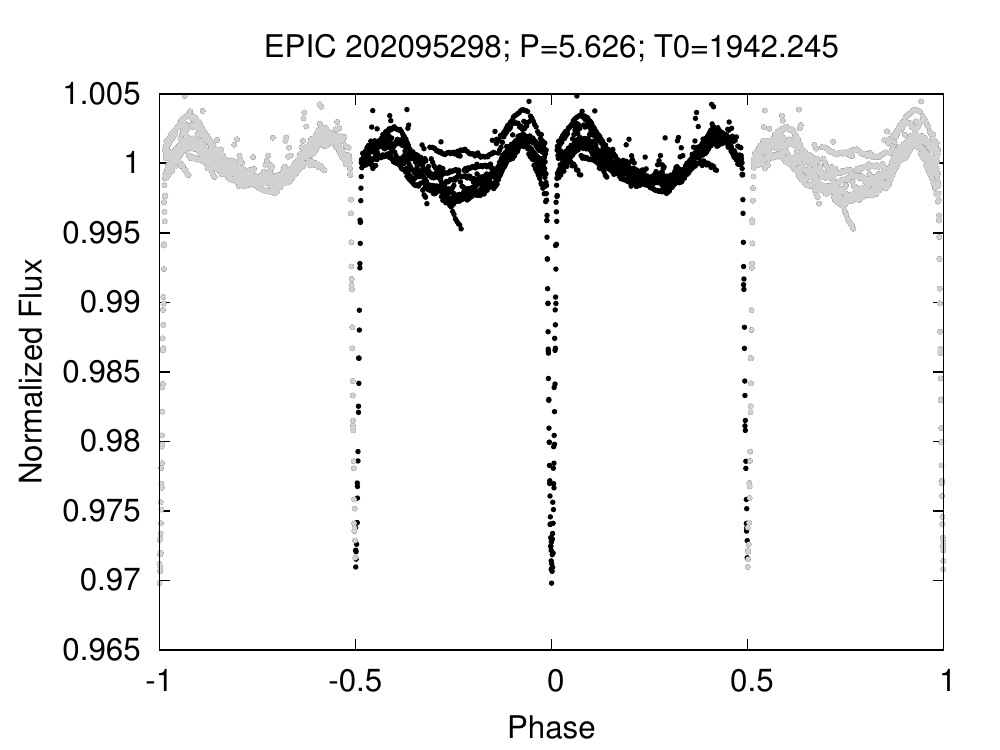} \\
\includegraphics[width=0.38\linewidth]{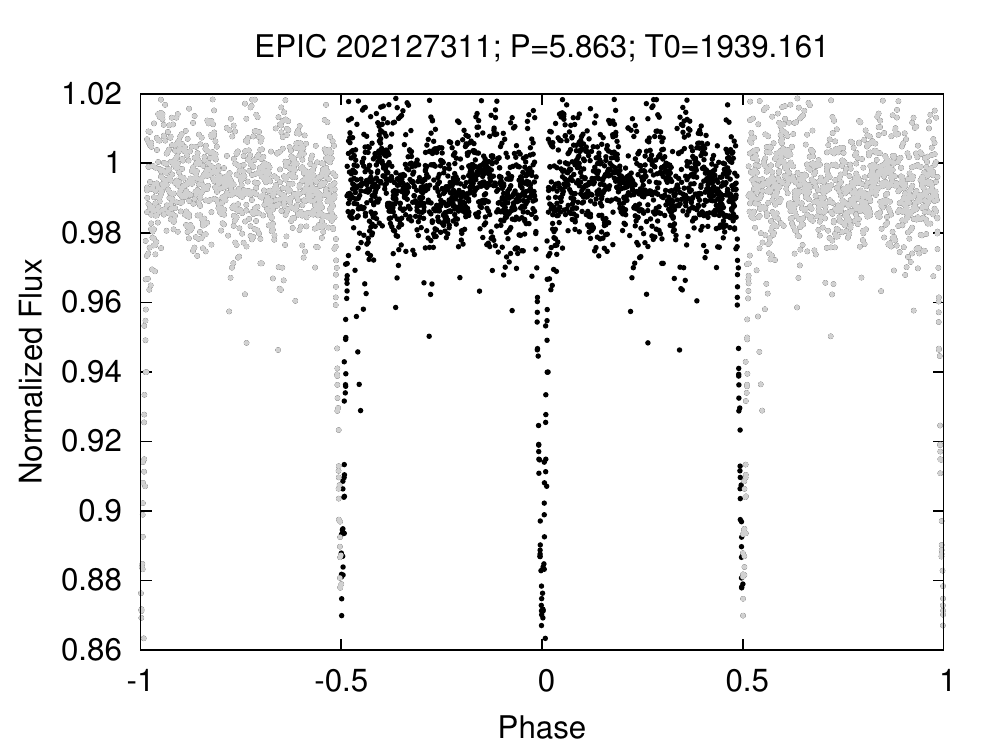} &
\includegraphics[width=0.38\linewidth]{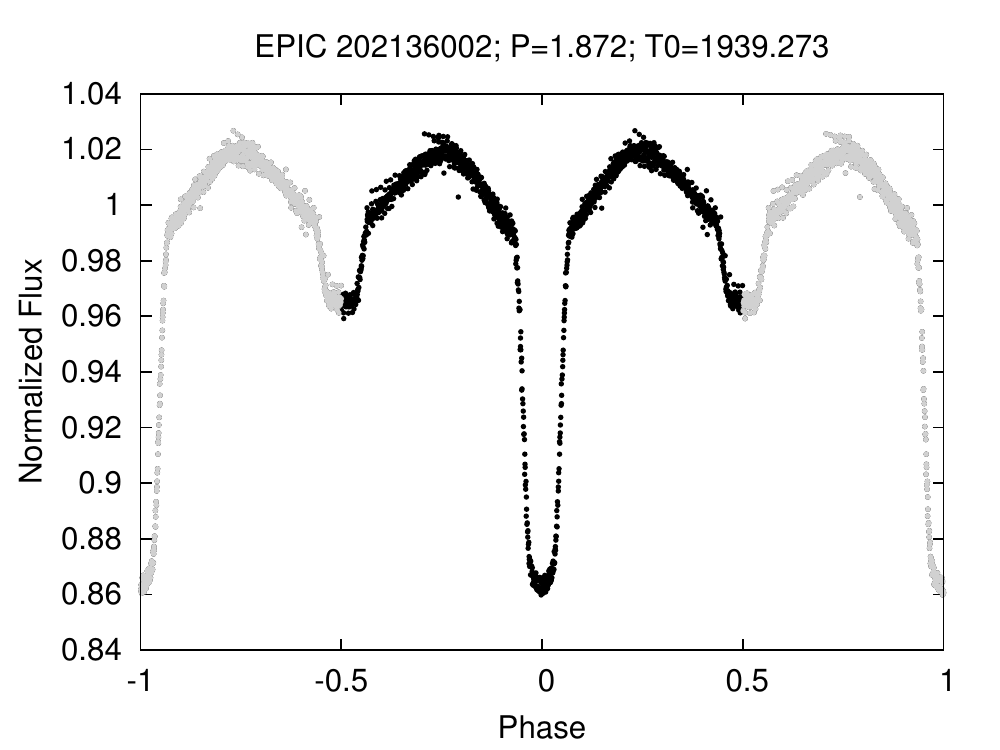} \\
\includegraphics[width=0.38\linewidth]{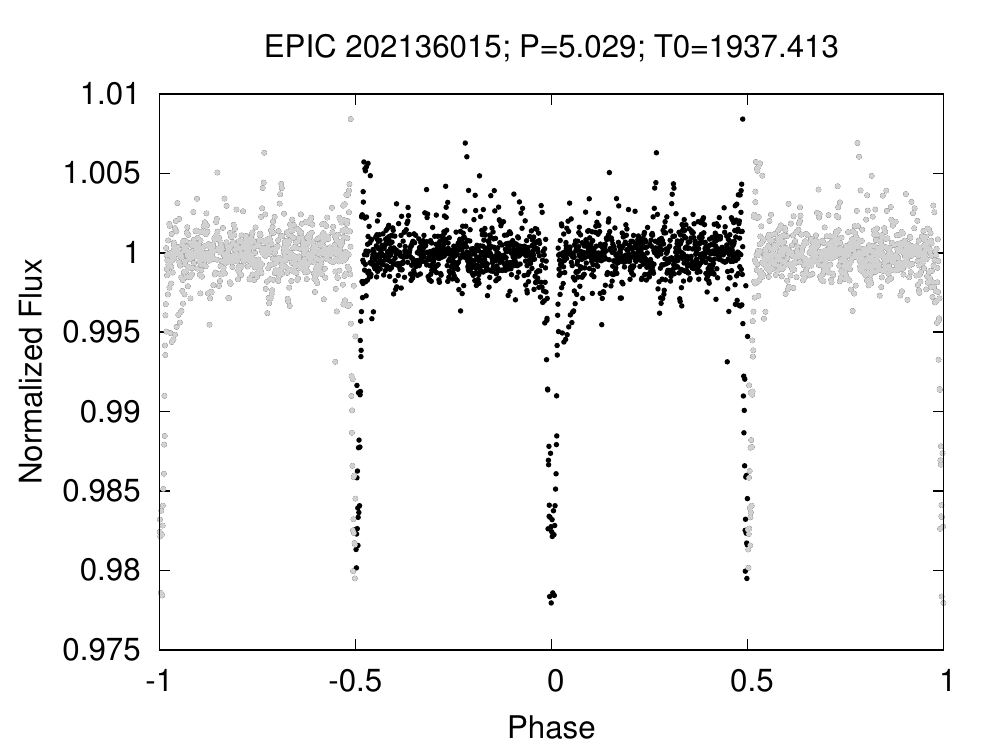} &
\includegraphics[width=0.38\linewidth]{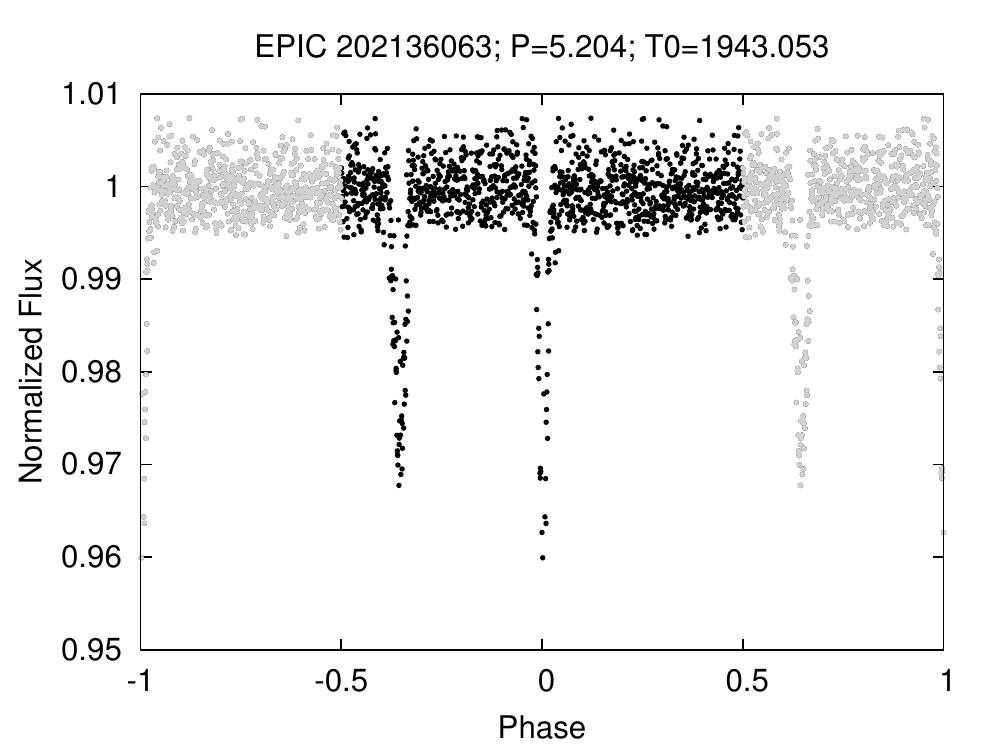} \\
\includegraphics[width=0.38\linewidth]{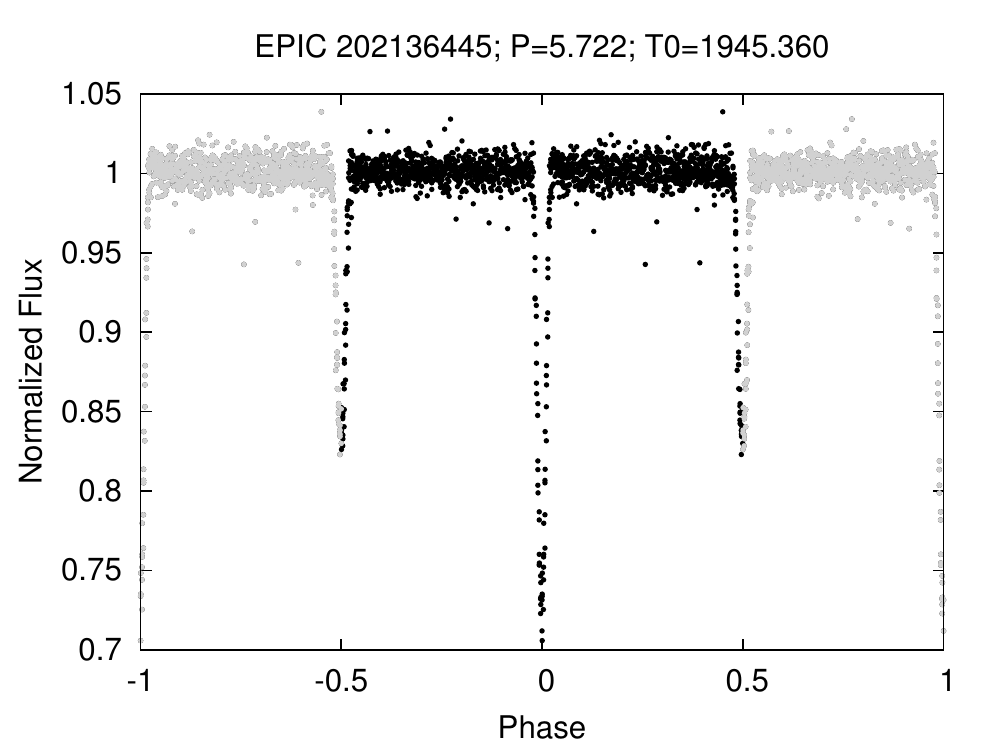} &
\includegraphics[width=0.38\linewidth]{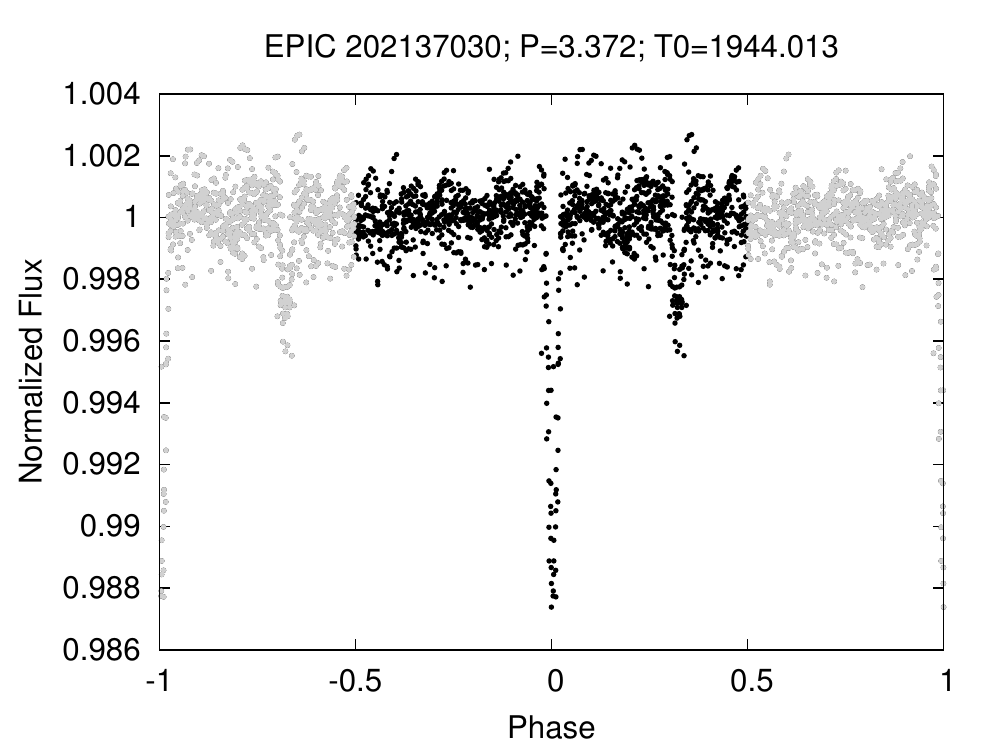} 
\end{tabular}
\end{center}
\caption{Plots of phased light curves for new eclipsing systems. Ephemerides are presented in Tables~\ref{tab:newEBs}, \ref{tab:m35eb}, and \ref{tab:nonEPIC}.  See Section~\ref{sec:pixel_level_analysis} for details.}
\label{fig:phasedplots_10}
\end{figure*}

\begin{figure*}
\begin{center}
\begin{tabular}{cc}
\includegraphics[width=0.38\linewidth]{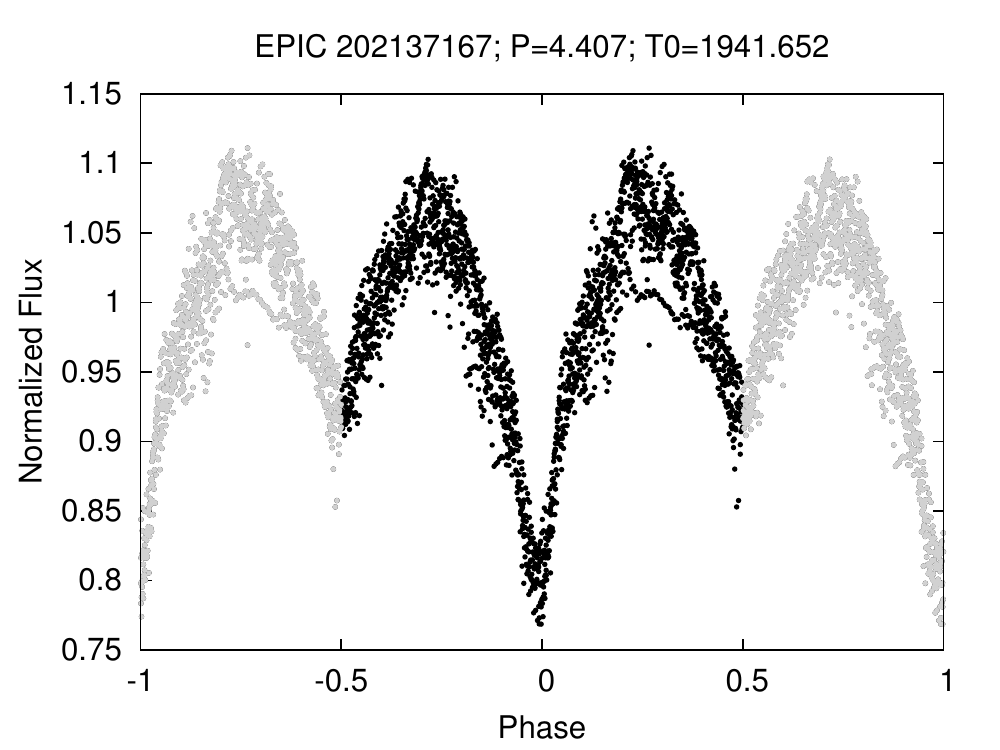} &
\includegraphics[width=0.38\linewidth]{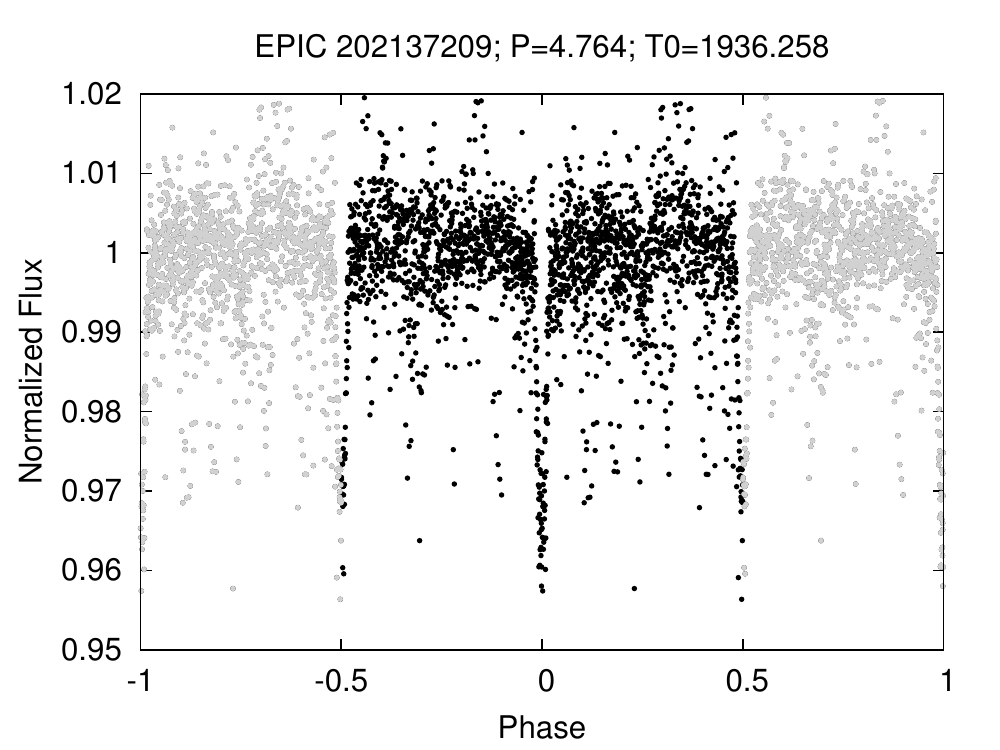} \\
\includegraphics[width=0.38\linewidth]{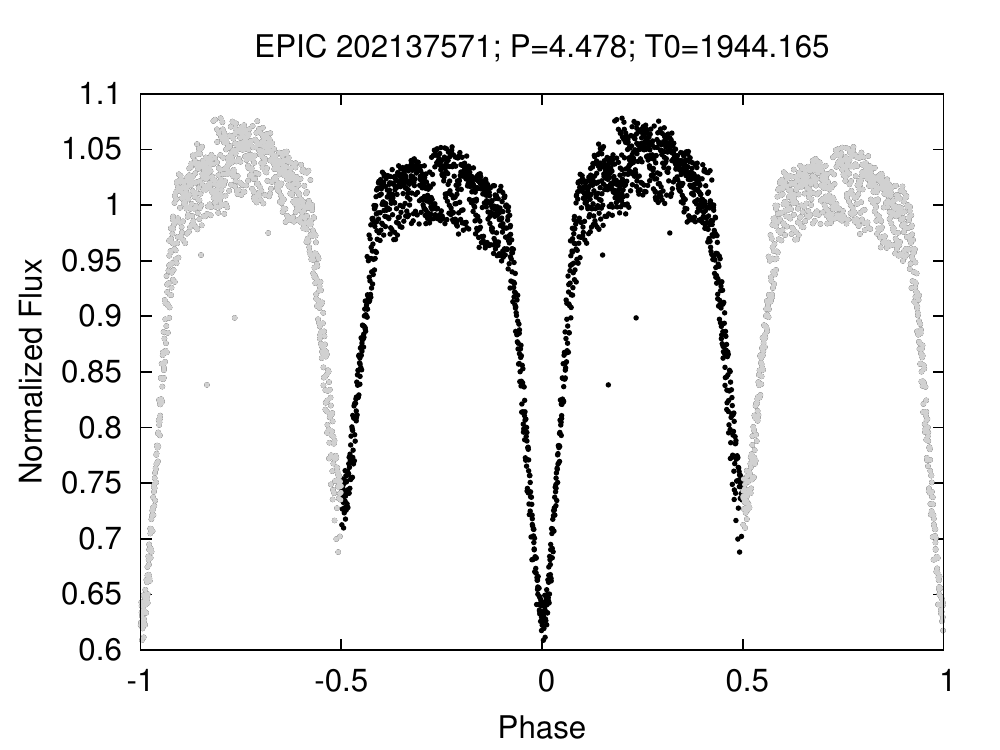} &
\includegraphics[width=0.38\linewidth]{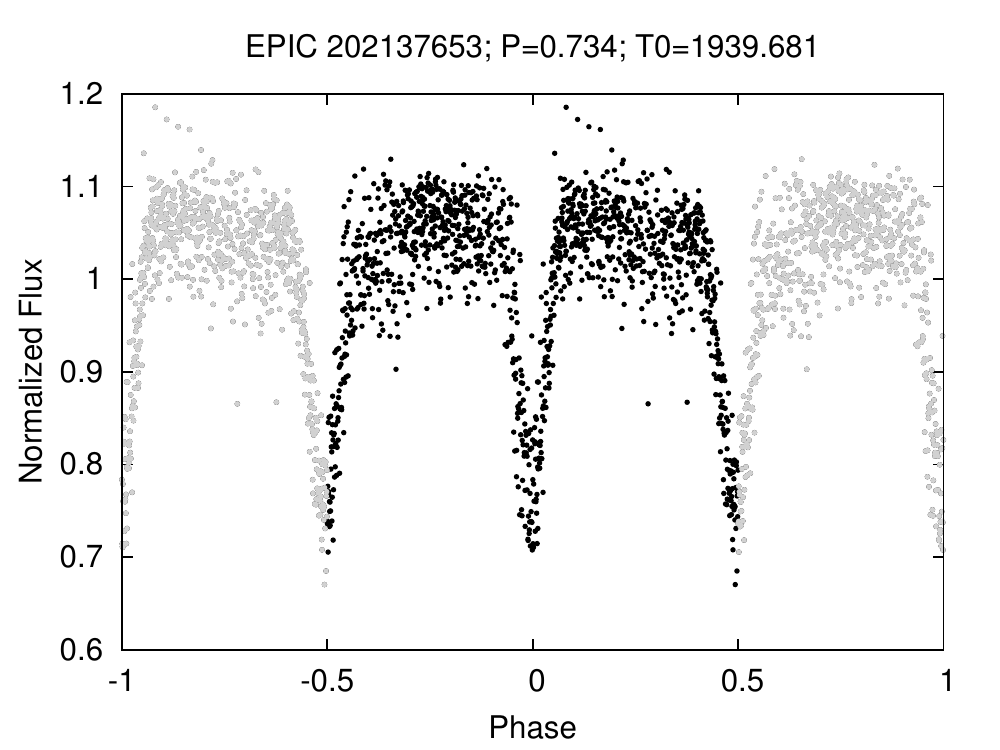} \\
\includegraphics[width=0.38\linewidth]{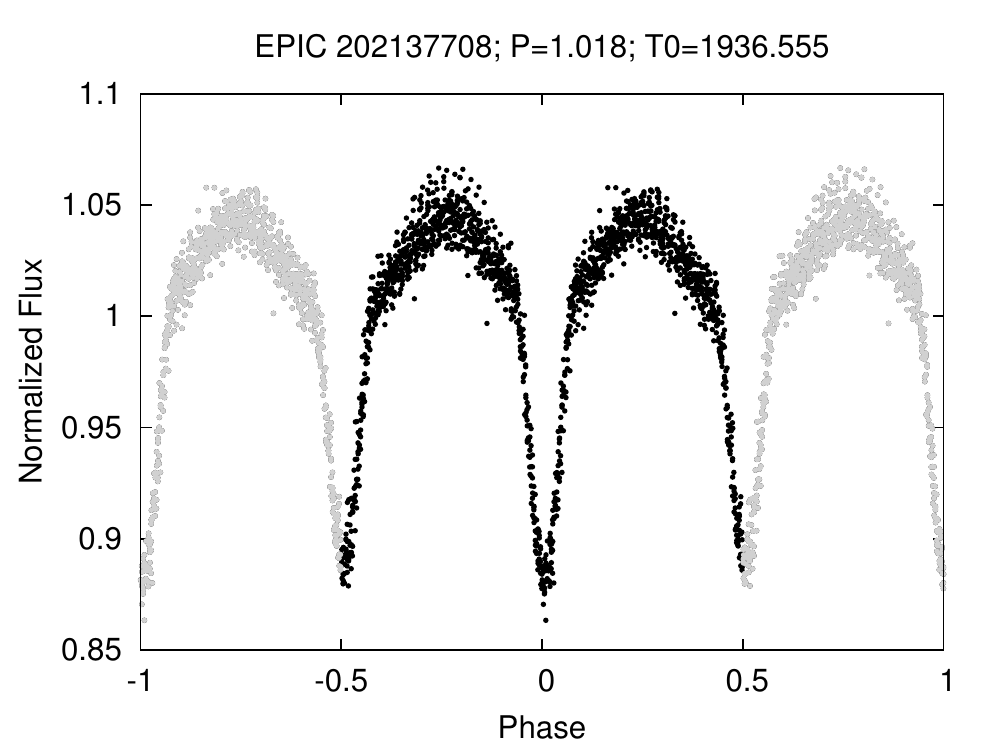} &
\includegraphics[width=0.38\linewidth]{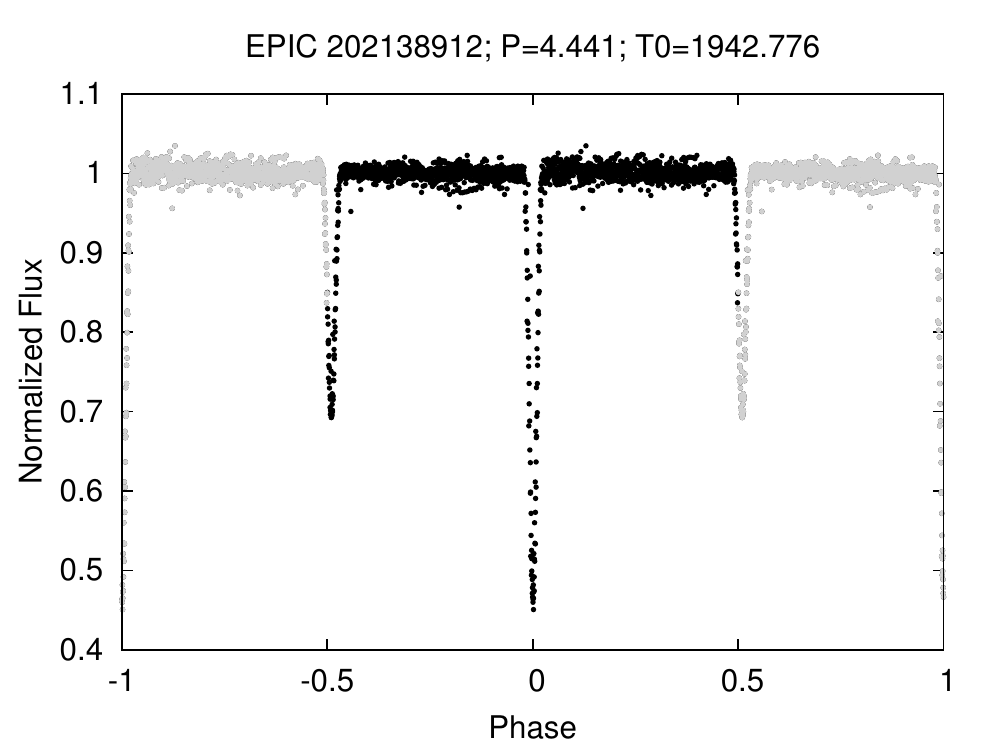} \\
\includegraphics[width=0.38\linewidth]{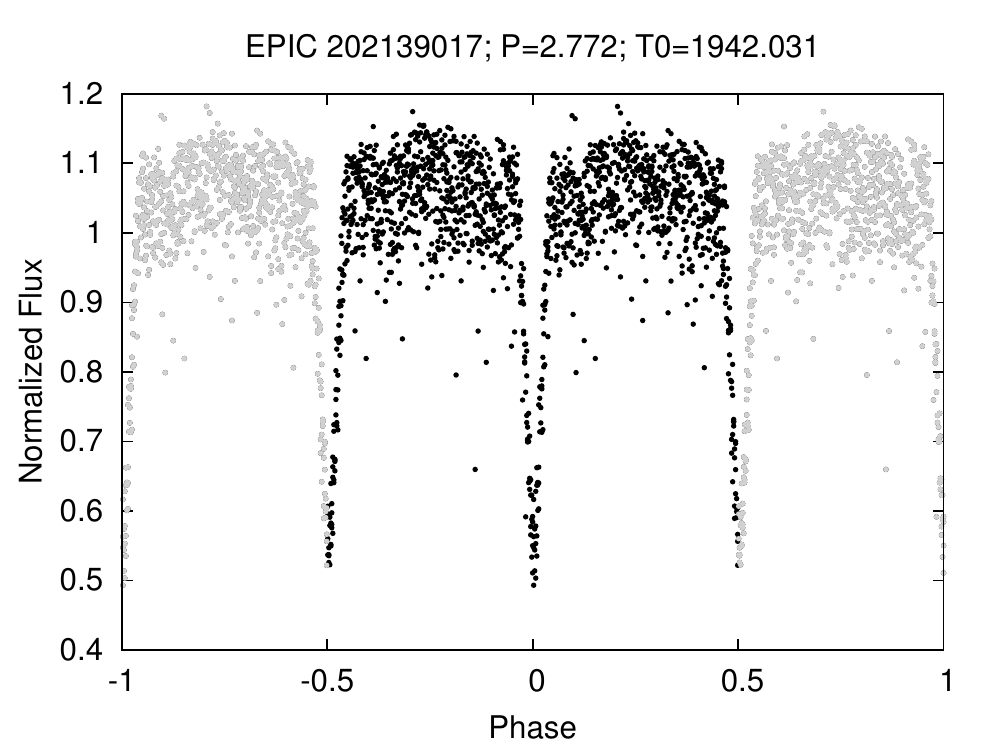} &
\includegraphics[width=0.38\linewidth]{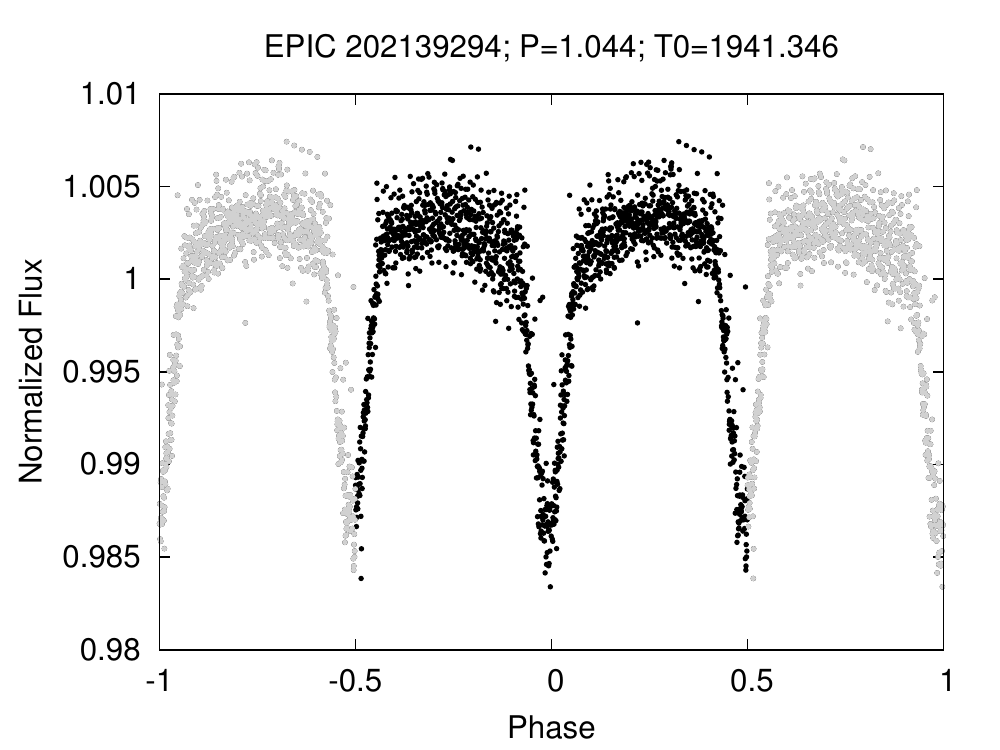} 
\end{tabular}
\end{center}
\caption{Plots of phased light curves for new eclipsing systems. Ephemerides are presented in Tables~\ref{tab:newEBs}, \ref{tab:m35eb}, and \ref{tab:nonEPIC}.  See Section~\ref{sec:pixel_level_analysis} for details.}
\label{fig:phasedplots_11}
\end{figure*}

\begin{figure*}
\begin{center}
\begin{tabular}{cc}
\includegraphics[width=0.38\linewidth]{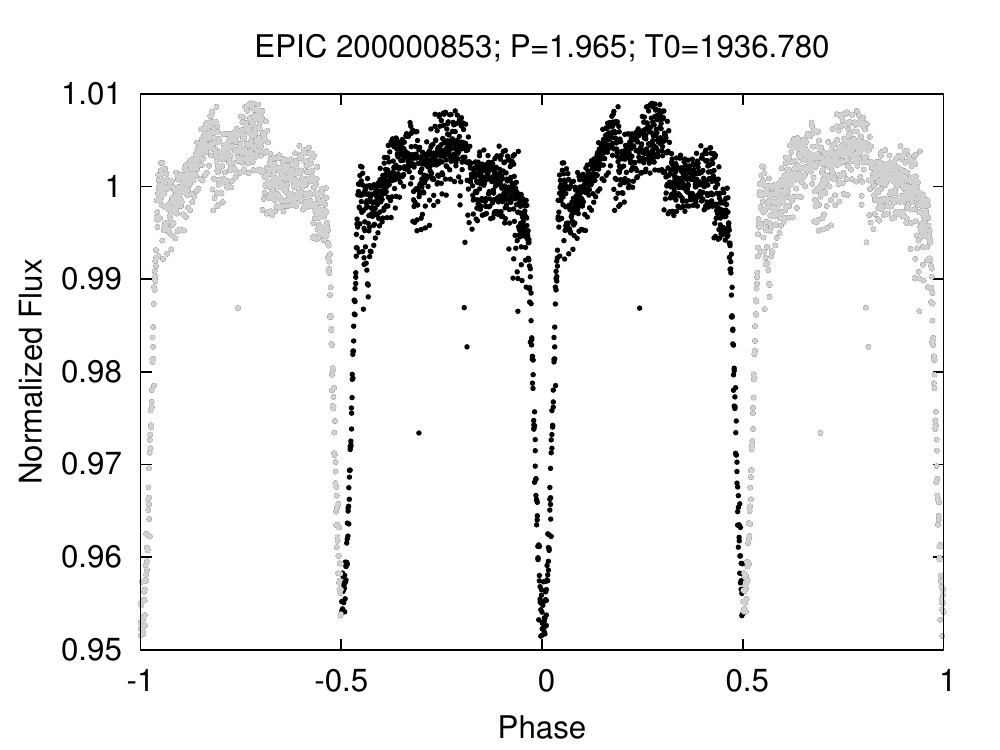} &
\includegraphics[width=0.38\linewidth]{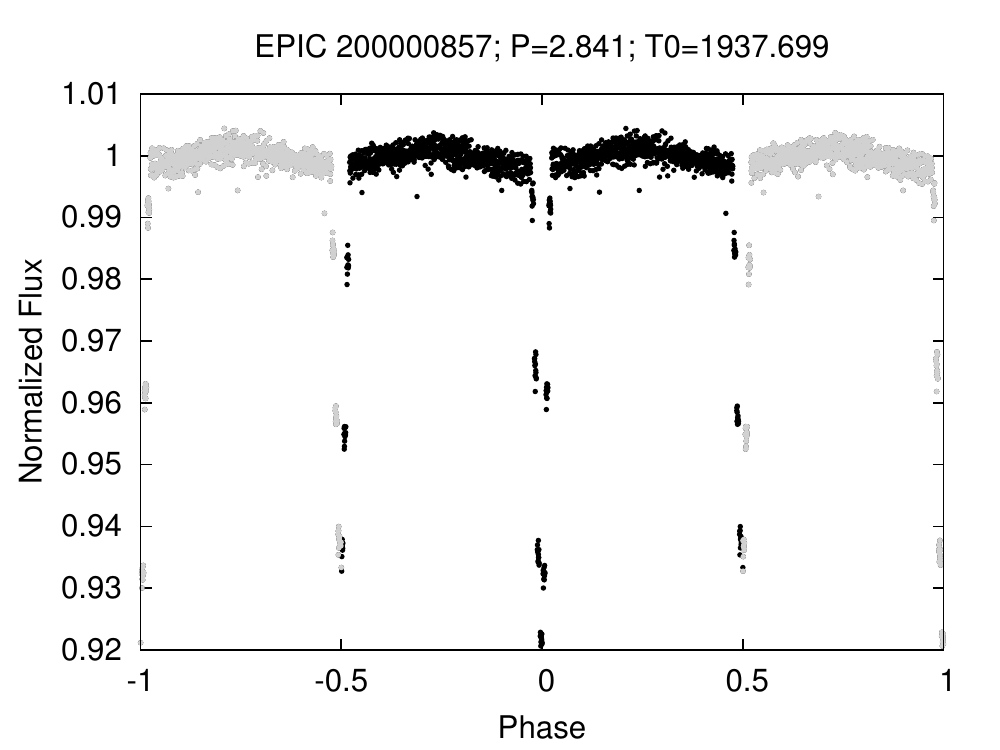} \\
\includegraphics[width=0.38\linewidth]{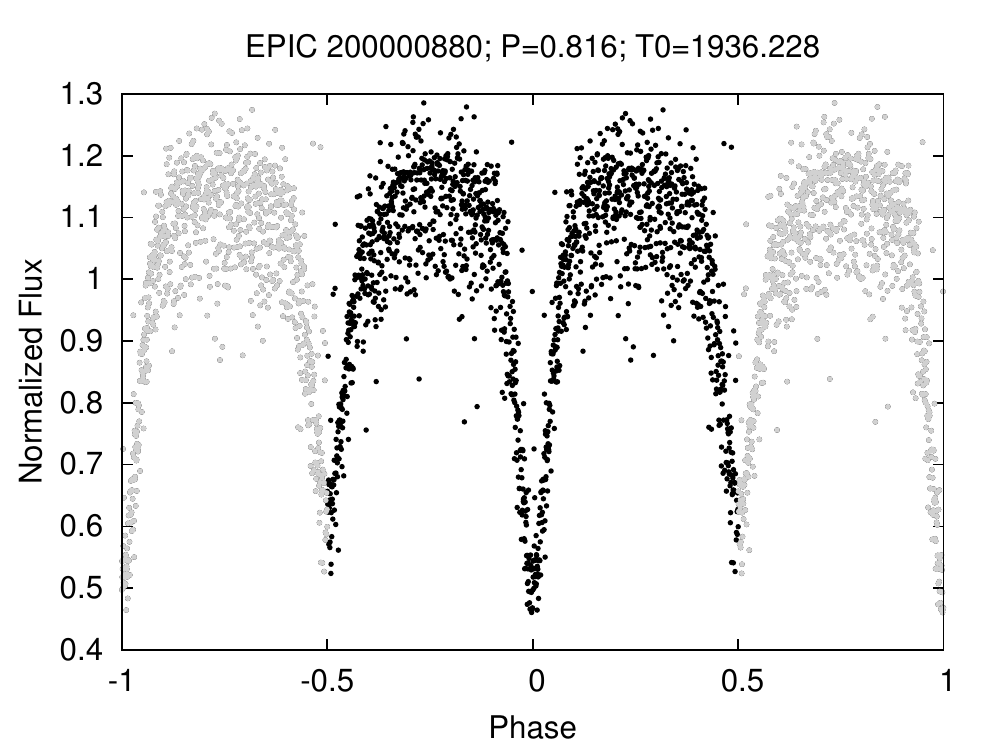} &
\includegraphics[width=0.38\linewidth]{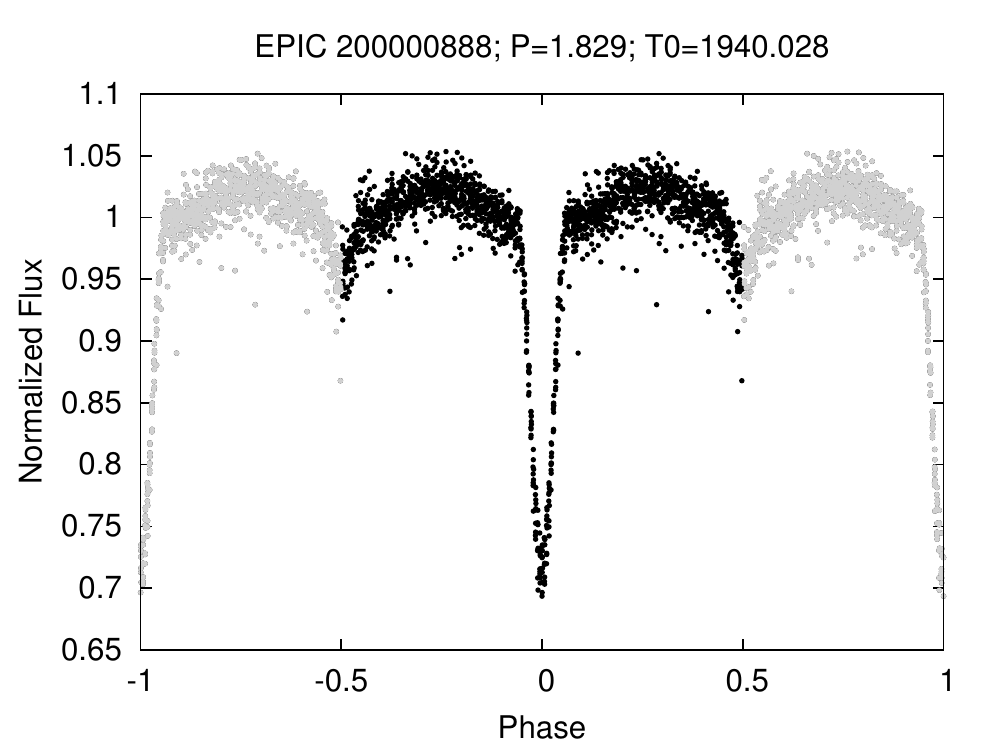} \\
\includegraphics[width=0.38\linewidth]{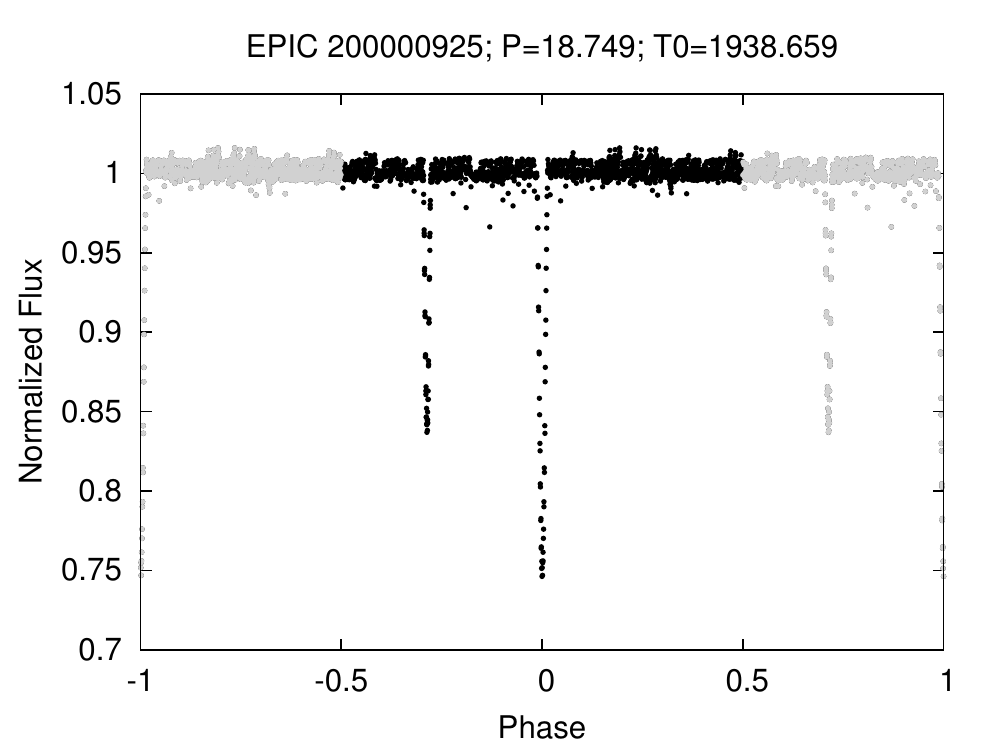} &
\includegraphics[width=0.38\linewidth]{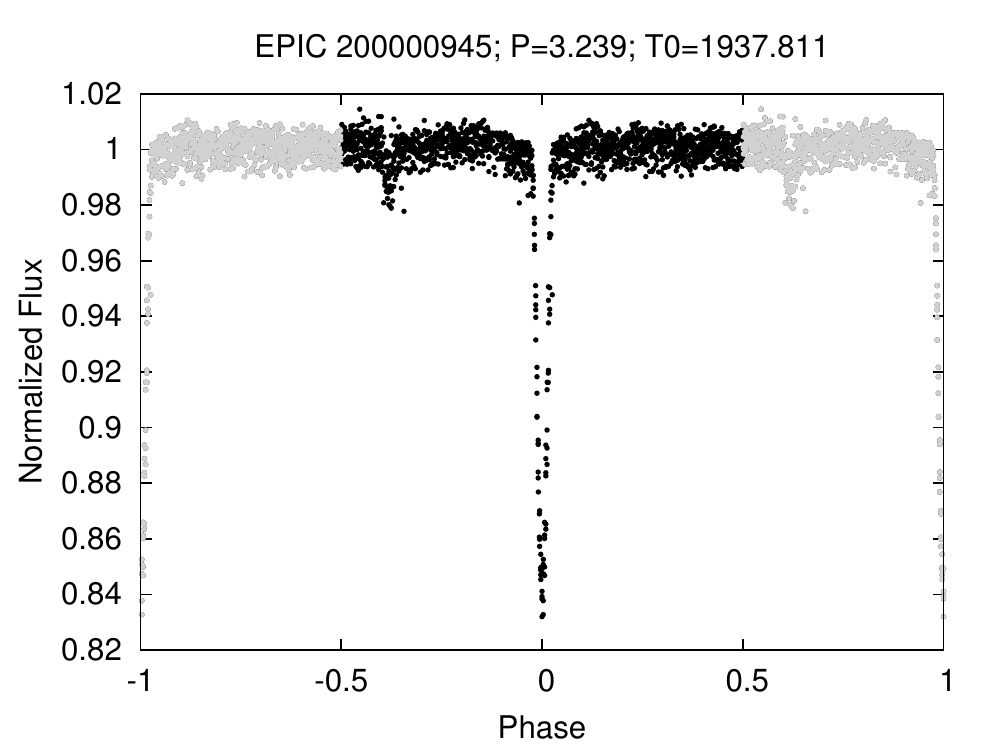} \\
\includegraphics[width=0.38\linewidth]{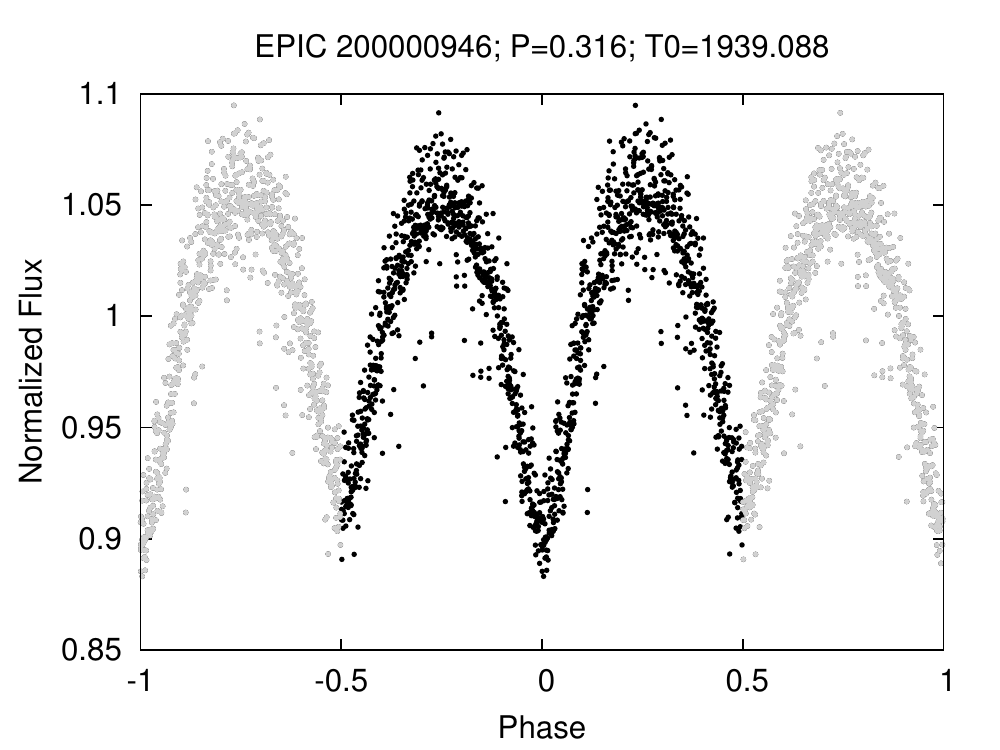} &
\includegraphics[width=0.38\linewidth]{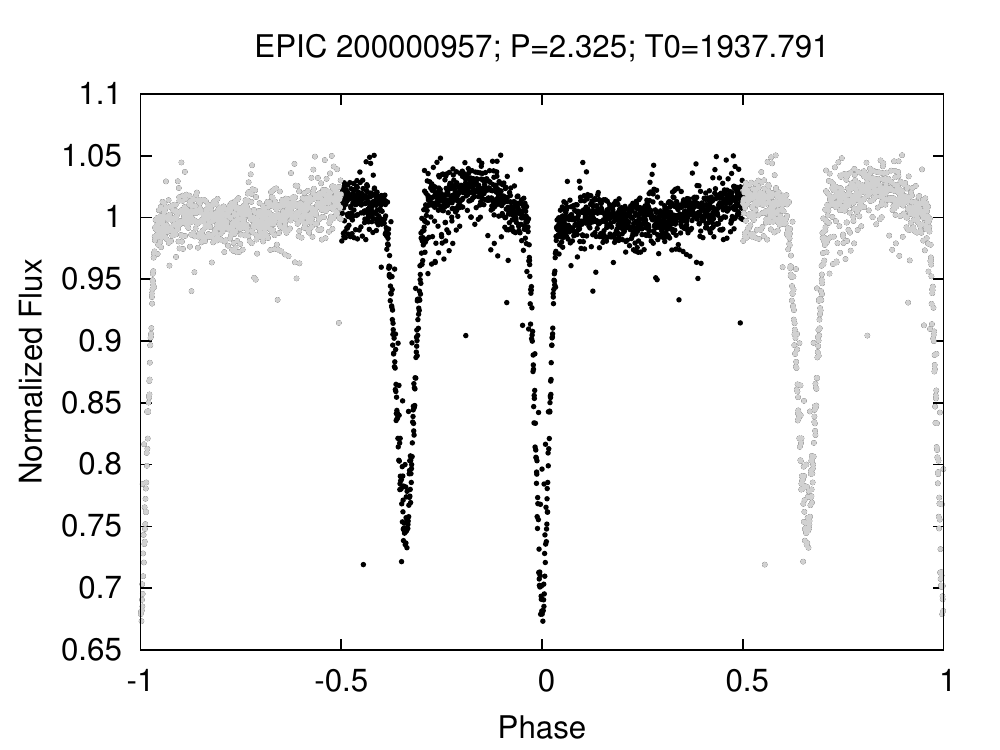} 
\end{tabular}
\end{center}
\caption{Plots of phased light curves for new eclipsing systems. Ephemerides are presented in Tables~\ref{tab:newEBs}, \ref{tab:m35eb}, and \ref{tab:nonEPIC}.  See Section~\ref{sec:pixel_level_analysis} for details.}
\label{fig:phasedplots_12}
\end{figure*}

We use these ephemerides to plot the EB phase curves, presented in Figures~\ref{fig:phasedplots_1} through \ref{fig:phasedplots_12}.  The phase of each system is arranged for illustrative purposes such that the secondary eclipse, if present, is visible. The phase folded light curves were then used to determine a system morphology.  Morphology values range between 0 (detached) and 1 (over-contact). Morphologies are determined by first fitting a chain of four quadratic functions to the shape of the phased light curve and then determining the morphology through application of a Local Linear Embedding \citet{mat12} routine. With the exception of Heartbeat binaries and EBs lacking three observed eclipses, morphologies are computed for the entire C0 sample.

All targets observed eclipsing in the C0 data set were crossed matched against all-sky catalog listings in The International Variable Star Index (VSX) \citep{wat06} and the SIMBAD database \citep{wen00} within a 20 arc second search radius using RA and Dec coordinates from \textsc{MAST}. Results for 2MASS source identifications were found to be in good agreement with the \textsc{K2-TESS} stellar properties catalog \citep{stas14}, which includes EPIC to 2MASS cross matches accurate to within one arc second. For M35 superstamp related targets, we also performed a cross check against recently published ground survey data from the Asiago Pathfinder for HARPS-N (APHN) program, which has performed a long term, high-precision (5 milli-mag) search of open clusters M35 and NGC 2158 for variable stars \citep{nar14}.

These cross matches allowed us to determine known stellar companions to our targets as well as alternate source identifications. We also checked and reviewed our results against the target proposal lists of the C0 Guest Observer (GO) proposals in order to determine which targets were known binary or multiple star systems. Our search of the GO proposals revealed that \totalebsknowneclipsing of the \totalebs eclipsing targets we identified and retained in our final sample had been requested for {\it K2} C0 observation by several different GO programs as known or suspected binaries. An additional balance of 158 EPICs (approximately 60\%) proposed by these binary star related GOs were not found to eclipse or possess ellipsoidal variations that could be attributed to a binary as per our culling criteria (Section~\ref{sec:Culling}). However a considerable portion of this balance may represent non-eclipsing objects (e.g., spectroscopic binaries, ELV binaries, cataclysmic variables, etc.) and thus the balance of non-detections in our {\it K2} C0 photometry is not unexpected.

Our analysis resulted in \totalebsnew new EBs detected in the C0 field. The orbital periods, ephemerides, morphologies, and cross matched identifications for all EBs in our sample are organized and presented in five tables. The \totalebsknowneclipsing previously identified or suspected EBs in C0 are cataloged in Table~\ref{tab:knownEBs}. The \totalebsnew new EB detections are presented in Table~\ref{tab:newEBs} (EPIC targets), Table~\ref{tab:m35eb} (M35 superstamp targets), and Table~\ref{tab:nonEPIC} (non-EPIC targets). The non-EPIC associated EBs were detected via their contamination of GO proposed target masks, creating a false positive EB signal. Due to the large size of C0 TPF stamps, we were able to extract and analyze light curves from these listed contaminators (see Section \ref{sec:Culling}).
A separate collection of false positive cases caused by direct pixel response function (PRF) contamination \citep{brys10} between pairs of observed EPIC targets are listed in Table~\ref{tab:blends}.

\begin{figure*}
  \begin{center}
      \includegraphics[width=\linewidth]{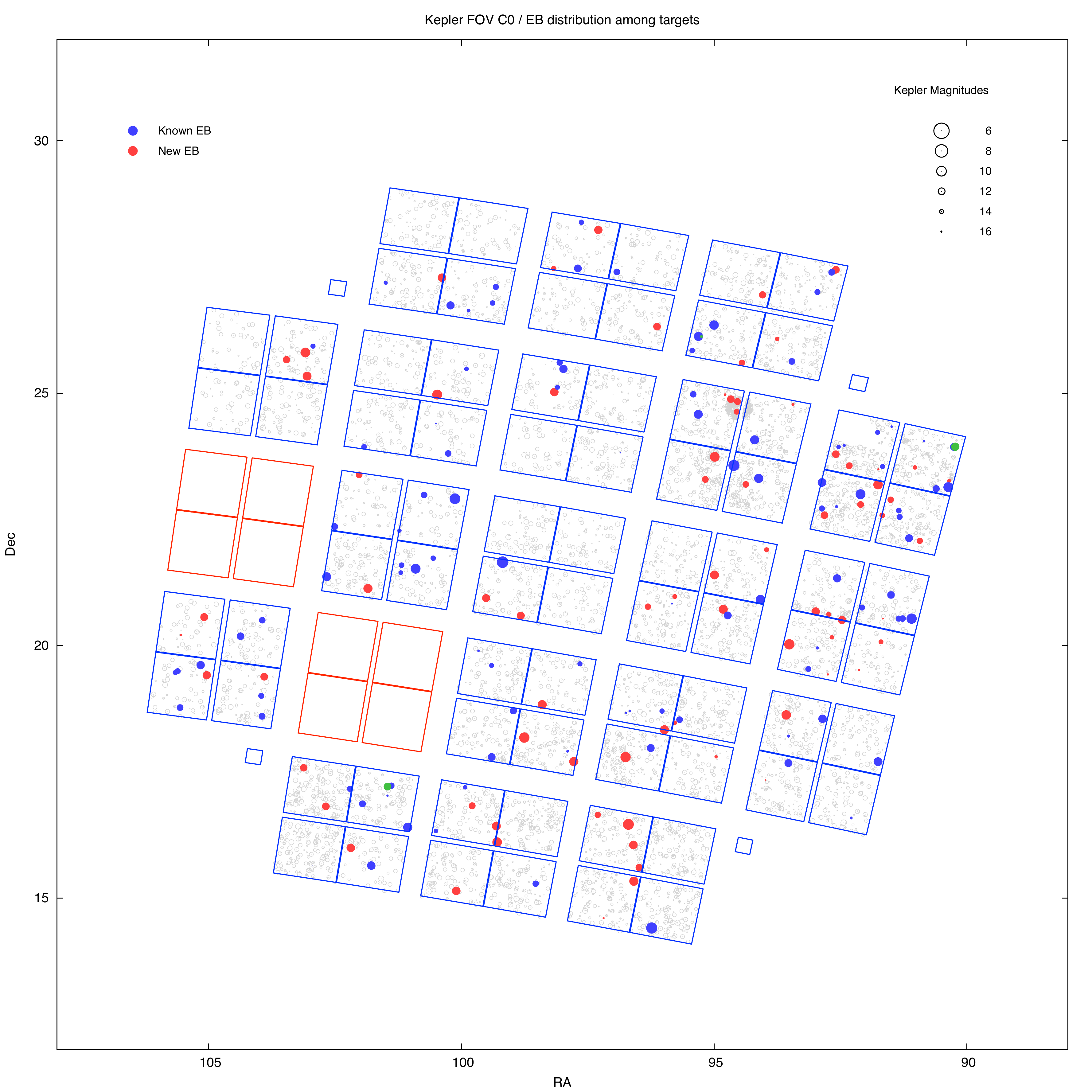}
  \end{center}
  \caption{K2 C0 field of view with positions plotted for of all eclipsing systems listed in Table~\ref{tab:knownEBs} (red), Table~\ref{tab:newEBs}, and Table~\ref{tab:nonEPIC} (blue). Data point size corresponds to apparent $K_p$~magnitude.  The three green data points correspond to the two exoplanet candidates and suspected false positive described in Section~\ref{sec:planet_candidates}.}
  \label{fig:c0_newEBs1}
\end{figure*}

The sky plotted distribution for all \totalebs~EBs in our final sample is presented in Figure~\ref{fig:c0_newEBs1}.

\section{Spectroscopic Observations}
\label{sec:spectroscopy}

We observed nine selected objects identified as interesting in our analysis (for details refer to Sections~\ref{sec:noteworthy_ebs} and \ref{sec:planet_candidates}) on the nights of December 15, 19, and 20, 2014 using the Observatoire Astronomique du Mont-M\'egantic's (OMM) 1.6-m telescope and long-slit spectrometer. The spectrometer employed a 1200-line/mm grating which yielded a dispersion of 0.89~\AA/pixel and a wavelength range of $\sim 1800$~\AA~(3750 -- 5600\AA).  Depending on the brightness of each object, up to three exposures were co-added to improve the signal to noise; the total integration time for each object ranged between 20 to 90 minutes. All of the spectra were background subtracted, corrected for the quantum efficiency of the CCD detector and the reflectance of the grating, and corrected for Rayleigh scattering in the atmosphere before being co-added.

\begin{figure*}                                                     
  \begin{center}                                        
    \begin{tabular}{cc}
        \includegraphics[trim=0.5in 3.1in 1in 3.1in, clip, width=0.45\linewidth]{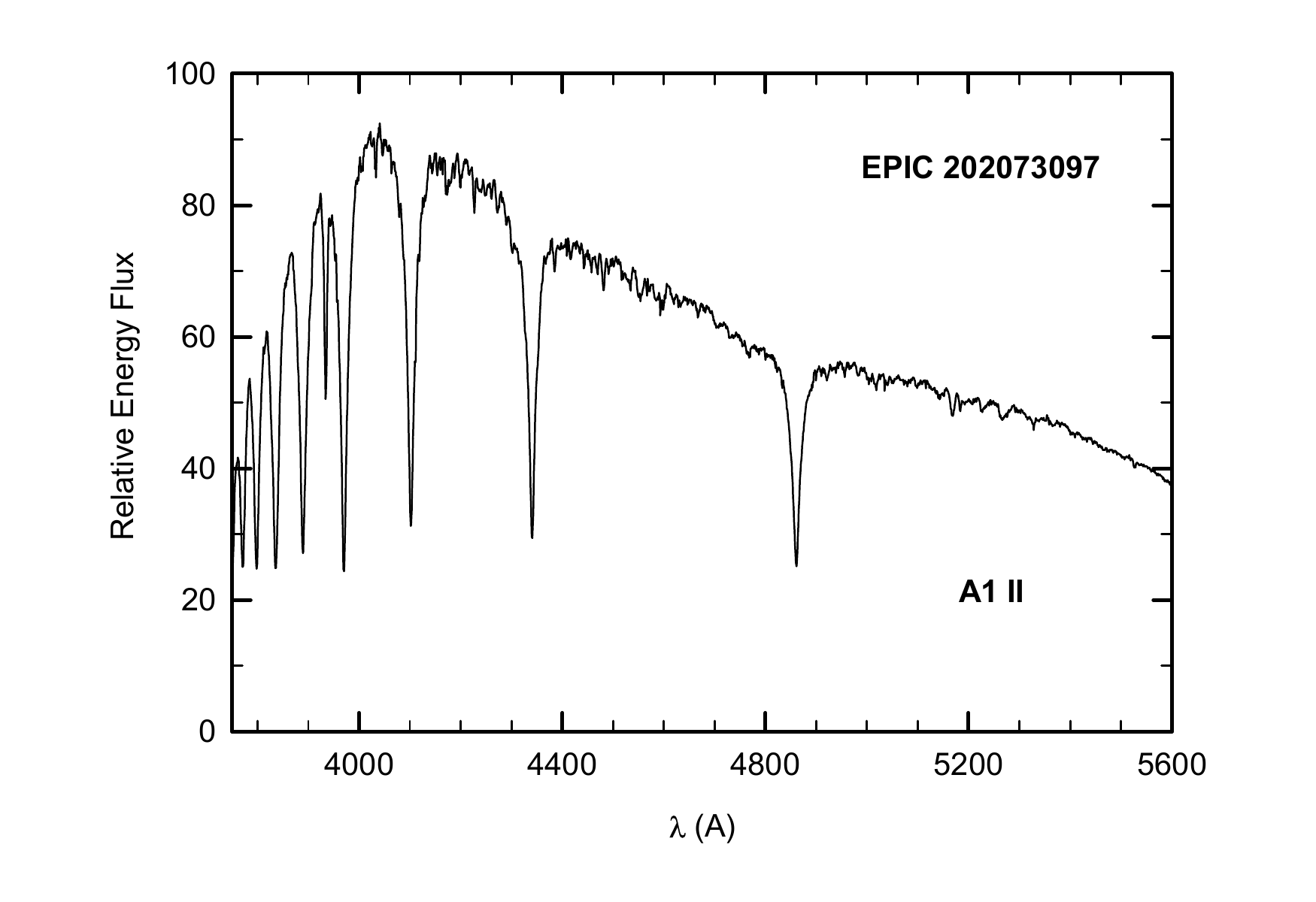} &
        \includegraphics[trim=0.5in 3.1in 1in 3.1in, clip, width=0.45\linewidth]{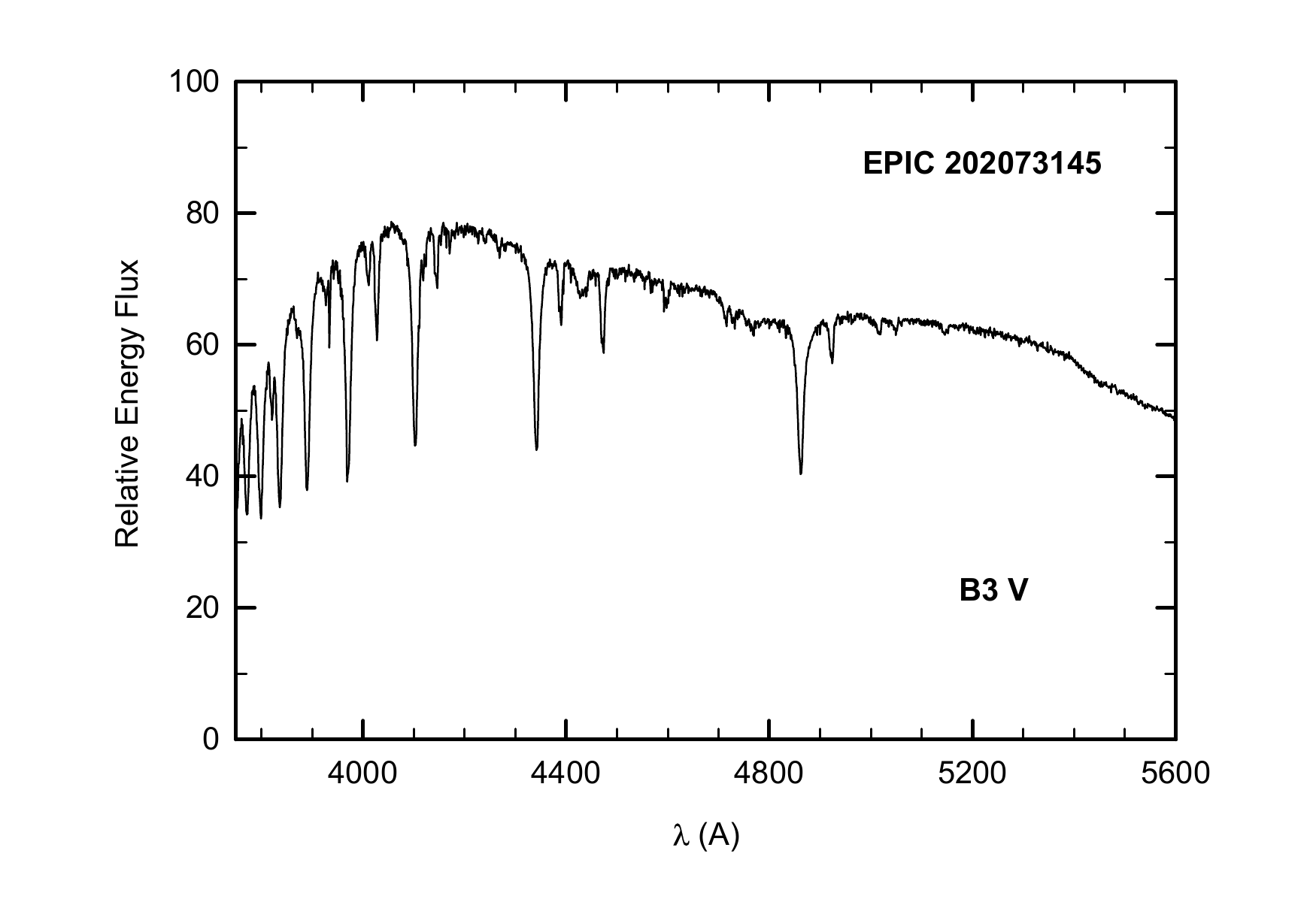} \\
         \includegraphics[trim=0.5in 3.1in 1in 3.1in, clip, width=0.45\linewidth]{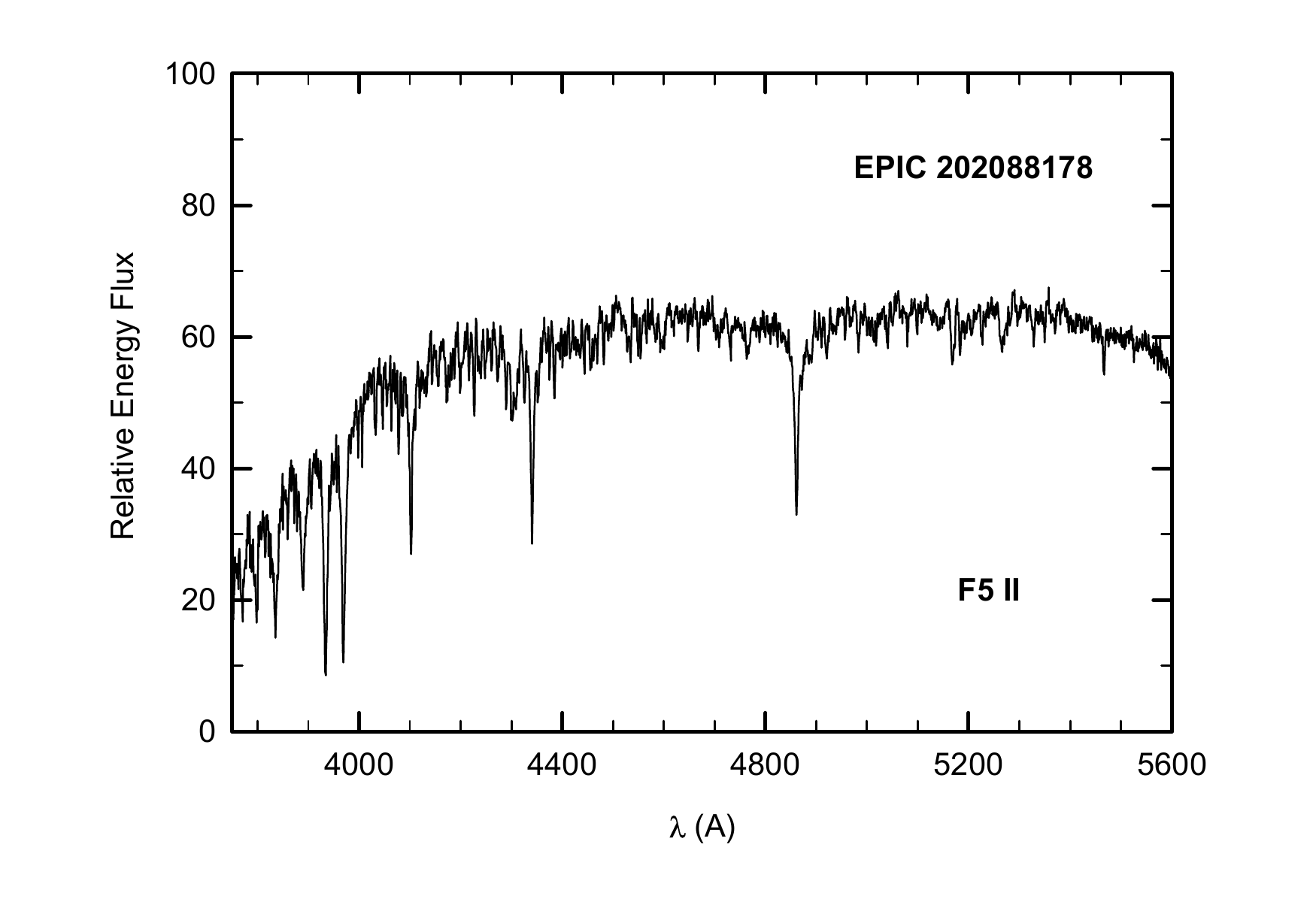} &	
        \includegraphics[trim=0.5in 3.1in 1in 3.1in, clip, width=0.45\linewidth]{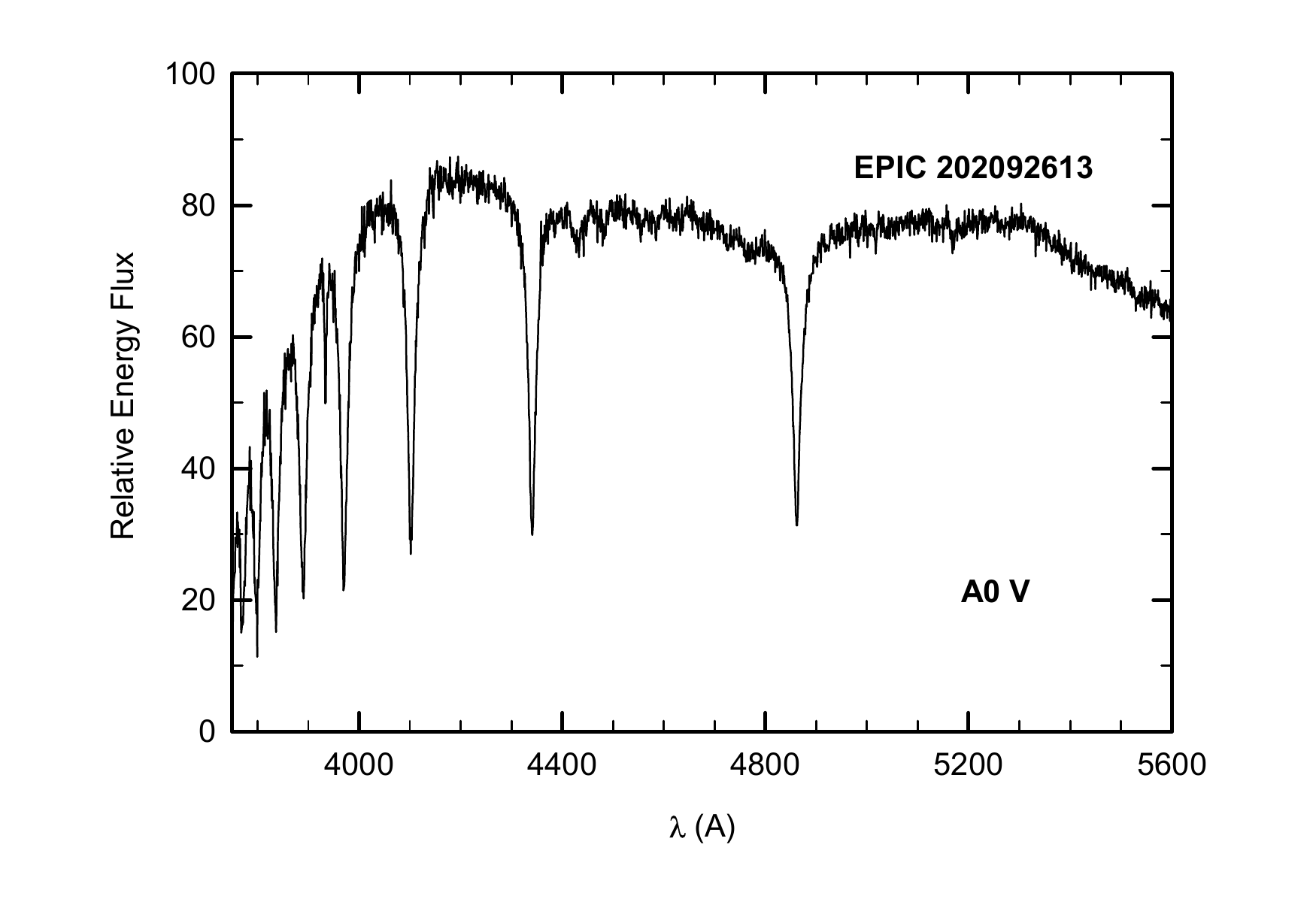} \\
        \includegraphics[trim=0.5in 3.1in 1in 3.1in, clip, width=0.45\linewidth]{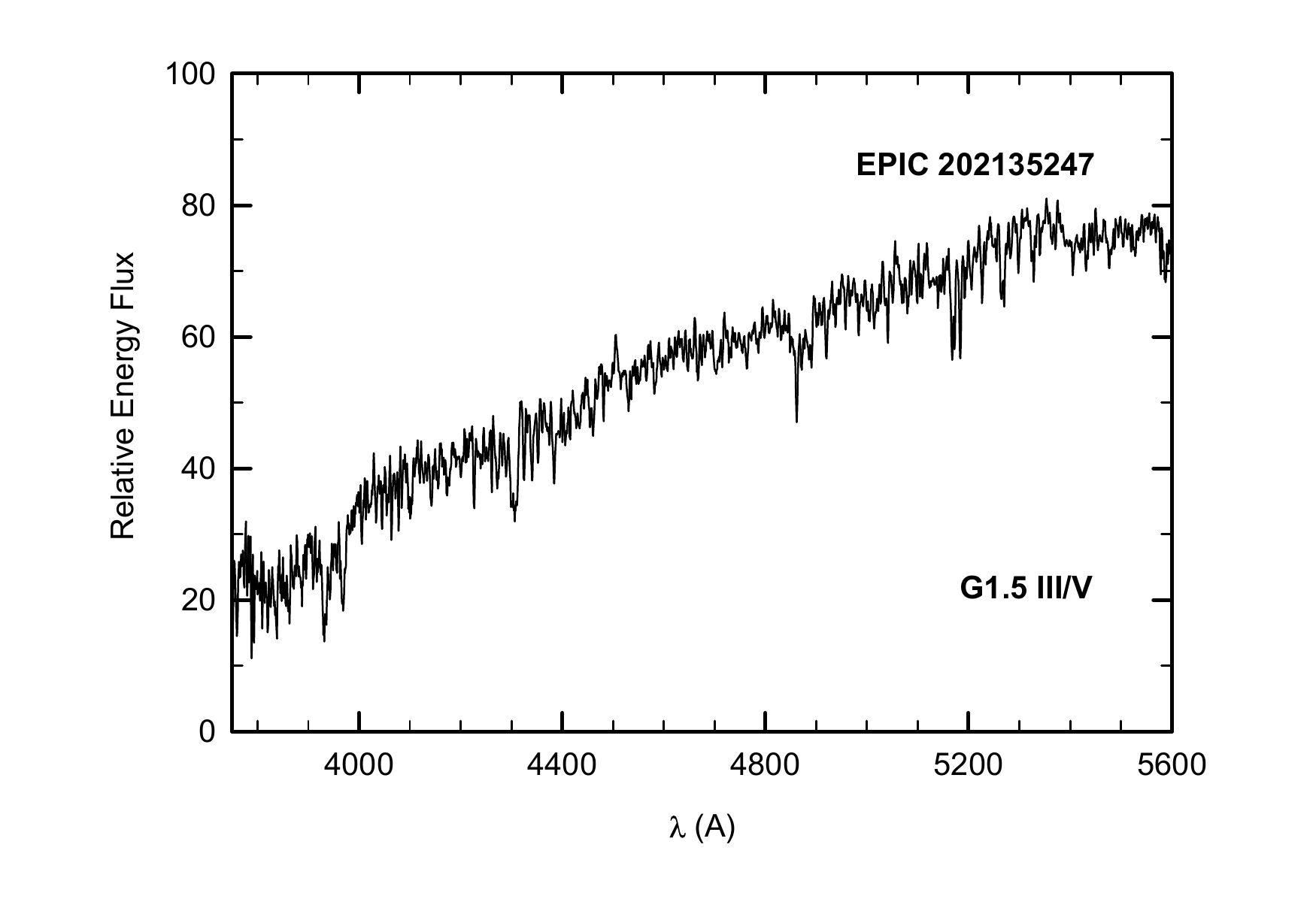} &	
        \includegraphics[trim=0.5in 3.1in 1in 3.1in, clip, width=0.45\linewidth]{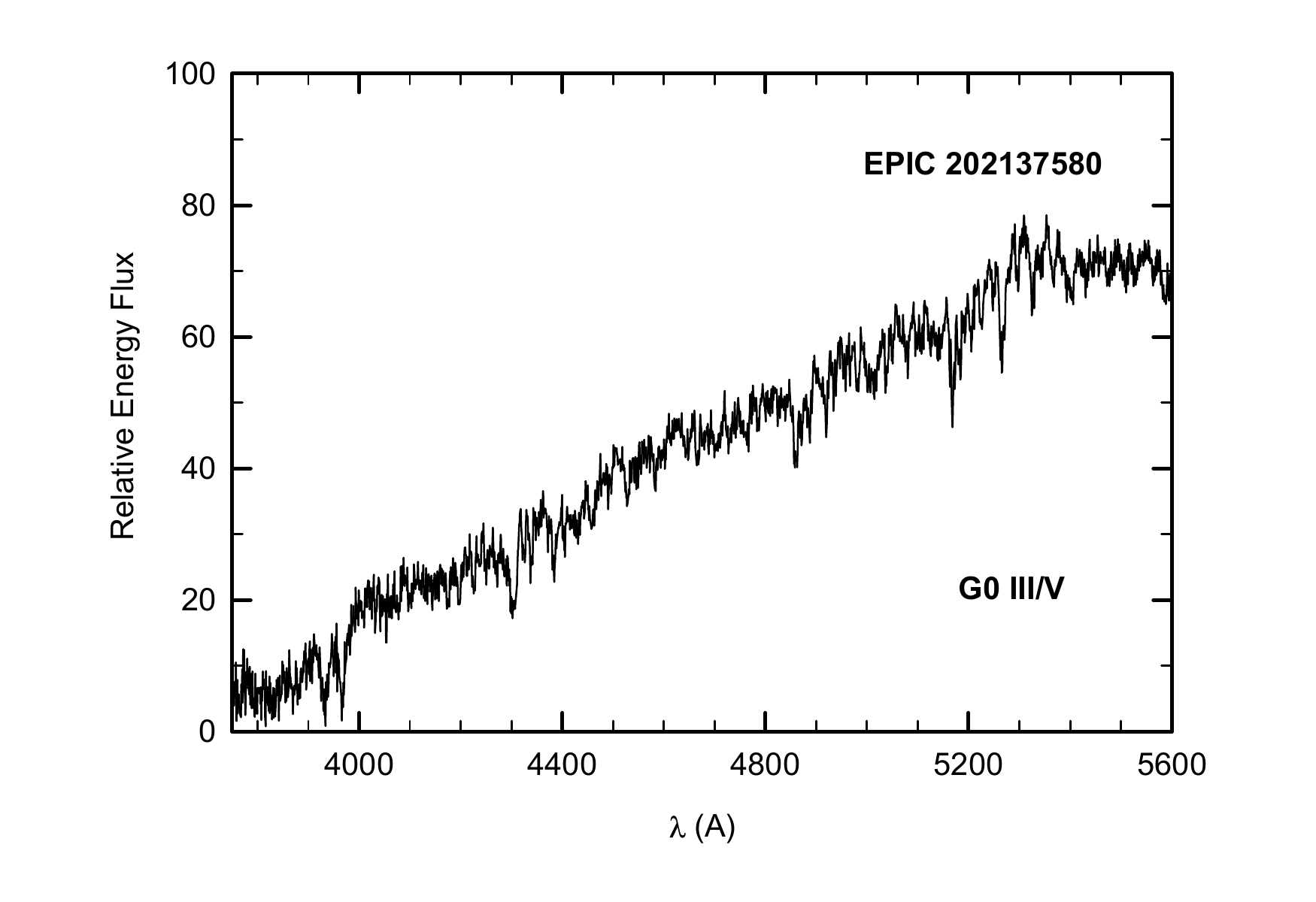} 
    \end{tabular}
 \end{center}
\caption{Spectra for selected EBs (EPIC 202073097, 202073145, 202088178, 202092613, 202135247, 202137580) described in Section~\ref{sec:spectroscopy} and \ref{sec:noteworthy_ebs}. Spectral type for each object is given in the bottom right of each panel.}
\label{fig:EBspectra1}
\end{figure*}

\begin{figure*}                                                     
  \begin{center}                                        
    \begin{tabular}{cc}
        \includegraphics[width=0.4\linewidth]{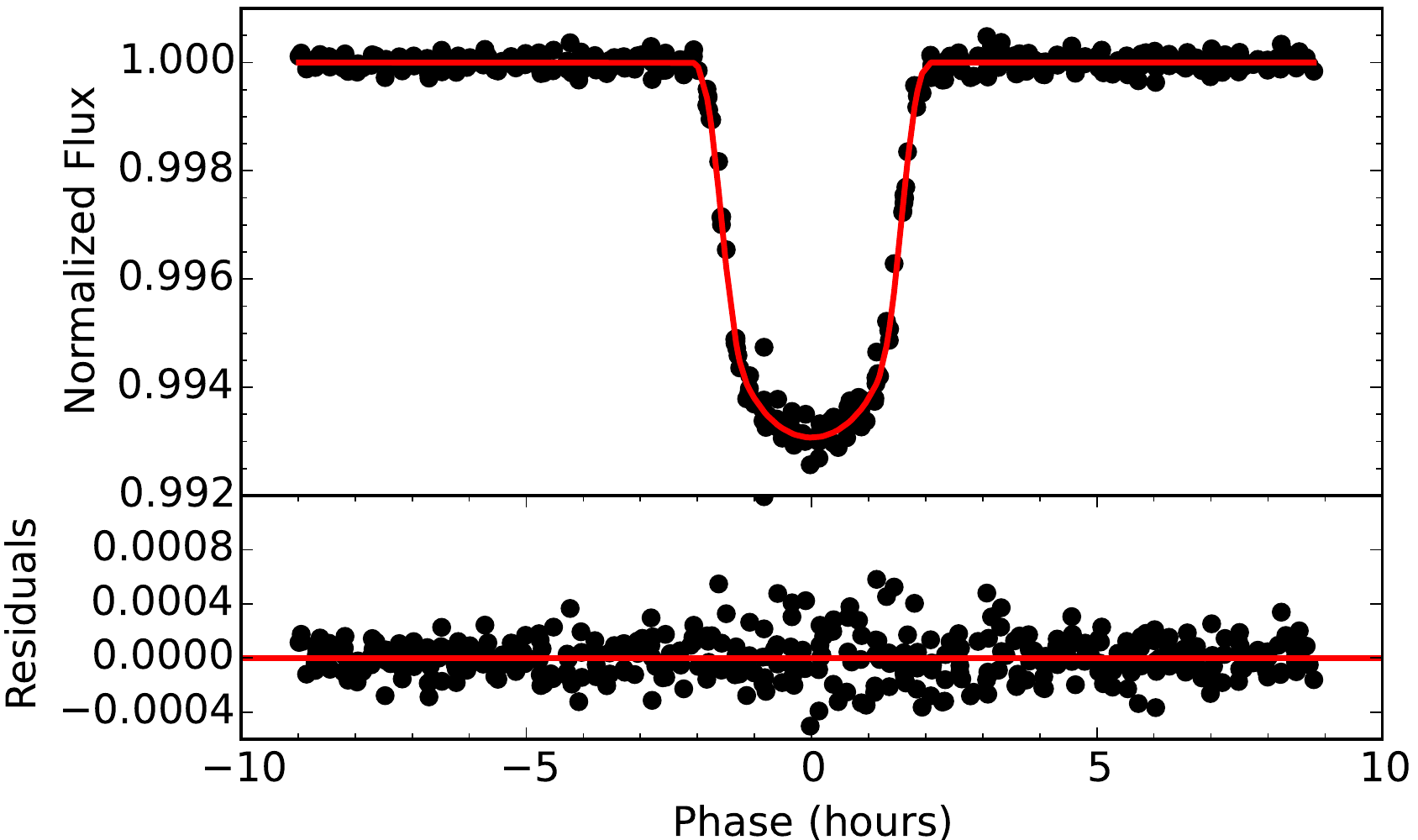}	&
        \includegraphics[trim=0.5in 3.4in 1in 3.4in, clip, width=0.4\linewidth]{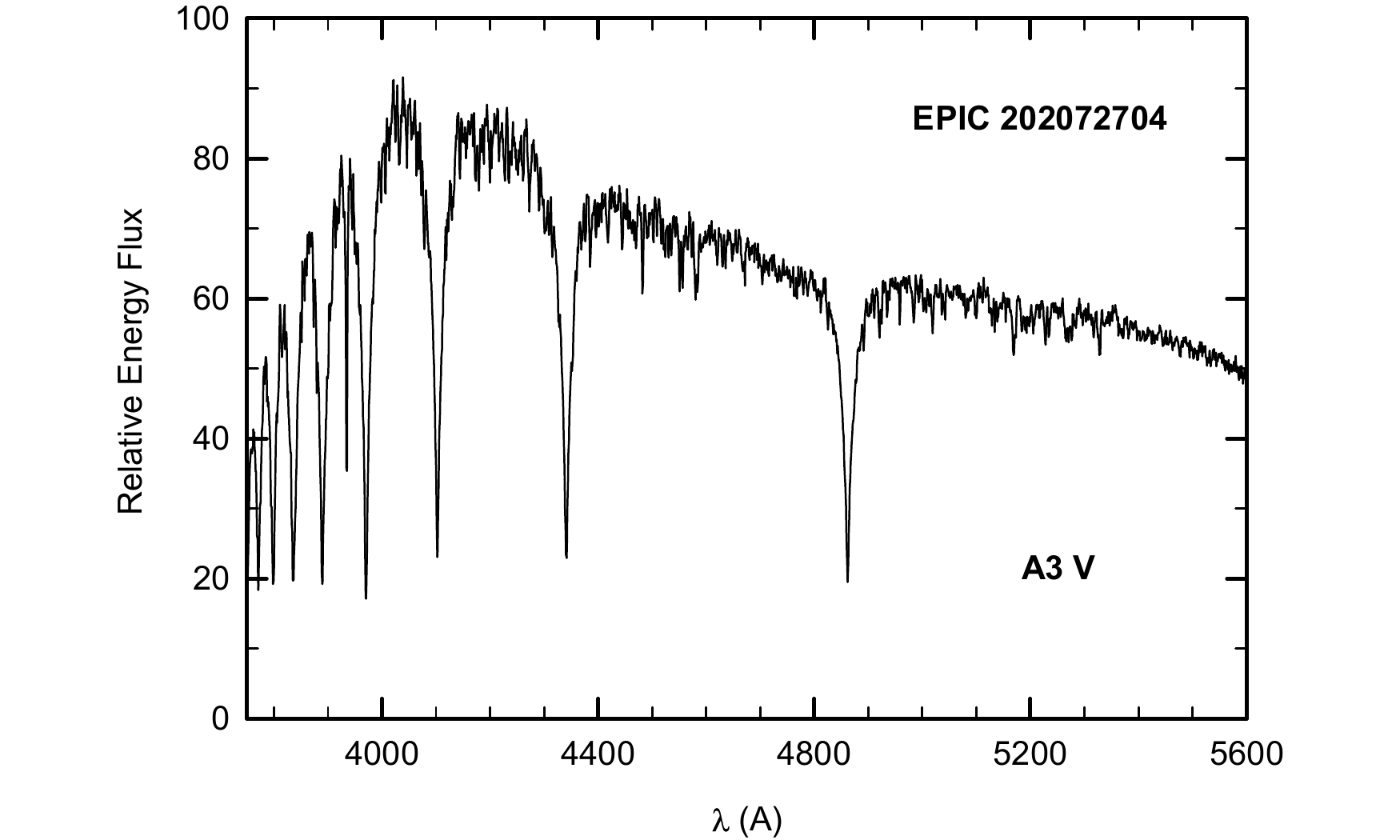} \\   
         \includegraphics[width=0.4\linewidth]{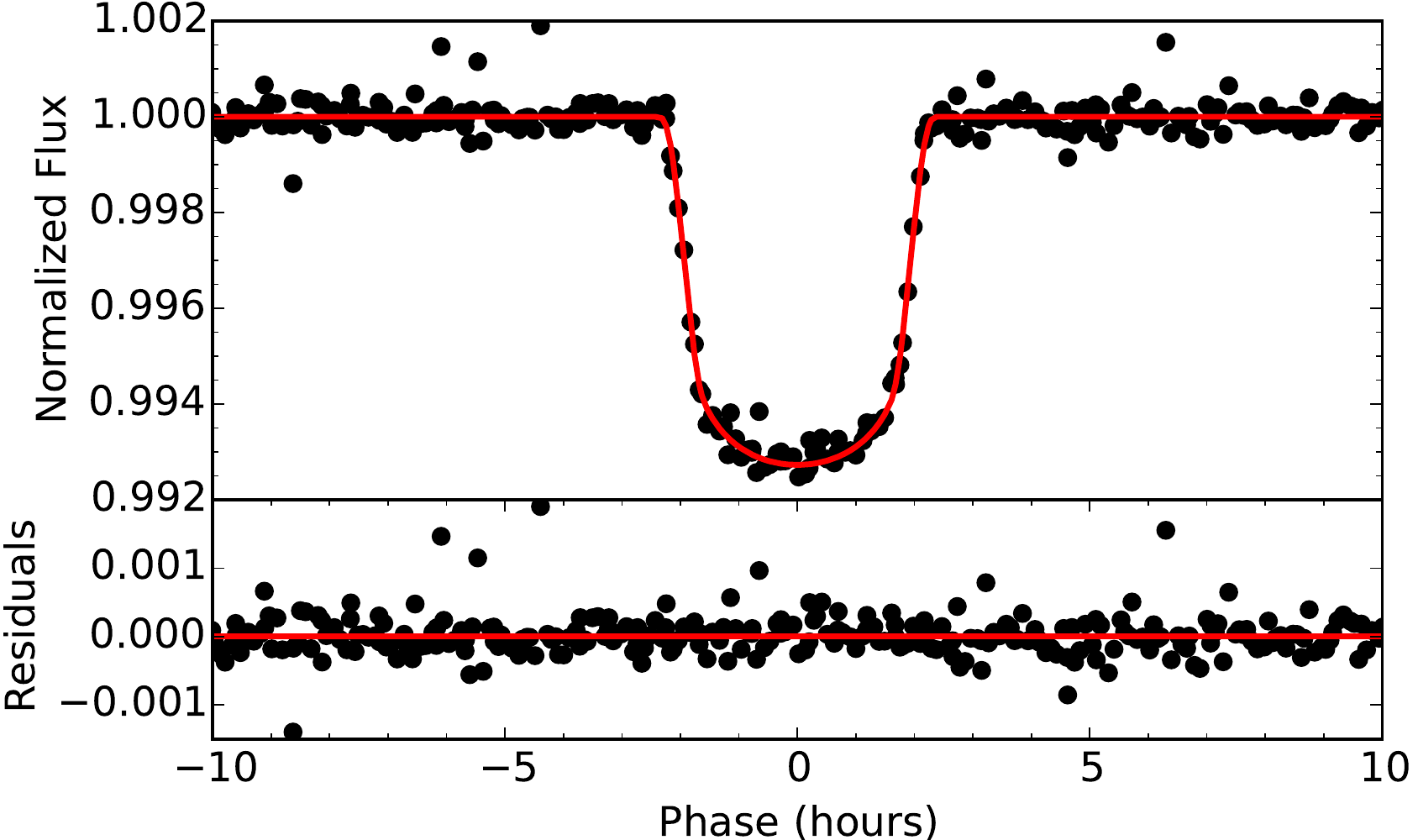}	&
        \includegraphics[trim=0.5in 3.4in 1in 3.4in, clip, width=0.4\linewidth]{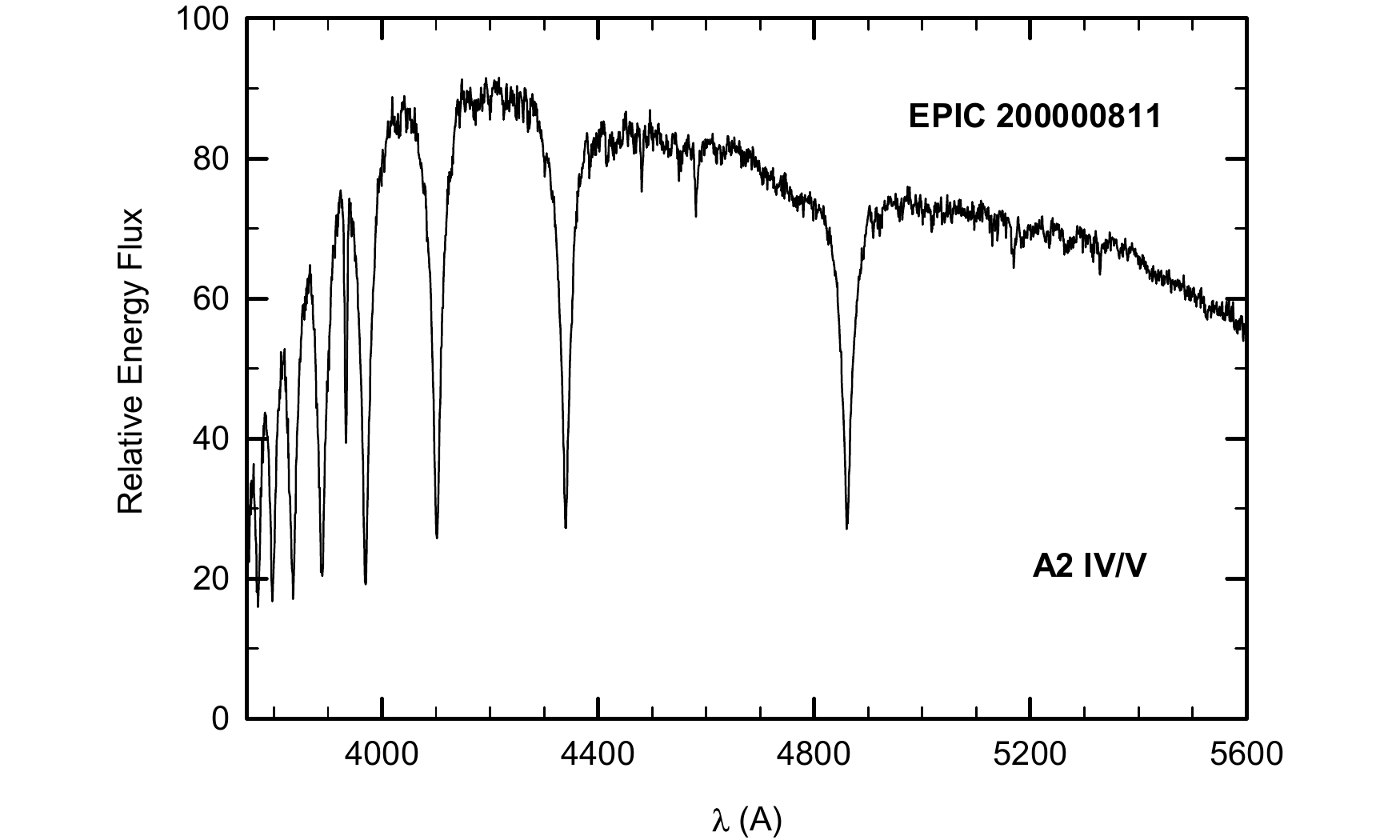} \\  
        \includegraphics[width=0.4\linewidth]{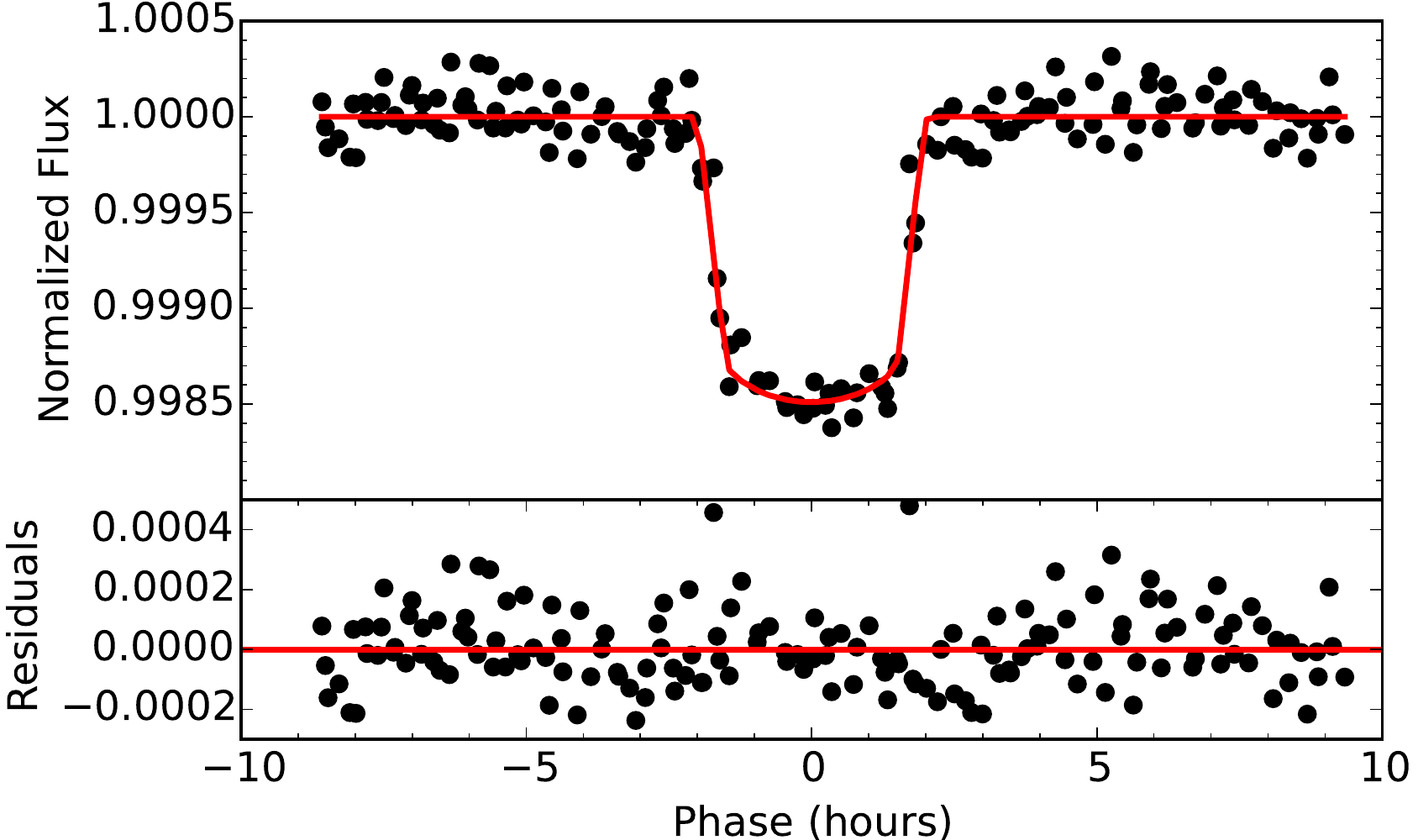}	&
        \includegraphics[trim=0.5in 3.4in 1in 3.4in, clip, width=0.4\linewidth]{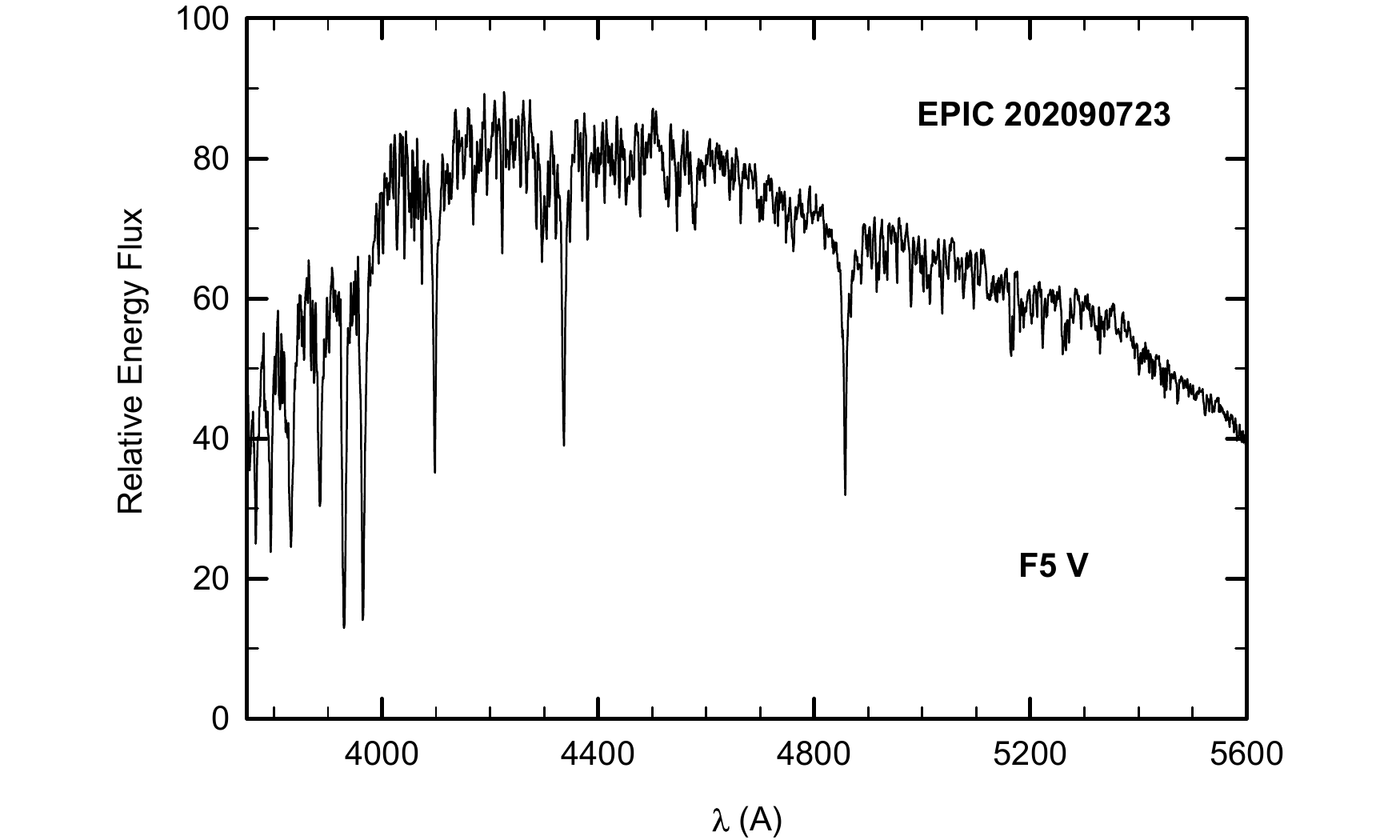} \\
    \end{tabular}
 \end{center}
\caption{{\it Left column}: Phase-folded transit light curves with TAP model over plotted in red. {\it Right column}: The associated OMM spectrum of each star with the object's spectral type in the bottom right of each panel.  The top two rows are the planet candidates identified in this paper; the bottom row is EPIC 202090723, a likely false positive.  See Section~\ref{sec:planet_candidates} for details.}
\label{fig:pcplots_1}
\end{figure*}

We classified the spectrum for each source by comparison to the Gray Digital Spectral Classification Atlas\footnote{https://ned.ipac.caltech.edu/level5/Gray/frames.html} \citep{gray09}.  We adopt a conservative uncertainty in the spectral classifications of 2 sub-classes.  After a match in spectral type and luminosity class was found, we used the reference table in \citet{boy13} to estimate the fundamental stellar properties for each classification. The spectra are presented in Figure~\ref{fig:EBspectra1} (EBs) and Figure~\ref{fig:pcplots_1} (planet candidates). The spectrum for each target is discussed where appropriate in the sections to follow. 

Nine stars for which we obtained spectroscopic \teff{} estimate overlap in the \textsc{K2-TESS} portal.  Eight of these nine stars have \teff{} estimates in the \textsc{K2-TESS} database \citep{stas14}. In all cases the photometric \textsc{K2-TESS} \teff{} determination is significantly cooler than the spectroscopic based estimate of \teff{}. Two of these stars are also present in the photometric catalog of all Tycho stars \citep{amm06}, also having similar \teff{} estimates as the {K2-TESS} estimates, though the \citet{amm06} errors are typically on the order of several thousands of degrees (e.g., EPIC 202073097 has a $T_{\rm eff} = 7418^{+4695}_{-447}$~K).  We suspect that the reason for this disagreement is that reddening is not accounted for in the \textsc{K2-TESS} Catalog.  Since the C0 field points to the galactic anti-center ($l \sim 190^{\circ} , b~\sim 5^{\circ}$) and thus much through the galactic plane, reddening makes considerable differences in the target colors which are used in the color-temperature relations by the \textsc{K2-TESS} Catalog\footnote{Dereddened $T_{\rm neff}$'s are now available in the \textsc{K2-TESS} portal.  These results generally perform better, though several hundreds of K differences from the spectroscopic values still remain in some cases.}.  As such, any of the temperatures listed in this paper (not estimated from our spectra) should be used with caution. It is also worth noting that neither method makes correction for binarity, however this task is not trivial with the small amount of data at hand. 

In any case, for this set of nine of spectroscopically observed stars, it appears that the \textsc{K2-TESS} dwarf/giant estimator method is reliable. Two stars that were observed spectroscopically were classified as likely giants (luminosity class III), and both of these were identified as giants also in the \textsc{K2-TESS} database on the basis of the reduced proper motion (see \citealt{stas14}).


\section{Ground-based Photometry}
\label{sec:kelt_phot}

With a time baseline of about 80 days, the K2 light curves can be used to identify EBs with periods out to $\sim 30$ days.  While K2 has greater detection capabilities than ground-based photometry, due to greater photometric precision and duty cycle, there is value in combining these results with those from ground-based photometric surveys and transit searches.  Projects including SuperWASP \citep{pol06, hel11}, HATNet \citep{bak04, bak13}, XO \citep{mcc05}, and KELT \citep{pep07,pep12} have accumulated observations over many years, and provide greater time baselines than K2.  The addition of longer time baselines to the K2 light curves makes it possible to search for eclipse timing variations or amplitude variations of the EBs.  

To explore the potential of that approach, we have cross-matched the K2 EBs in Tables~\ref{tab:knownEBs} and \ref{tab:newEBs} with data from the KELT survey.  KELT is a wide-field, small-aperture photometric survey, with a 23 degree x 23 degree field of view, and a nonstandard broad $V+I$ filter.  Two of the currently reduced fields from KELT overlap with K2 C0 north of Dec~$= +19^{\circ}$.  The KELT light curves were reduced and extracted according to the procedures described in \citet{siv12}, and include about 7700 to 8600 observations from Oct 2006 to March 2013.  Additional observations after that time have been gathered but not yet reduced.  Typical photometric precisions are 0.5\% rms around $V=9^{th}$ magnitude, and 2\% rms around $V=12^{th}$ magnitude. North of $+19$ degrees, KELT has light curves for all K2 C0 targets brighter than $12^{th}$ magnitude, and has light curves for at least 50\% of the K2 C0 targets in the range $12 < V < 13$.

\begin{figure*}
  \begin{center}                	
        \includegraphics[width=\linewidth]{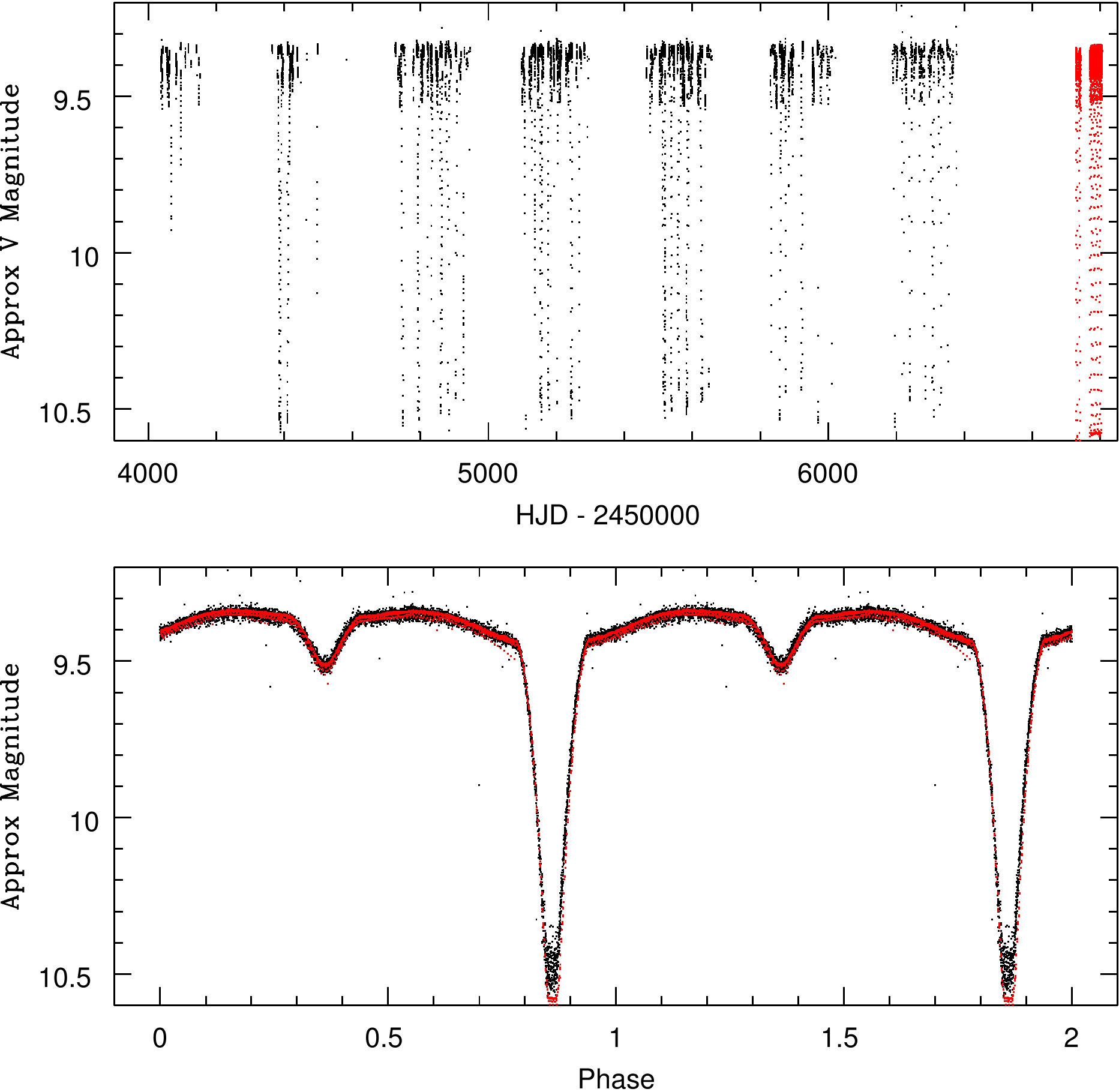} 
 \end{center}
\caption{Light curves for EPIC\,202060135 from data taken with KELT (black) and K2 (red).  The top panel displays the full light curves, showing the full power of the 7-year baseline from KELT.  The bottom panel shows the phased light curve.  The KELT light curve appears slightly shallower than the K2 light curve in the primary eclipse due to the difference in bandpasses. For details, see Section~\ref{sec:kelt_phot}.}
\label{fig:KELT_K2_full}
\end{figure*}

The long time baseline and good photometric precision of the KELT light curves allow us to measure the properties of many of the eclipsing binaries found in K2. A full analysis is ongoing, but we show an example of combining the KELT and K2 data in Figure~\ref{fig:KELT_K2_full}. Table~\ref{tab:KELT_phot} displays the overlap of targets with K2 C0 and the currently available KELT data in hand.  Potential applications include searching for eclipse depth, duration, and timing variations indicative of outer companions, or placing upper limits on such variations.

\section{Noteworthy EBs}
\label{sec:noteworthy_ebs}

\begin{figure*}
  \begin{center}
     \includegraphics[width=0.7\linewidth]{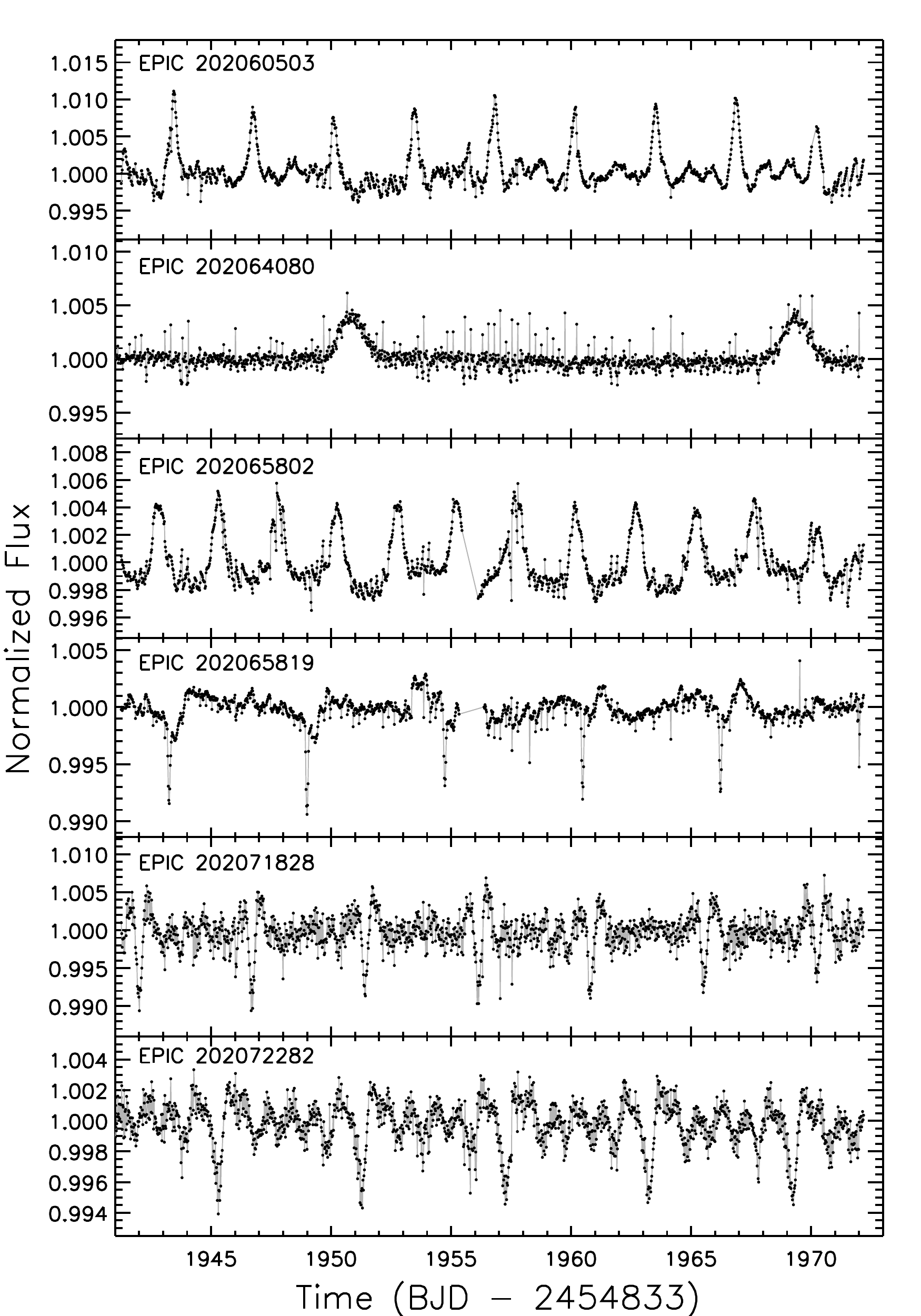}
  \end{center}
  \vspace{-0.1in}
  \caption{Light curves for 
  six Heartbeat binaries detected in our C0 sample.  See Section~\ref{sec:noteworthy_ebs} and Table~\ref{tab:newEBs} for details. }
  \label{fig:Heartbeats}
\end{figure*}

Noteworthy EBs within our final sample include ten detached eclipsing binaries with periods longer than 20 days (EPICs: 202064253, 202071945, 202072282, 202084588, 202086225, 202090938, 202091197, 202091404, 202092842, 202071645) and six candidates possessing single eclipses of significant depth but for which no period is yet determined (EPICs: 202060921, 202071902, 202072917, 202085278, 202135247, 202137580). We have also discovered six \textquoteleft Heartbeat binaries\textquoteright~(e.g., \citealt{welsh11}; EPICs: 202060503, 202064080, 202065802, 202065819, 202071828, 202072282); see Figure~\ref{fig:Heartbeats}.  The most eccentric binaries, identified from the phase difference of the observed primary and secondary minimas, are presented in Table~\ref{tab:ecc}.

\begin{figure*}
  \begin{center}
      \includegraphics[width=0.7\linewidth]{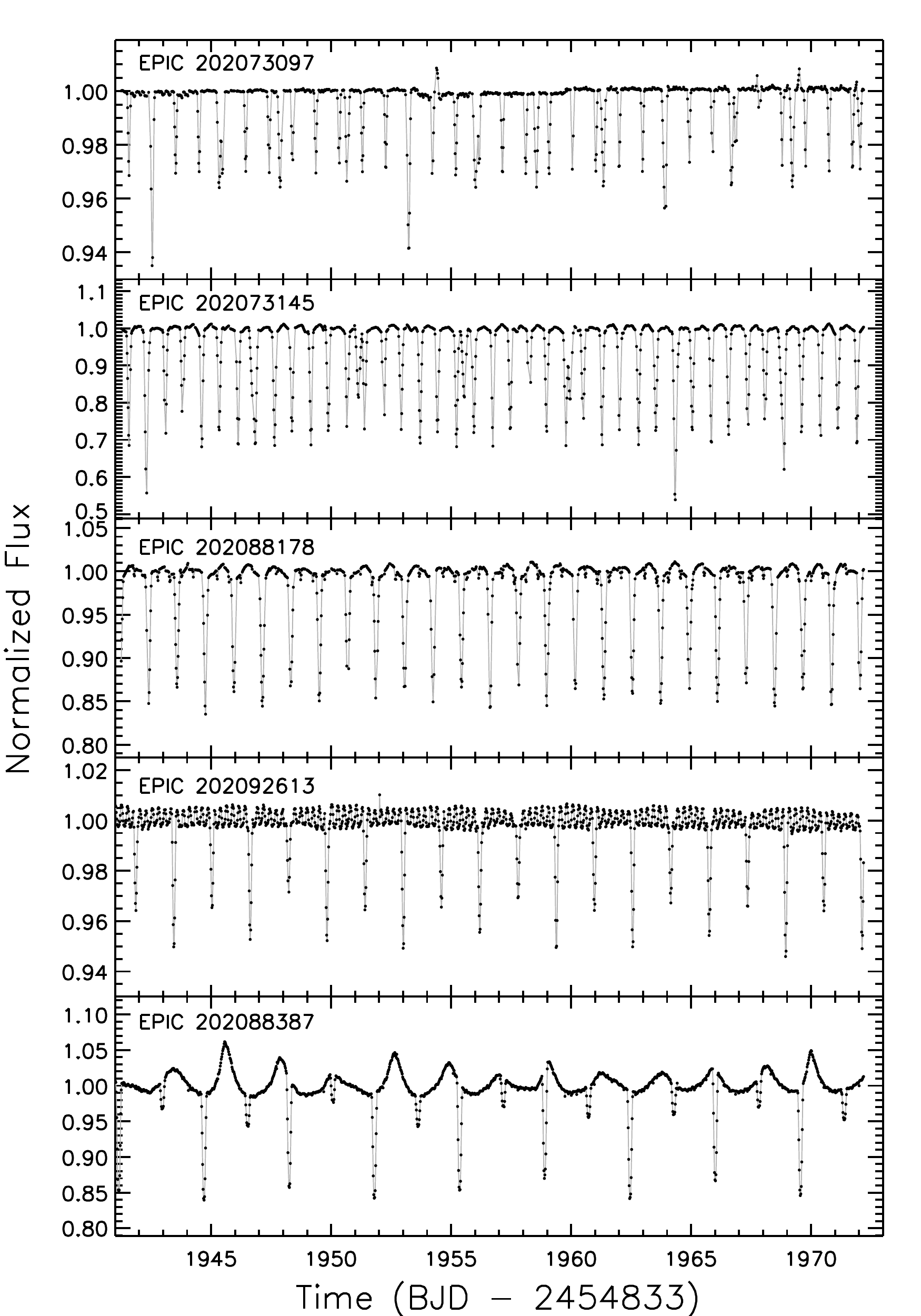}
  \end{center}
  \caption{Full light curves for notable C0 EBs. From top to bottom: EPIC 202073097, 202073145, 202088178, 202092613, and 202088387. See Section~\ref{sec:noteworthy_ebs} for details.}
  \label{fig:DetachedEbs}
\end{figure*}

\begin{figure*}
  \begin{center}
      \includegraphics[width=0.7\linewidth]{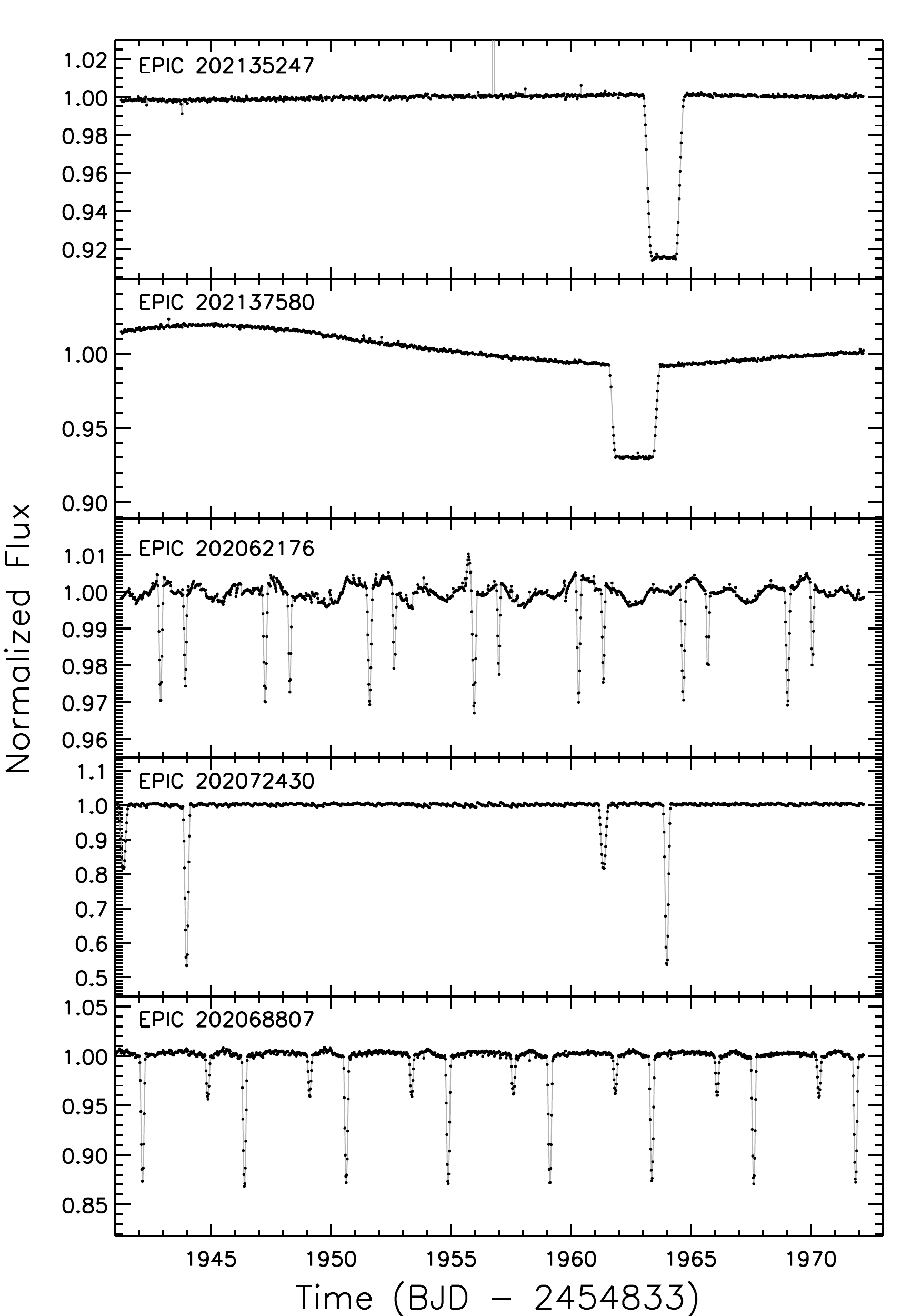}
  \end{center}
  \caption{Full light curves for notable C0 EBs. From top to bottom: EPIC 202135247, 202137580, 202062176, 202072430, and 202068807. See Section~\ref{sec:noteworthy_ebs} for details.}
  \label{fig:DetachedEbs2}
\end{figure*}

The following is a brief discussion on ten notable EBs in our sample.  Their full light curves are presented in Figures~\ref{fig:DetachedEbs} and \ref{fig:DetachedEbs2}. Figure~\ref{fig:EBspectra1} shows the spectrum of the object, if available (Section~\ref{sec:spectroscopy}).

\begin{itemize}

\item {\bf EPIC 202073097} A GO proposed EB with a period of 0.97~days and associated with a point source of {\it Kepler} magnitude $K_p=9.6$.  Alternate designations for EPIC\,202073097 include HD\,251042 and 2MASS 06042191+2032032. The C0 light curve shows a second set of eclipses of similar depth and duration with a period of 5.33~days. The five day period is not observed in extractions from the halo pixel sources within the TPF for EPIC 202073097. It is unclear if this anomaly represents two independent binaries or a gravitationally bound pair of binaries in a quadruple system. Spectral analysis indicates a late A1 or A0 component is present, possibly evolved as \ion{Fe}{2} and \ion{Si}{2} are observed. This is consistent with the The Henry Draper Catalogue classification of A2 \citep{can93}.

\item {\bf EPIC 202073145} A GO proposed EB with a period of 1.52 days, 30\% depth, and $K_p=11.3$. The light curve also displays an additional anomalous signal with a depth of 15\% and a period of 4.44 days.  Alternate designations for EPIC\,202073145 include TYC 1323-169-1 and 2MASS 06185463+2036038. It is unclear solely from a pixel-level review if the longer period represents a close background contaminator or a second binary in a quadruple system. The \textsc{K2-TESS} Catalog lists $T_{\rm eff} = 6208$~K for this source; our spectrum indicates a considerably hotter B3V type star, which is assumed to be un-evolved due to the absence of \ion{O}{2}.

\item {\bf EPIC 202088178} A new EB from this paper with a period of 2.37 days, 15\% depth, and $K_p = 13.3$.  EPIC\,202088178 is also known as 2MASS 06230702+1828149. An additional series of eclipses are observed with a shorter period of 0.49 days, and a depth of only 0.5\%. A pixel-level review was inconclusive but cannot rule out with confidence that the light curve contains two unresolved binary systems. The \textsc{K2-TESS} Catalog lists $T_{\rm eff} = 5270$~K for this source.  Our OMM spectrum indicates the star has a true temperature much higher with a spectral type of F5~II. 

\item {\bf EPIC 202092613} A new EB with two periodic signals: a 3.18~day period with a primary eclipse depth of 5\%, and a 0.25-day periodic modulation with an amplitude of 0.5\%. .  The additional short period signal is not resonant with the 3.18 day period, and we suspect the cause of this 0.5\% amplitude modulation is attributed to either pulsations or star spots common to one of the binary components. We classify the OMM spectrum of this object as a A0V based on Balmer line strengths, weak \ion{Mg}{2} ratio and lack of \ion{Si}{2} in the OMM spectrum.  If the observed short period variability is driven by star spots, it could indicate a rapid rotation period of $\sim 6$~hours. However, given the stellar effective temperature and ratio of eclipse depths it is more likely that the short period signal represents pulsations.   

\item {\bf EPIC 202088387} A new EB with a period of $3.55$~days, depth of 15\%, and $K_p=12.9$.  Alternate designations for EPIC\,202088387 include 2MASS 06064169+2234547 and SDSS J060641.69+223454.8. The light curve also exhibits asymmetrical brightening events of $\sim 2$~days duration with an amplitude $\sim 5$\%, which we assume to be associated with nearly co-rotating starspots or pulsations in one of the stellar components. The \textsc{K2-TESS} Catalog $T_{\rm eff} =5652$~K; we do not have a spectrum of this object to confirm this temperature.  

\item {\bf EPIC 202135247} A new EB ($K_p = 14.4$; also known as 2MASS 06195142+1747474) with unknown period -- a single, flat-bottomed, 9\% eclipse depth with a duration of 35 hours is observed. The \textsc{K2-TESS} Catalog lists $T_{\rm eff}=3919$~K.  Our OMM spectrum suggests an earlier-type primary component of G1.5. While the G-band strengths in our spectrum could be representative of either a luminosity class of V or III, the eclipse profile suggests an evolved star and annular eclipse of a smaller stellar companion. 

\item {\bf EPIC 202137580} A new EB ($K_p = 13.1$; also known as 2MASS 06031044+2330174) with unknown period. The light curve has a single 7\% flat bottomed eclipse with a 37~hour duration.  While it has a similar eclipse profile to EPIC~202135247 (above) which may suggest an evolved component, the observed impact durations are considerably different as evidenced by shorter ingress and egress coupled to a longer disc passage of 28~hours. With our OMM spectrum, we classify the primary component as G0, much hotter than $T_{\rm eff}=3755$~K from the \textsc{K2-TESS} Catalog.   The evolutionary state of this target is unclear as the strength of the G-band in our spectrum rules out a supergiant, but does support either luminosity class V or III. An unexplained long term trend in the light curve is visible and may be associated with orbital modulation and a period of under $\sim 100$~days. 

\item {\bf EPIC 202062176} An eccentric (Table~\ref{tab:ecc}), new EB also known as 2MASS 06095262+2030273 having a period of 4.35~days, eclipse depth of $\sim 3.5$\%, and $K_p = 11$. Independently proposed by two GO programs for massive stars and a third for exoplanet detection, this EPIC may be a member of NGC~2174/2175. 

\item {\bf EPIC 202072430} Also known as TYC 1330-2152-1 and 2MASS 06402451+1508324, EPIC 202072430 is one of the ten longest period eclipsing binaries identified in our sample with $P= 20.03$~days. This eccentric (Table~\ref{tab:ecc}) EB has a $K_p = 10.9$ and 45\% primary eclipse depths. Two orbital cycles are captured in the C0 data; the light curve exhibits additional quasi-periodic variability in the out of eclipse regions on the order of $\sim 8$~hours.

\item {\bf EPIC 202068807} EPIC 202068807 ($K_p=12.01$), also known as TYC 1878-947-1 and 2MASS 06204185+2317264 is a new, eccentric (Table~\ref{tab:ecc}), EB with a 4.24 day period.  The light curve has primary and secondary eclipse depths of $\sim 12$ and 4\%, respectively. This eccentric binary displays out-of-eclipse asymmetrical ellipsoidal variation at the 0.5\% level.

\end{itemize}

\section{Exoplanet Candidates}               %
\label{sec:planet_candidates}

We report the detections of transiting planet candidates in the {\it K2} C0 data set: 
EPIC 202072704, and 2MASS 06101557+2436535 (no EPIC assignment as of this writing). The light curves for these objects were initially detected via the processing methods described in Section~\ref{sec:C0_data}. After a detailed search of the TPF to rule out obvious contamination effects from other sources, the light curves were normalized and detrended specifically for analysis using the Transit Analysis Package \citep{gaz11} in order to make a preliminary validation of their sub-stellar nature. HAT-P-54b, recently discovered by \citep{bakos14}, was also identified in our analysis, but we do not elaborate upon it here.

In an effort to better characterize the candidate host stars in our orbital fits, we obtained a spectrum of each of the new planet candidate host stars (excluding HAT-P-54b) with the OMM spectrograph (see Section~\ref{sec:spectroscopy} for details). The spectra reveal that the hosts are early-type stars. Our  best-fit transit model retains a sub-stellar nature for each of the transiting companions, and we consider them new {\it K2} objects of interest worthy of further follow up. The candidates common to these systems are of particular interest as there are currently few confirmed examples in the literature of exoplanets transiting early-type stars (e.g., WASP-33 \citealt{cam10}, and Kepler-13 \citealt{shp11, shp14}).

The orbital solutions for each C0 candidate are shown in Table~\ref{tab:PCs} and the phase-folded transit fits are shown in Figure~\ref{fig:pcplots_1} with the spectrum of each point source.

\begin{itemize}

\item {\bf EPIC 202072704} This host star candidate has a $K_p=11.4$, with alternate designations including HD 263309 and 2MASS 06455102+1712250. We identify twelve transit events of 3.16 hours duration, depths of 6588~ppm, and a period of 2.65~days. There are no apparent contaminating sources within the TPF halo. The OMM spectrum suggests the host star is an A3V, consistent with the Henry Draper Catalogue classification of A5 \citep{can93}. Using the tables from \citet{pec13}, we assume the host has a nominal radius of $1.7$~R$_{\odot}$. A best fit transit model based on these parameters indicates the transiting object has a radius of $1.26$~\rjup~($14.1$~\rearth)~and orbital radius of 0.04~AU. 

\item {\bf 2MASS 06101557+2436535} Located within the first listed TPF (200000811) for the M35 star cluster super stamp, a transit signal is recovered from a $V=12.6$  point source associated with pixel coordinates corresponding to column number 52 and row number 36. This source has been cross matched to 2MASS 06101557+2436535, as well as KIC 27058 from the original Kepler Input Catalog, although it lies out of the {\it Kepler} field of view \citep{bro11}. The light curve for this source displays five transits with a period of 7.5559~days, and depths of 6928~ppm. We are able to recover two additional transits of acceptable quality in the early coarse point data. Using the OMM spectrum, we classify this source as an A2IV/V star. Assuming a stellar radius of 1.8~R$_{\odot}$ \citep{pec13}, our best fit transit model results in a planet radius of $1.37$~\rjup~($14.50$~\rearth)~object orbiting at a separation of 0.08~AU. Further stellar characterization is needed to determine the evolutionary state of the host star, as its radius is quite uncertain. 

\end{itemize}

In addition to these planet candidates, EPIC 202066192 and EPIC 202090723 were initially noted as potential candidates.  However, EPIC 202066192, was ultimately rejected as a false positive due to a V-shaped eclipse profile, high impact parameter, and early-type A0III-IV OMM spectral classification.  
EPIC 202090723's spectral type is determined from our OMM spectrum to be F5V -- perfect agreement with the F5V classification from \citet{mcc67}.  Although the best fit transit model resulted in a planet radius of $\sim 0.5$~\rjup, a single high-resolution TRES spectrum showed signs of a secondary in the line profile (D. Latham, private communication). Note that our spectral classification infers a source about a thousand degrees hotter than the \textsc{K2-TESS} Catalog estimate of $T_{\rm eff} = 5442$~K, similar to other discrepancies identified in this paper.

\section{Eclipse Timing Variations (ETVs)}
\label{sec:etvs}

Eclipse Timing Variations (ETVs) can be induced by the gravitational presence of a third object in a system \citep{con14a}. Particularly so for short period binaries, the hierarchical triple system occurrence rate is estimated to be 40\% or greater \citep{toko06}. The Kepler Eclipsing Binary Catalog provides comprehensive eclipse timings from which over 100 confirmed and candidate triple systems have been identified (see also \citealt{rap13,bork15}). 

Using the periods and ephemerides derived in Section~\ref{sec:pixel_level_analysis}, we analyzed the C0 light curves for a deviation from a strictly linear ephemeris within the threshold of {\it K2}'s 30 minute observation cadence. This analysis resulted in several C0 EBs that display intriguing low amplitude trends and sinusoidal variations. However, these observed effects could be driven by interactions that do not require a third object (e.g. dynamical mass transfer, quadrupole coupling, apsidal motion; \citealt{con14a}). We are further burdened by the limited baseline for C0, in that most third body ETVs exhibit an O-C residual period greater than 45~days \citep{rap13}. This makes it unlikely for a full ETV cycle to have been serendipitously captured for a given target. We conclude that our analyses returned no high confidence ETV periods that clearly suggest the presence of a third body companion.

\section{Summary and Discussion}
\label{sec:summary}

Despite the new challenges inherent in working with currently available {\it K2} data, the overall quality of these light curves, when properly decorrelated, is still well suited for the detection of eclipsing binaries.  A total of \totalebs eclipsing binaries are recovered by visual inspection of the {\it K2} C0 data set. For the \totalebsnew newly identified EBs, we provide periods, ephemerides, and cross-matched catalog identifications. Due to the shorter baseline of C0 compared to the {\it Kepler} data set, the balance of detected eclipsing systems is unsurprisingly dominated by short period binaries, with a steep rise in  EBs with periods of less than five days. Of \totalebs systems identified in the {\it K2} C0 data, 58 have periods in excess of five days. The majority of these (42 systems) are previously undetected with ground photometry (Table~\ref{tab:newEBs}). The Kepler Eclipsing Binaries Catalog has been expanded to include all EPIC EBs\footnote{\url{keplerebs.villanova.edu/k2}} in the C0 field with three or more observed eclipses. The EPICs contaminated by neighboring EBs are cataloged as false positives in the Kepler Eclipsing Binary Catalog, having identifiers associated with the EPIC target that led to their detection. 

\begin{figure*} 
  \begin{center}
    \includegraphics[width=0.8\linewidth]{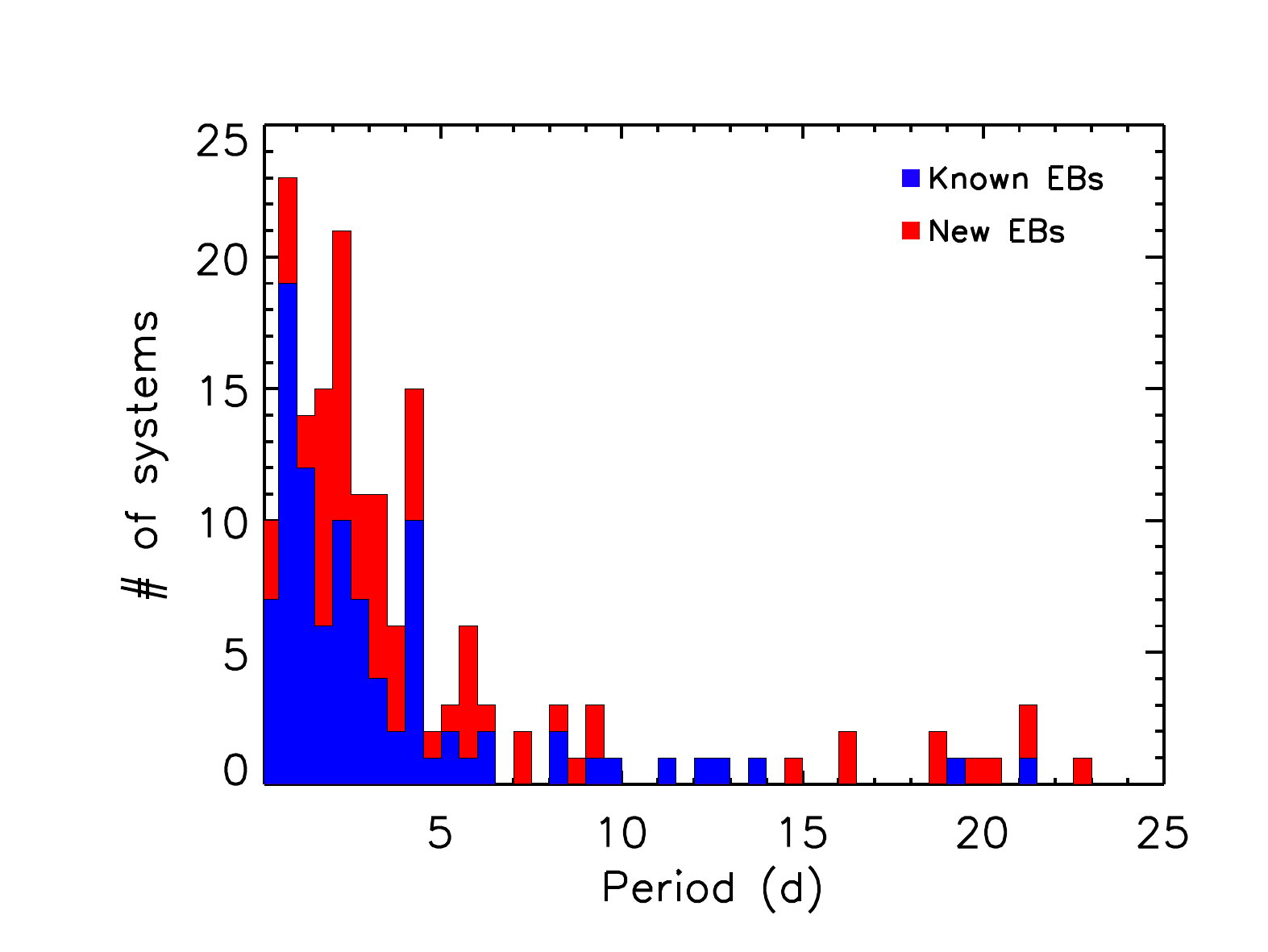}
  \end{center}
  \caption{Period distribution for new and known K2 C0 eclipsing binaries with periods under 25 days.}
  \label{fig:period_distro}
\end{figure*}

The period distribution for the K2 C0 EBs with orbital periods less than 25 days is displayed in Figure \ref{fig:period_distro}. Due to unknown selection biases it is not possible to evaluate incompleteness or draw firm conclusions regarding the fractional occurrence of EBs found in the C0 data set. Overall occurrence rate for all EBs in our {\it K2} C0 sample is noted at 2.5\%, which is slightly higher than the 1.6\% determined for {\it Kepler}. Despite this, we detect an underrepresented range of EBs with periods of less than one day; the cause of this shortage is currently unknown but likely to be influenced by GO target selection bias and the self-imposed limitations of our own selection processes. We further recognize that the decorrelation methods utilized in our survey have played a major role in our recovery ratios and overall completeness. While \textsc{KepSFF} and \textsc{KepPCA} provide an excellent foundation from which to reduce {\it K2} data, intrinsic astrophysical signals can be obfuscated or removed by such solutions and care must be taken in their application. Alternative approaches to mitigating the effect of spacecraft pointing jitter are being explored, such as those described by \citet{agr15} and \citet{for15}.

There are compelling incentives that encourage the community to analyze each incoming {\it K2} Campaign data set in a diligent and timely manner in order to characterize new eclipsing binary and exoplanet candidates, particularly in cases where the host star is bright or close. Many such systems are within reach from the ecliptic and {\it K2} stands uniquely poised to provide high precision photometry for a limited selection of targets with overlapping coverage common to the upcoming TESS \citep{ric14} mission, which will begin an all sky survey little more than one year after the currently scheduled conclusion of {\it K2} operations. For such systems meeting TESS selection criteria, the {\it K2} campaign baseline of 80 days will provide expanded pre-coverage that should serve to make prospective detections of long period objects (e.g., objects that transit only once or twice in a {\it K2} baseline) whose signals might conceivably manifest themselves as transient dimming events during ecliptic-orientated TESS observations. While {\it K2} observations alone cannot fully characterize such long period events, their initial identification will allow potential spectroscopic or photometric follow-up from ground based or in-orbit facilities.

In addition, any M or K dwarfs local to Earth (e.g. within 20 parsecs) that can be shown to be binaries or blended binaries via {\it K2} photometry (or subsequent ground based follow-up) constitute a limited yet important early sample from a population of high priority targets for the upcoming JWST mission. Owing to their reduced stellar radii, transits and eclipses occurring in such systems can yield significantly higher final average SNR values and in some cases allow limited transmission spectroscopy. However, by consequence, the deeper transits noted for such stellar types must be carefully vetted against the host of false positive possibilities. The JWST Continuous Viewing Zone will have some degree of overlap with {\it K2} ecliptic fields, but the instrument will also possess a finite supply of thruster fuel. Thus, it is critical that no exoplanet follow-up observations be squandered on a previously unidentified EB or BGEB.

{\it Note Added in Manuscript:} Since our paper was essentially completed, we have learned of a related study of the {\it K2} Field 0 by \citep{arm15}. The \textsc{K2 Variable Catalog II} lists 2619 Field 0 variable objects including 137 EBs, but used a qualitatively different detrending algorithm than we have employed and does not include full ephemerides or cross matched identifications.

\acknowledgments

Funding for the {\it Kepler} and {\it K2} Discovery missions is provided by NASAs Science Mission Directorate. TSB, DAF, JW, and JRS acknowledge support provided through NASA grant ADAP12-0172 and ADAP14-0245. LN thanks the Natural Sciences and Engineering Research Council (NSERC) of Canada for financial support, and the staff at the Observatoire Astronomique du Mont-M\'egantic for their technical assistance. Some of the data presented in this paper were obtained from the Mikulski Archive for Space Telescopes (MAST). STScI is operated by the Association of Universities for Research in Astronomy, Inc., under NASA contract NAS5-26555. Support for MAST for non-HST data is provided by the NASA Office of Space Science via grant NNX13AC07G and by other grants and contracts. This research has made use of the NASA Exoplanet Archive, which is operated by the California Institute of Technology, under contract with the National Aeronautics and Space Administration
under the Exoplanet Exploration Program. This research has made use of the VizieR catalogue access tool, CDS, Strasbourg, France. The original description of the VizieR service was published in \citet{och00}. This research has made use of the SIMBAD database, operated at CDS, Strasbourg, France \citep{wen00}. This research has made use of the International Variable Star Index (VSX) database, operated at AAVSO, Cambridge, Massachusetts, USA. This publication makes use of data products from the Two Micron All Sky Survey, which is a joint project of the University of Massachusetts and the Infrared Processing and Analysis Center/California Institute of Technology, funded by the National Aeronautics and Space Administration and the National Science Foundation. This publication makes use of data products from the Wide-field Infrared Survey Explorer, which is a joint project of the University of California, Los Angeles, and the Jet Propulsion Laboratory/California Institute of Technology, funded by the National Aeronautics and Space Administration. The Digitized Sky Surveys were produced at the Space Telescope Science Institute under U.S. Government grant NAG W-2166. The images of these surveys are based on photographic data obtained using the Oschin Schmidt Telescope on Palomar Mountain and the UK Schmidt Telescope. The plates were processed into the present compressed digital form with the permission of these institutions.

The authors gratefully acknowledge Ball Aerospace, the Kepler/K2 team and the Guest Observer community for making the {\it K2} mission possible.



{\it Facilities:} \facility{Kepler/K2}, \facility{Observatoire Astronomique du Mont-M\'egantic's (OMM)}




\bibliographystyle{plainnat}


\clearpage

\clearpage

\clearpage

\clearpage

\clearpage

\clearpage

\clearpage

\clearpage

\clearpage

\clearpage

\clearpage

\clearpage

\clearpage
  

\clearpage

\clearpage


\clearpage

\clearpage

\clearpage

\clearpage

\clearpage


\clearpage



\clearpage
\begin{landscape}

\begin{deluxetable}{lcccccccc}
\tabletypesize{\scriptsize}
\tablecaption{Previously Known EBs in Campaign 0\label{tab:knownEBs}}
\tablewidth{0pt}
\tablehead{
\colhead{EPIC} & 
\colhead{$K_{\rm P}$}   & 
\colhead{RA} & 
\colhead{DEC} &
\colhead{morph} &
\colhead{Period} &
\colhead{BJD$_0$} &
\colhead{Cross matched} &
\colhead{ }\\
\colhead{ } & 
\colhead{(mag)} &
\colhead{(deg)} & 
\colhead{(deg)} &
\colhead{} &
\colhead{(days)} &
\colhead{(BJD-2454833)} &
\colhead{identification} &
\colhead{Note Flag}
}
\startdata

202059360	&	 \phn 9.00	&	 \phn 96.23357	&	14.40689	&	0.63	&	3.674801 	&	1942.5111	&	CCDM J06249+1 (**); 2MASS 06245606+1424249	&		\\
202059377	&	 \phn 8.70	&	 \phn 99.18429	&	21.65157	&	0.55	&	1.576882 	&	1942.7053	&	HD 46882 (Star); 2MASS 06364422+2139056	&		\\
202059416	&	 \phn 9.70	&	 \phn 102.71083	&	16.23806	&	0.59	&	0.900783 	&	1941.6122	&	HD 264884 (Star); 2MASS 06505062+1614177	&		\\
202060124	&	 \phn 9.10	&	 \phn 94.60670	&	23.57145	&	0.65	&	4.285046	&	1943.0287	&	V* LT Gem (EB*); 2MASS 06182560+2334172	&		\\
202060135	&	 \phn 9.89	&	 \phn 90.36692	&	23.14098	&	0.55	&	2.865379 	&	1943.2567	&	V* RW Gem (EB*Algol); 2MASS 06012805+2308274	&		\\
202060541	&	 10.27	&	 \phn 94.08259	&	20.91274	&	0.76	&	4.049160 	&	1944.4214	&	HD 254205 (Star); 2MASS 06161955+2054502	&		\\
202060752	&	 10.80	&	 \phn 92.86670	&	23.23170	&	0.74	&	0.770391 	&	1941.4077	&	ASAS J061128+2313.9; 2MASS 06112770+2313532	&		\\
202061814	&	 13.40	&	 \phn 91.76612	&	24.22436	&	0.82	&	1.021701 	&	1941.9754	&	[GMB2010] WO (SB); 2MASS 06070387+2413277	&		\\
202062154	&	 11.50	&	 \phn 93.64370	&	18.47454	&	0.84	&	13.643500 \phn 	&	1944.09	&	V* V2789 Ori (EB*betLyr); 2MASS 06143448+1828283	&		\\
202063828	&	 10.60	&	 102.66519	&	21.36553	&	0.57	&	1.243528 	&	1943.1855	&	V* AF Gem (EB*Algol); 2MASS 06503974+2121596	&		\\
202064026	&	 11.20	&	 \phn 96.25542	&	17.97021	&	0.52	&	4.067909 	&	1944.0496	&	V* BO Gem (EB*Algol); 2MASS 06250115+1758032	&		\\
202064535	&	 11.30	&	 \phn 97.69583	&	27.47333	&	0.80	&	0.508341 	&	1941.5144	&	ASAS J063047+2728.4; 2MASS 06304688+2728258	&		\\
202064549	&	 13.10	&	 \phn 97.62533	&	28.38753	&	0.13	&	9.763444 	&	1944.8127	&	2MASS 06302924+2823149	&		\\
202065543	&	 10.40	&	 101.06046	&	16.40108	&	0.59	&	1.446499 	&	1941.5102	&	V* V382 Gem (EB*Algol); 2MASS 06441514+1623599	&		\\
202071266	&	 12.50	&	 \phn 91.33042	&	22.54782	&	0.78	&	2.593621 	&	1944.18	&	V* DS Gem (Orion\_V*); 2MASS 06051954+2232527	&		\\
202071842	&	 9.90	&	 100.90493	&	21.52375	&	0.30	&	4.168098 	&	1943.0289	&	TYC 1342-1581-1 (Star); 2MASS 06433718+2131255	&		\\
202072451	&	 11.30	&	 \phn 99.40000	&	17.79239	&	0.25	&	11.306934 \phn 	&	1949.92	&	V0400 Gem (EB*Algol); 2MASS 06373603+1747326	&		\\
202072929   &    10.50  &     \phn 97.98333  &  19.66989    &   0.50    &   3.05    &   1943.20 &   V* AY Gem (EB*Algol); 2MASS 06315619+1940115    &   \\
202072932	&	 10.50	&	 \phn 95.30965	&	26.12376	&	0.39	&	3.133688 	&	1941.5201	&	TYC 1882-433-1 (Star); 2MASS 06211420+2607209	&		\\
202072933	&	 12.20	&	 100.73799	&	22.98607	&	0.06	&	3.144429 	&	1944.0016	&	2MASS 06425680+2259128	&		\\
202072941	&	 14.80	&	 \phn 96.66667	&	18.70383	&	0.54	&	12.447767 	&	1950.07	&	V* DU Gem (EB*Algol); 2MASS 06263978+1842228	&		\\
202072958	&	 10.50	&	 \phn 94.19764	&	24.07841	&	0.50	&	4.024547 	&	1942.6683	&	TYC 1878-1258-1 (Star); 2MASS 06164724+2404389	&		\\
202072961	&	 \phn 9.30	&	 100.12564	&	22.91052	&	0.40	&	4.097024 	&	1945.3702	&	HD 261572 (Star); 2MASS 06402979+2254445	&		\\
202072963	&	 11.40	&	 104.37154	&	20.18725	&	0.41	&	4.209922 	&	1941.5103	&	TYC 1352-1219-1 (Star); 2MASS 06572939+2011151	&		\\
202072971	&	 15.50	&	 \phn 96.74292	&	18.66639	&	0.58	&	4.399157 	&	1944.2988	&	V* DV Gem (EB*Algol); 2MASS 06265777+1839547	&		\\
202072978	&	 12.20	&	 100.26213	&	23.80780	&	0.54	&	4.885995 	&	1944.6566	&	CCDM J06410+2 (*in**); 2MASS 06410235+2348198	&		\\
202072988	&	 10.60	&	 \phn 91.75583	&	17.70056	&	0.52	&	5.322461 	&	1946.1201	&	V* CP Ori (EB*Algol); 2MASS 06070185+1741581	&		\\
202072991	&	 10.20	&	 \phn 95.00300	&	26.34982	&	0.47	&	5.495907 	&	1945.3442	&	V* V396 Gem (EB*Algol); 2MASS 06200072+2620593	&		\\
202073004	&	 15.20	&	 101.46042	&	17.02611	&	0.57	&	6.154338 	&	1946.518	&	V* KU Gem (EB*Algol); 2MASS 06455049+1701347	&		\\
202073035	&	 12.20	&	 103.93750	&	20.50333	&	0.83	&	21.4100 \phn 	&	1945.59	&	ASAS J065545+2030.2; 2MASS 06554469+2030131	&		\\
202073040	&	 12.20	&	 \phn 91.27080	&	20.53500	&	0.55	&	2.121410 	&	1941.4715	&	TYC 1321-16-1 (Star); 2MASS 06050489+2032128	&		\\
202073043	&	 13.70	&	 \phn 99.92500	&	17.19211	&	\nodata	&	52.266500 \phn 	&	1949.445	&	V* EU Gem (EB*Algol); 2MASS 06394202+1711355	&	2	\\
202073061	&	 14.60	&	 \phn 93.52667	&	18.20556	&	0.06	&	2.482188 	&	1943.79	&	V* V644 Ori (EB*Algol); 2MASS 06140639+1812098	&		\\
202073063	&	 11.50	&	 \phn 91.50028	&	21.00122	&	0.45	&	2.412447 	&	1942.85	&	TYC 1325-414-1 (Star); 2MASS 06055967+2100001	&		\\
202073067	&	 13.80	&	 101.22500	&	22.27639	&	0.61	&	2.349227 	&	1943.206	&	V* IP Gem (EB*betLyr); 2MASS 06445396+2216348	&		\\
202073068	&	 12.80	&	 \phn 91.34558	&	22.67711	&	0.58	&	2.334819 	&	1942.5926	&	TYC 1864-1836-1 (Candidate\_EB*); 2MASS 06052293+2240375	&		\\
202073074	&	 12.50	&	 \phn 91.34580	&	20.53500	&	0.54	&	2.243308 	&	1942.6813	&	2MASS 06052247+2032120 &		\\
202073088	&	 14.50	&	 \phn 92.96083	&	19.95389	&	0.57	&	0.824149 	&	1941.3464	&	V* V668 Ori (EB*Algol); 2MASS 06115063+1957141	&		\\
202073096	&	 11.80	&	 \phn 98.97030	&	18.70835	&	0.93	&	1.960230 	&	1942.4908	&	TYC 1333-46-1 (Star); 2MASS 06355255+1842337	&		\\
202073097	&	  9.60	&	 \phn 91.09233	&	20.53502	&	0.36	&	0.972411 	&	1941.9621	&	HD 251042 (Star); 2MASS 06042191+2032032	&	3	\\
202073117	&	 12.70	&	 \phn 93.13750	&	19.53667	&	0.59	&	1.762277 	&	1942.9571	&	NSVS 9739376 (Candidate\_EB*); 2MASS 06123367+1932132	&		\\
202073121	&	 15.70	&	 \phn 96.85250	&	23.82528	&	0.52	&	4.201499 	&	1944.51	&	V* HU Gem(EB*Algol); 2MASS 06272454+2349310	&		\\
202073124	&	 13.40	&	 \phn 99.40333	&	19.60750	&	0.55	&	1.677702 	&	1941.5202	&	V* TZ Gem (EB*Algol); 2MASS 06373732+1936236	&		\\
202073144	&	 11.10	&	 100.21498	&	26.74048	&	0.56	&	1.550610 	&	1941.7757	&	TYC 1888-310-1 (Star); 2MASS 06405105+2644286	&		\\
202073145	&	 11.30	&	 \phn 94.72917	&	20.59833	&	0.54	&	1.516638 	&	1942.2704	&	TYC 1323-169-1 (Star); 2MASS 06185463+2036038	&	3	\\
202073160	&	 13.20	&	 \phn 97.65417	&	19.64122	&	0.57	&	1.412923 	&	1942.47	&	V* CK Gem  (EB*Algol); 2MASS 06303710+1938280	&		\\
202073161	&	 14.80	&	 \phn 97.89708	&	17.90667	&	0.54	&	1.403637 	&	1942.3089	&	V* CM Gem (EB*betLyr); 2MASS 06313532+1754237	&		\\
202073174	&	 12.00	&	 103.94292	&	18.59806	&	0.65	&	1.288439 	&	1941.7416	&	TYC 1335-728-1 (Star); 2MASS 06554632+1835554	&		\\
202073175	&	 12.60	&	 \phn 92.07080	&	20.75330	&	0.54	&	1.281167 	&	1941.8027	&	ASAS J060817+2045.2; 2MASS 06081675+2045150	&		\\
202073185	&	 13.10	&	 \phn 96.02920	&	18.70330	&	0.62	&	0.612481 	&	1941.7653	&	V* GZ Gem (EB*Algol); 2MASS 06240682+1842112	&		\\
202073186	&	 10.40	&	 \phn 94.11483	&	23.31200	&	0.55	&	1.224829 	&	1941.7959	&	CCDM J06165+2 (*IN**); 2MASS 06162816+2318506	&		\\
202073203	&	 12.90	&	 101.18333	&	21.59333	&	0.61	&	1.198015 	&	1942.501	&	ASAS J064444+2135.6; 2MASS 06444368+2135371	&		\\
202073207	&	 14.70	&	 \phn 92.57667	&	22.75333	&	0.58	&	1.050329 	&	1942.2679	&	V* BS Gem (EB*Algol); 2MASS 06101841+2245114	&		\\
202073210	&	 14.70	&	 \phn 92.28667	&	16.58583	&	0.53	&	1.025071 	&	1942.05	&	V* V667 Ori (EB*betLyr); 2MASS 06090828+1635064	&		\\
202073217	&	 12.40	&	 101.38521	&	17.22515	&	0.60	&	0.941135 	&	1941.319	&	V* EY Gem (EB*betLyr); 2MASS 06453218+1713361	&		\\
202073218	&	 13.00	&	 100.55833	&	21.73000	&	0.64	&	0.921221 	&	1941.4036	&	ASAS J064214+2143.8; 2MASS 06421396+2143506	&		\\
202073232	&	 12.30	&	 \phn 98.52917	&	15.28167	&	0.80	&	0.838008	&	1941.667	&	ASAS J063407+1516.9; 2MASS 06340720+1516541	&		\\
202073235	&	 12.00	&	 101.95687	&	16.86322	& 0.60	&	0.819121 	&	1941.4962	&	V* FG Gem (EB*Algol); 2MASS 06474902+1651469	&		\\
202073238	&	 11.80	&	 \phn 90.60747	&	23.10947	&	0.64	&	0.791658 	&	1941.5186	&	ASAS J060226+2306.5; 2MASS 06022579+2306340	&		\\
202073248	&	 12.80	&	 101.92500	&	23.93661	&	0.65	&	0.757302 	&	1941.69	&	V* V383 Gem (EB*Algol); 2MASS 06474171+2356022	&		\\
202073253	&	 11.20	&	 \phn 93.53042	&	17.67444	&	0.74	&	0.733801 	&	1941.62	&	ASAS J061407+1740.5 (Candidate\_EB*); 2MASS 06140718+1740269	&		\\
202073262	&	 14.90	&	 \phn 90.84625	&	24.05083	&	0.59	&	2.246169 	&	1942.6555	&	V* DR Gem (EB*WUMa); 2MASS 06032303+2403028	&		\\
202073266	&	 12.10	&	 \phn 96.92732	&	27.40250	&	0.94	&	0.689607 	&	1941.9902	&	TYC 1887-1247-1 (Star); 2MASS 06274249+2724083	&		\\
202073267	&	 12.50	&	 \phn 92.86917	&	22.71556	&	0.62	&	0.684031 	&	1941.52	&	ASAS J061129+2242.9 (Star); 2MASS 06112855+2242578	&		\\
202073270	&	 10.90	&	 \phn 92.85417	&	18.54989	&	0.62	&	0.659306 	&	1941.4588	&	V* V392 Ori (EB*Algol); 2MASS 06112516+1832596	&		\\
202073276	&	 12.00	&	 102.50708	&	22.35778	&	0.76	&	0.644164 	&	1941.66	&	ASAS J065002+2221.5 (Candidate\_EB*); 2MASS 06500165+2221277	&		\\
202073297	&	 15.40	&	 \phn 95.83333	&	20.83286	&	0.62	&	0.527435 	&	1941.46	&	V* GR Gem (EB*betLyr); 2MASS 06232025+2049578	&		\\
202073307	&	 12.60	&	 \phn 92.95463	&	27.00535	&	0.72	&	0.505216 	&	1941.7346	&	2MASS 06114893+2700259	&	\\
202073319	&	 10.90	&	 101.78330	&	15.64330	&	0.76	&	0.462071 	&	1941.4898	&	V* V405 Gem (EB*WUMa); 2MASS 06470785+1538369	&		\\
202073346	&	 12.50	&	 \phn 99.31727	&	27.10678	&	0.75	&	0.384243 	&	1941.459	&	TYC 1888-1317-1 (Star); 2MASS 06371598+2706167	&		\\
202073348	&	 12.10	&	 \phn 93.46083	&	25.62861	&	0.77	&	0.320736	&	1942.9913	&	ASAS J061351+2537.7 (Candidate\_EB*); 2MASS 06135085+2537402	&		\\
202073353	&	 12.00	&	 \phn 92.67190	&	27.39488	&	0.77	&	0.371250 	&	1941.5099	&	2MASS 06104125+2723415	&		\\
202073361	&	 11.10	&	 \phn 90.25184	&	23.93759	&	0.75	&	0.361833 	&	1941.41	&	TYC 1864-1065-1 (Candidate EB*); 2MASS 06010044+2356153	&		\\
202073397	&	 12.60	&	 103.95833	&	19.00333	&	0.94	&	0.345527 	&	1941.3673	&	ASAS J065550+1900.2; 2MASS 06555020+1900166	&		\\
202073438	&	 15.00	&	 \phn 99.66250	&	19.89447	&	0.52	&	2.405896 	&	1943.0241	&	V* OQ Gem (EB*Algol); 2MASS 06383922+1953397	&		\\
202073440	&	 13.20	&	 \phn 91.66625	&	23.54556	&	0.40	&	2.624142 	&	1941.72	&	V* LQ Gem (EB*); 2MASS 06063990+2332435	&		\\
202073442	&	 13.70	&	 \phn 92.53767	&	23.93611	&	0.95	&	3.867399 	&	1943.69	&	[GMB2010] WO (SB); 2MASS 06100903+2356099	&		\\
202073445	&	 14.50	&	 \phn 92.42750	&	23.96817	&	0.47	&	2.835402 	&	1942.0075	&	[GMB2010] WO (SB); 2MASS 06094259+2358053	&		\\
202073476	&	 15.00	&	 \phn 91.48542	&	24.33889	&	0.34	&	\nodata 	&	1957.76	&	V* HN Gem (EB*Algol); 2MASS 06055652+2420201	&	1	\\
202073489	&	 12.60	&	 101.76169	&	15.62306	&	0.73	&	0.348725 	&	1941.547	&	V* V404 Gem (EB*WUMa); 2MASS 06470269+1537288	&		\\
202073490	&	 13.50	&	 101.19833	&	21.44417	&	0.58	&	2.029235 	&	1942.9043	&	V* OR Gem (EB*Algol); 2MASS 06444669+2126378	&		\\
202073495	&	 15.50	&	 100.50292	&	24.39889	&	0.53	&	2.860568 	&	1943.0098	&	V* KO Gem (EB*Algol); 2MASS 06420065+2423568	&		\\
202083021	&	 16.00	&	 102.95375	&	15.64889	&	0.60	&	2.933567 	&	1940.9997	&	V* KX Gem (EB*Algol); 2MASS 06514889+1538558	&		\\
202087711   &    11.00    &   98.69724  &   18.87943     &  0.41 &   2.29    &   1942.495      &   TYC 1337-48-1; 2MASS 06344733+1852459    &   \\
202103762	&	 13.20	&	 102.93668	&	25.93092	&	0.48	&	1.326493 	&	1941.4718	&	2MASS 06514459+2555550	&		\\
202122545	&	 13.30	&	 \phn 98.09915	&	25.11887	&	0.04	&	19.2700 \phn 	&	1950.8918	&	2MASS 06322356+2507167	&	2	\\
202126825	&	 13.09	&	 105.66026	&	19.46655	&	0.21	&	9.206803 	&	1947.8098	&	2MASS 07023846+1927595	&		\\
202126847	&	 12.42	&	 102.20130	&	17.16191	&	0.24	&	8.342127 	&	1945.2552	&	2MASS 06484823+1709502	&		\\
202126851	&	 13.96	&	 101.49694	&	27.18665	&	0.32	&	4.480119 	&	1945.9267	&	2MASS 06455926+2711119	&		\\
202126864	&	 12.65	&	 105.61144	&	19.49339	&	0.28	&	6.187675 	&	1946.1115	&	2MASS 07022674+1929362	&		\\
202126871	&	 13.62	&	 100.50464	&	16.32700	&	0.87	&	8.431037 	&	1942.6508	&	2MASS 06420111+1619371	&		\\
202126877	&	 11.02	&	 \phn 97.98012	&	25.48210	&	0.02	&	\nodata 	&	1955.44	&	TYC 1883-1721-1 (Star); 2MASS 06315522+2528555	&	1	\\
202126878	&	 12.25	&	 \phn 95.41234	&	24.97781	&	0.29	&	5.559292 	&	1946.2296	&	TYC 1882-1294-1 (Star); 2MASS 06213896+2458401	&		\\
202126880	&	 12.96	&	 \phn 95.43424	&	25.84569	&	0.29	&	3.460849 	&	1941.51	&	2MASS 06214421+2550444; SDSS J062144.21+255044.4	&		\\
202126886	&	 14.25	&	 \phn 99.85710	&	26.63629	&	0.45	&	2.941136 	&	1942.2333	&	2MASS 06392557+2638008	&		\\
202126887	&	 13.03	&	 \phn 99.38054	&	26.78655	&	0.07	&	12.85570 \phn 	&	1944.2557	&	2MASS 06373133+2647115	&		\\

\enddata
\tablecomments{Table flags are as follows: 
1) Single eclipse observed in C0,
2) Two primary eclipses observed in C0 (period unconfirmed),
3) Stellar $T_{\rm eff}$ estimated from OMM observation (Section~\ref{sec:spectroscopy})}

\end{deluxetable}

\end{landscape}


\begin{landscape}

\begin{deluxetable}{lcccccccc}
\tabletypesize{\scriptsize}

\tablecaption{New EPIC EBs in K2 Campaign 0\label{tab:newEBs}}
\tablewidth{0pt}
\tablehead{
\colhead{EPIC} & 
\colhead{$K_{\rm P}$}   & 
\colhead{RA} & 
\colhead{DEC} & 
\colhead{morph} & 
\colhead{Period}    &
\colhead{BJD$_0$}   &
\colhead{Cross matched} &
\colhead{Note}  \\
\colhead{ } & 
\colhead{(mag)} &
\colhead{(deg)} & 
\colhead{(deg)} & 
\colhead{}    &
\colhead{(days)}    &
\colhead{(BJD-2454833)} &
\colhead{identification}    &
\colhead{flag}  
}
\startdata

202060506	&	 13.69	&	 \phn 91.02725	&	23.52939	&	0.56	&	 1.959777	&	1942.7863	&	2MASS 06040653+2331457; SDSS J060406.54+233145.7	&		\\
202060523	&	 10.20	&	 \phn 94.98682	&	23.73928	&	0.13	&	 16.409028 	&	1946.9955	&	HD 255133; 2MASS 06195683+2344214	&		\\
202060551	&	 10.35	&	 \phn 91.75614	&	23.18982	&	0.17	&	 16.496998	&	1946.2014	&	HD 251696; 2MASS 06070147+2311234	&		\\
202060577	&	 10.5	&	 \phn 93.07996	&	25.58141	&	0.60	&	 1.019488	&	1939.4459	&	2MASS 06121919+2534530	&		\\
202060800	&	 11.58	&	 \phn 92.59105	&	23.79414	&	0.31	&	 3.262032 	&	1940.6815	&	HD 252612; 2MASS 06102186+2347390	&		\\
202060911	&	 12.60	&	 \phn 92.10217	&	22.79439	&	0.31	&	 7.024977	&	1945.8409	&	GSC 01877-0048; 2MASS 06082451+2247398	&		\\
202060921	&	 12.00	&	 \phn 92.32424	&	23.56279	&	0.03	&	 \nodata 	&	1971.3295	&	TYC 1877-964-1; 2MASS 06091781+2333462	&	1	\\
202061000	&	 12.40	&	 \phn 90.93004	&	22.07661	&	0.36	&	 5.869057 	&	1943.0212	&	GSC 01325-0100; 2MASS 06034320+2204357	&		\\
202062176	&	 11.00	&	 \phn 92.46929	&	20.50456	&	0.36	&	 4.354250 	&	1942.8988	&	2MASS 06095262+2030273	&		\\
202063160	&	  9.30	&	 \phn 96.69498	&	16.45886	&	0.45	&	 9.069795 	&	1943.0525	&	HD 45193; 2MASS 06264647+1627422	&		\\
202064080	&	 11.40	&	 \phn 92.81720	&	22.58563	&	\nodata	&	 18.6700	&	1950.8607	&	HD 252893; 2MASS 06111653+2235072	&	3	\\
202064253	&	 12.30	&	 \phn 91.50423	&	22.89176	&	0.06	&	 21.1403	&	1944.993	&	TYC 1864-1626-1; 2MASS 06060100+2253303	&		\\
202065802	&	 11.53	&	 \phn 94.66820	&	24.88376	&	\nodata	&	 2.488563 	&	1942.8072	&	2MASS 06184036+2453015; SDSS J061840.14+245257.0	&	3	\\
202065819	&	 11.90	&	 \phn 94.53534	&	24.83398	&	\nodata	&	 5.744089 	&	1943.2566	&	2MASS 06180848+2450023; SDSS J061808.48+245002.2	&	3	\\
202066394	&	 14.30	&	 \phn 95.43298	&	25.03245	&	0.34	&	 3.186362 	&	1941.7972	&	2MASS 06214391+2501569; SDSS J062143.91+250156.8	&		\\
202066699	&	 14.40	&	 \phn 98.24756	&	26.86962	&	0.50	&	 2.215812 	&	1942.0412	&	2MASS 06325941+2652106; SDSS J063259.41+265210.6	&		\\
202066811	&	 14.50	&	 \phn 103.36132	&	16.83309	&	0.54	&	 1.696599 	&	1942.2817	&	2MASS 06532672+1649592	&		\\
202068686	&	 12.21	&	 \phn 94.37093	&	23.19523	&	0.17	&	 8.753366 	&	1947.1591	&	TYC 1878-625-1; 2MASS 06172901+2311431	&		\\
202068807	&	 12.02	&	 \phn 95.17472	&	23.29332	&	0.35	&	 4.244843 	&	1942.1365	&	TYC 1878-947-1; 2MASS 06204185+2317264	&		\\
202071293	&	 10.80	&	 \phn 96.59682	&	16.05121	&	0.68	&	 0.572925 	&	1941.6329	&	2MASS 06262323+1603043	&		\\
202071505	&	 10.10	&	 \phn 99.29125	&	16.10936	&	0.31	&	 3.555458	&	1942.7771	&	TYC 1329-1160-1; 2MASS 06370987+1606445	&		\\
202071579	&	 10.40	&	 \phn 96.58702	&	15.33108	&	0.43	&	 3.127251 	&	1941.4203	&	TYC 1328-1566-1; 2MASS 06262107+1519429	&		\\
202071631	&	 10.50	&	 \phn 98.40078	&	18.83125	&	0.58	&	 3.473756 	&	1937.0563	&	TYC 1337-314-1; 2MASS 06333619+1849524	&		\\
202071635	&	 10.20	&	 \phn 93.57181	&	18.62702	&	0.20	&	 6.270290 	&	1942.4393	&	TYC 1318-530-1; 2MASS 06141723+1837373	&		\\
202071731	&	 10.60	&	 \phn 99.30961	&	16.42521	&	0.51	&	 2.740255	&	1942.0820	&	TYC 1329-456-1; 2MASS 06371430+1625307	&		\\
202071828	&	 10.90	&	 \phn 98.16272	&	25.02194	&	\nodata	&	 4.708328 	&	1942.0007	&	TYC 1883-2053-1; 2MASS 06323905+2501189	&	3	\\
202071902	&	 10.40	&	 \phn 101.84678	&	21.13419	&	0.01	&	 \nodata 	&	1958.5707	&	TYC 1342-1936-1; 2MASS 06472322+2108030	&	1	\\
202071945	&	 9.60	&	 \phn 93.50981	&	20.02771	&	\nodata	&	 22.920600 \phn 	&	1956.1625	&	BD+20 1321; 2MASS 06140273+2001306	&	2	\\
202071994	&	 9.60	&	 \phn 96.75253	&	17.78967	&	0.43	&	 4.051073	&	1945.0441	&	TYC 1332-561-1; 2MASS 06270059+1747227	&		\\
202072061	&	 9.90	&	 \phn 100.47882	&	24.97612	&	\nodata	&	 2.100424	&	1942.6534	&	TYC 1897-1168-1; 2MASS 06415491+2458339	&		\\
202072282	&	 10.60	&	 \phn 94.99049	&	21.40122	&	\nodata	&	 5.977524	&	1939.3031	&	TYC 1327-1570-1; 2MASS 06195750+2123545	& 3	\\
202072430	&	 10.90	&	 100.10215	&	15.14237	&	0.11	&	 20.007618	&	1943.9820	&	TYC 1330-2152-1; 2MASS 06402451+1508324	&		\\
202072485	&	 10.90	&	 102.18925	&	15.99120	&	0.09	&	 14.6300	&	1943.2469	&	TYC 1331-1925-1; 2MASS 06484542+1559283	&		\\
202072486	&	 11.20	&	 105.08763	&	20.56381	&	0.49	&	 2.921628 	&	1943.3058	&	TYC 1352-289-1; 2MASS 07002042+2033458	&		\\
202072502	&	 11.20	&	 \phn 98.82478	&	20.59244	&	0.58	&	 1.923701 	&	1941.4794	&	TYC 1337-283-1; 2MASS 06351795+2035328	&		\\
202072563	&	 11.00	&	 105.03774	&	19.41262	&	0.50	&	 2.123748 	&	1942.4788	&	TYC 1352-1558-1; 2MASS 07000906+1924455	&		\\
202072596	&	 11.20	&	 \phn 99.51074	&	20.93931	&	0.35	&	 3.979679 	&	1943.87	&	TYC 1341-702-1; 2MASS 06380257+2056215	&		\\
202072917	&	 11.90	&	 \phn 99.78832	&	16.82623	&	0.02	&	 \nodata   \phn 	&	1953.6261	&	TYC 1329-752-1; 2MASS 06390919+1649345	&	1	\\
202083510	&	 16.50	&	 105.54264	&	20.20912	&	0.17	&	 3.307592	&	1943.22	&	2MASS 07021029+2012333	&		\\
202083650	&	 15.80	&	 \phn 93.98480	&	17.33772	&	0.40	&	 2.460416 	&	1942.5311	&	2MASS 06155635+1720157	&		\\
202083924	&	 14.00	&	 \phn 90.35016	&	23.26361	&	0.58	&	 0.378380	&	1941.3662	&	2MASS 06012405+2315497; SDSS J060124.07+231548.8	&		\\
202084588	&	 15.30	&	 \phn 92.13496	&	19.51602	&	\nodata	&	 28.78800 	&	1959.6731	&	2MASS 06083199+1930591	&	2	\\
202085014	&	 14.70	&	 \phn 93.43762	&	24.77931	&	0.65	&	 0.547850	&	1936.8511	&	2MASS 06134502+2446455; SDSS J061345.04+244645.7	&		\\
202086225	&	 12.30	&	 \phn 96.30833	&	20.77100	&	\nodata	&	 25.6485 	&	1945.9332	&	2MASS 06251404+2046206	&	2	\\
202086291	&	 12.30	&	 \phn 94.44968	&	25.60257	&	0.54	&	 2.434963 	&	1943.563	&	2MASS 06174792+2536092; SDSS J061747.92+253609.1	&		\\
202086627	&	 11.70	&	 \phn 96.48285	&	15.60354	&	0.19	&	 7.151257 	&	1944.6892	&	2MASS 06255588+1536127	&		\\
202087553	&	 13.70	&	 \phn 92.66988	&	20.16510	&	0.23	&	 9.001198 	&	1943.8183	&	2MASS 06104077+2009543	&		\\
202088178	&	 13.60	&	 \phn 95.77925	&	18.47082	&	0.53	&	 2.370616 	&	1942.3929	&	2MASS 06230702+1828149	&	4	\\
202088191	&	 11.70	&	 \phn 94.03926	&	26.95089	&	0.44	&	 1.661616	&	1942.623	&	TYC 1885-208-1; 2MASS 06160906+2656555	&		\\
202088387	&	 12.90	&	 \phn 91.67373	&	22.58188	&	0.41	&	 3.550170	&	1944.6967	&	2MASS 06064169+2234547; SDSS J060641.69+223454.8	&		\\
202091197	&	 11.60	&	 \phn 103.11794	&	17.57922	&	\nodata	&	 29.62270 	&	1950.01	&	TYC 1335-1000-1; 2MASS 06522830+1734451	&	2	\\
202091203	&	 15.20	&	 \phn 92.75029	&	19.42760	&	0.34	&	 5.214161	&	1942.7558	&	2MASS 06110014+1925298	&		\\
202091278	&	 11.70	&	 103.45979	&	25.66458	&	0.18	&	 19.8183	&	1948.4041	&	TYC 1898-2293-1; 2MASS 06535034+2539524	&	2	\\
202091404	&	 \phn 9.60	&	 \phn 98.75453	&	18.17719	& 0.11	&	 21.0842 	&	1949.9074	&	TYC 1333-449-1; 2MASS 06350172+1810345	&	2	\\
202091514	&	 10.00	&	 103.08750	&	25.80658	&	0.20	&	 8.312728 	&	1941.3878	&	TYC 1898-2911-1; 2MASS 06522100+2548236	&		\\
202091545	&	 10.90	&	 100.38259	&	27.28787	&	0.52	&	 1.857788 	&	1942.9097	&	TYC 1888-1789-1; 2MASS 06413182+2717163	&		\\
202092480	&	 10.60	&	 103.05227	&	25.33909	&	0.31	&	 5.687839 	&	1945.0349	&	TYC 1898-973-1; 2MASS 06521237+2520109	&		\\
202092613	&	 13.30	&	 \phn 93.96165	&	21.89863	&	0.37	&	 3.187483 	&	1943.44	&	2MASS 06155077+2154029	&	4	\\
202092842	&	 13.10	&	 \phn 98.17189	&	27.47062	&	\nodata	&	  36.9350   	&	1906.605	&	2MASS 06324125+2728142; SDSS J063241.25+272814.2	&	2	\\
202093968	&	 10.20	&	 \phn 96.16126	&	21.12574	&	0.35	&	 2.47000 	&	1941.5364	&	CCDM J06246+2; 2MASS 06243870+2107326	&		\\
202094117	&	 11.40	&	 \phn 96.12867	&	26.31839	&	0.42	&	 1.634486 	&	1942.45	&	TYC 1886-455-1; 2MASS 06243088+2619061	&		\\
202094234	&	 13.70	&	 \phn 93.75359	&	26.07596	&	0.57	&	 2.420777 	&	1941.57	&	2MASS 06150153+2604278	&		\\
202095298	&	 10.50	&	 \phn 94.81912	&	20.72068	&	0.27	&	 5.627027 	&	1942.2452	&	TYC 1327-300-1; 2MASS 06191658+2043144	&		\\
202135247	&	 14.40	&	 \phn 94.96192	&	17.79667	&	0.26	&	 \nodata   	&	1963.8686	&	2MASS 06195142+1747474 	&	1,4	\\
202137167	&	 13.40	&	 \phn 91.69799	&	20.07536	&	0.70	&	 4.395594	&	1941.6524	&	2MASS 06064813+2004326	&		\\
202137209	&	 15.50	&	 \phn 91.66285	&	20.53502	&	0.47	&	 2.381835 	&	1936.2585	&	2MASS 06063908+2032060	&		\\
202137571	&	 15.30	&	 \phn 91.74892	&	23.49284	&	0.63	&	 4.477240	&	1944.1655	&	2MASS 06065974+2329342; SDSS J060659.73+232934.3	&		\\
202137580	&	 13.10	&	 \phn 90.79350	&	23.50484	&	0.35	&	\nodata	&	1962.645	&	2MASS 06031044+2330174 &	1,4	\\
202139294	&	 11.40	&	 \phn 92.59146	&	27.44547	&	0.61	&	 1.043312	&	1941.346	&	2MASS 06102194+2726436	&		\\

\enddata
\tablecomments{
Table flags are as follows: 
1) Single primary eclipse observed in C0, 
2) Two primary eclipses observed in C0 (period unconfirmed),
3) Heartbeat binary candidate, 
4) Stellar $T_{\rm eff}$ estimated from OMM observation (Section~\ref{sec:spectroscopy})
} 

\end{deluxetable}

\clearpage

 \end{landscape}


\clearpage
\begin{landscape}

\begin{deluxetable}{lccccccccc}

\tabletypesize{\scriptsize}
\tablecaption{M35 Superstamp EBs in Campaign 0\tablenotemark{a}\label{tab:m35eb}}
\tablewidth{0pt}
\tablehead{
\colhead{2MASS ID} & 
\colhead{$K_{\rm P}$}   & 
\colhead{RA} & 
\colhead{DEC} & 
\colhead{Period}    &
\colhead{BJD$_0$}   &
\colhead{Superstamp} &
\colhead{RA} & 
\colhead{Dec} &
\colhead{ Note Flag } \\
\colhead{} & 
\colhead{(mag)} &
\colhead{(deg)} & 
\colhead{(deg)} & 
\colhead{(days)}    &
\colhead{(BJD-2454833)} &
\colhead{TPF}    &
\colhead{(deg)}  &
\colhead{(deg)} &
\colhead{ } 
}

 \startdata

2MASS 06081216+2431193 & 13.18 & 92.0504 & 24.5219 & 2.51 & 1937.11 & 200000819 & 92.0776 & 24.5320  & 2 \\
2MASS 06093832+2417534 & 10.87 & 92.40956 & 24.29817 & 1.96 & 1936.7795 & 200000853 & 92.3814 & 24.3117 &  \\
2MASS 06083089+2415122 & 11.30 & 92.1285 & 24.25325 & 2.84 & 1937.6988 & 200000857 & 92.1450 & 24.2630 & 2 \\
2MASS 06101142+24241552 & 14.51 & 92.5474 & 24.4044 & 0.81 & 1936.2278 & 200000880 & 92.5454 & 24.4019 & 2 \\
2MASS 06094437+2434194 & 13.52 & 92.43516 & 24.57205 & 1.84 & 1937.28 & 200000884 & 92.4462 & 24.5513  \\
2MASS 06085327+2428371  & 12.99 & 92.2219 & 24.4771 & 1.82 & 1940.0357 & 200000888 & 92.2094 & 24.5027 &  \\
2MASS 06092044+2415155 & 16.31 & 92.3351 & 24.2544 & 0.52 & 1939.1598 & 200000924 & 92.3357 & 24.2457 &  2 \\
2MASS 06090042+2414108  & 12.93 & 92.2516 & 24.2364	 & 18.74 & 1938.6592 & 200000925 & 92.2766 & 24.2336 & 1 \\
2MASS 06092929+2407028 & 16.30 & 92.3718 & 24.1176	& 0.39 & 1939.1086 & 200000940 & 92.3624 & 24.1380 &  2 \\
2MASS 06101502+2408460 & 13.72 & 92.5623 & 24.1462 & 3.23 & 1937.8113 & 200000945 & 92.5528 & 24.1205 & 2 \\
2MASS 06100186+2405498 & 13.43 & 92.5075 & 24.0972 & 0.31 & 1939.088 & 200000946 & 92.4938 & 24.1084 & 2 \\
2MASS 06095723+2403294 & 14.06 & 92.4882 & 24.0582 & 2.32 & 1937.7909 & 200000957 & 92.5071 & 24.0545 & 2 \\
2MASS 06083223+2359391 & 14.12 & 92.1341 & 23.9942 & 3.93 & 1940.4535 & 200000963 & 92.1533 & 23.9815 & 2 \\

\enddata
\tablenotetext{a}{These EBs occupy C0 M35 TPFs; the point source 2MASS ID and RA/Dec is supplied followed by the corresponding superstamp TPF and its center pixel coordinates.}
\tablecomments{
Table flags are as follows: 
1) Two primary eclipses observed in C0 (period unconfirmed),
2) Listed in \citealt{nar14},
}

\end{deluxetable}

\clearpage

 \end{landscape}


\clearpage
\begin{landscape}

\begin{deluxetable}{lccccccccc}

\tabletypesize{\scriptsize}
\tablecaption{Non-EPIC EBs in Campaign 0 \label{tab:nonEPIC}}
\tablewidth{0pt}
\tablehead{
\colhead{2MASS ID\tablenotemark{a}} & 
\colhead{$K_{\rm P}$}   & 
\colhead{RA} & 
\colhead{DEC} & 
\colhead{Period}    &
\colhead{BJD$_0$}   &
\colhead{Contaminated} &
\colhead{RA} & 
\colhead{Dec} &
\colhead{ Note Flag } \\
\colhead{} & 
\colhead{(mag)} &
\colhead{(deg)} & 
\colhead{(deg)} & 
\colhead{(days)}    &
\colhead{(BJD-2454833)} &
\colhead{EPIC}    &
\colhead{(deg)}  &
\colhead{(deg)} &
\colhead{ } 
}

 \startdata

2MASS 06115706+2040109  &  12.02 & 92.98780 & 20.67400 & 6.81 & 1942.5413 & 202060198 & 92.98965 & 20.67244 &  \\
2MASS 06105541+2037042  &  9.93 & 92.72550 & 20.61270 & 3.34 & 1943.4607 & 202060503 & 92.73293 & 20.61952 & 3 \\
2MASS 07003695+1937177  &  11.78 & 105.1540417 & 19.62186 & 4.49 & 1944.2174 & 202064550 & 105.15858 & 19.61459 &  \\
2MASS 06181254+2438032  &  11.33 & 94.54720  & 24.63440 & 7.83 & 1944.5744 & 202065879 & 94.5522 & 24.63424 & \\
2MASS 06083223+2359391  &  13.28 & 97.77700 & 17.70110	& 23.79	& 1945.9438	& 202071645 & 97.77472 & 17.70121 & 2 \\
2MASS 06404939+2404128  &  9.65 & 100.20554 & 24.07038 & 7.80 & 1941.2548 & 202072624 & 100.21330 & 24.07368 & \\
2MASS 06553695+1923084  &  14.45 & 103.91137 & 19.38180 & 7.60 & 1944.4453 & 202072756 & 103.90250 &	19.38504 & \\
2MASS 06235708+1820182  &  12.94 & 95.98780 & 18.33850 & 4.91 & 1943.226 & 202083222 & 95.9835 & 18.3345 & \\
2MASS 06143505+2343535  &  14.82 & 93.646041 & 23.73158 & 0.65 & 1936.3811 & 202083688 & 93.646041 & 23.73158 & \\
2MASS 06262323+1603043  &  9.36  & 96.5968 & 16.0513 & 0.57    & 1939.3339 & 202084063 & 96.59837 & 16.06072 & \\
2MASS 06262360+1603385  &  15.40 & 97.18890 & 14.60340 & 2.03 & 1943.0423 & 202084843 & 97.18514 & 14.60301 & \\
2MASS 06190294+1828199  &  14.38 & 94.76225	& 18.47227 & 11.24 & 1948.14 & 202085157 & 94.75625 & 18.47464 & \\	
2MASS 06184418+1821189  &  13.37 & 94.68404	& 18.35525 & \nodata & 1948.1254 & 202085278 & 94.68155 & 18.35271 & 1 \\
2MASS 06480648+2323022  &  16.15 & 102.02680	& 23.3839 & 2.56 & 1941.4388 & 202087552 & 102.02209 & 23.38287 &  \\
2MASS 06230585+2058156  &  13.55 & 95.78300 & 20.97161 & 0.90 & 1942.0308 & 202090938 & 95.77686	& 20.97275 & 2 \\
2MASS 06302365+2622128  &  13.91 & 97.5985 & 26.37016 &	0.44 & 1946.7301 & 202092874 & 97.60019 & 26.36851 & \\
2MASS 06290799+2814236  &  12.89 & 97.28320	& 28.23990 & 5.13 &	1945.208 & 202095074 & 97.28940 & 28.23426 & \\
2MASS 07021637+1846346  &  11.50 & 105.5682083 & 18.77650 & 2.41 & 1942.305 & 202126863 & 105.56551 & 18.77254 & \\
2MASS 06504495+1648335  &  15.48 & 102.68690 & 16.80910	& 2.92 & 1942.0928 & 202127311 & 102.6836 & 16.81334 & \\
2MASS 06240232+1613363  &  11.87 & 96.00954	& 16.2267 & 1.87 & 1939.2725 & 202136002 & 96.0155 & 16.2275 & \\
2MASS 06235881+1622092  &  12.73 & 95.9949 & 16.3690 & 2.51	& 1937.4132 & 202136015 & 95.98770 & 16.36856 & \\
2MASS 06402291+1559465  &  13.19 & 97.29830	& 16.6421 & 5.20 &	1943.0525 & 202136063 & 97.30045 & 16.64588 & \\
2MASS 06185304+1840323  &  13.68 & 100.09530 & 15.9961 & 5.74 & 1945.3597 & 202136445 & 100.09492 & 15.99292 & \\
2MASS 06185243+1840339  &  11.35 & 94.71887 & 18.67119 & 3.37 & 1944.0126 & 202137030 & 94.72101 & 18.67565 & \\
2MASS 06083624+2347222  &  13.10 & 92.15095	& 23.78966 & 6.14 &	1944.5813 & 202137637 & 92.15124 & 23.79245 & \\
2MASS 06080534+2352495  &  15.05 & 92.0220 & 23.8805 & 0.73	& 1939.5807	& 202137653 & 92.02741 & 23.88216 & \\
2MASS 06245004+1813328  &  14.33 & 96.20833 & 18.22583 & 1.01 & 1936.5550 & 202137708 &  96.21516 & 18.23077 & \\
2MASS 06190708+2458234  &  13.65 & 94.77945	& 24.97306 & 4.45 &	1942.7764 & 202138912 &  94.78377 & 24.97246 & \\
2MASS 06172371+2539374  &  13.75 & 94.34867 & 25.66047 & 2.77 & 1942.0306 & 202139017 & 94.34667 & 25.66506 & \\

\enddata
\tablenotetext{a}{These EBs occupy C0 TPF halo pixels; their eclipses contaminate the noted EPIC target}
\tablecomments{
Table flags are as follows: 
1) Single primary eclipse observed in C0, 
2) Two primary eclipses observed in C0 (period unconfirmed),
3) Heartbeat binary candidate. 
}

\end{deluxetable}

\clearpage

 \end{landscape}


\clearpage

\begin{deluxetable}{lc}

\tabletypesize{\scriptsize}
\tablecaption{Blended EBs in Campaign 0 \label{tab:blends}}
\tablewidth{0pt}
\tablehead{
\colhead{EPIC ID (EB)} & 
\colhead{EPIC ID (blend)}
}
\startdata

202073348 & 202073377 \\
202062450 & 202072972 \\
202071279 & 202073364 \\
202126863 & 202126867 \\
202060198 & 202070263 \\
202073489 & 202073366 \\
202073489 & 202073362 \\ 
202065879 & 202066041 \\
202065879 & 202065929 \\

\enddata

\end{deluxetable}

\clearpage


\clearpage
\begin{landscape}

\begin{deluxetable}{lcccccc}

\tabletypesize{\scriptsize}
\tablecaption{K2 C0 Eclipsing Binaries Observed by KELT-North\label{tab:KELT_phot}}
\tablewidth{0pt}
\tablehead{
\colhead{KELT} & 
\colhead{EPIC}   & 
\colhead{$K_{\rm P}$} & 
\colhead{RA} & 
\colhead{Dec}    &
\colhead{Period} &
\colhead{BJD$_0$}    \\
\colhead{ID} & 
\colhead{} &
\colhead{(mag)} & 
\colhead{(deg)} & 
\colhead{(deg)}    &
\colhead{(days)} &
\colhead{(BJD-2454833)}    
}

 \startdata

KN 05E09313	 &  202072963 &	11.40 &	104.37154 &	20.18725 &	4.209922 	&	1941.5103 \\
KN 04E010922 &	202073097 &	9.60	  &  91.09233 &	20.53502 &	0.972411 	&	1941.9621 \\
KN 04E006640 &	202060541 &	10.30 &	94.08259 &	20.91274 &	4.049160 	&	1944.4214 \\
KN 04E102374 &	202071266 &	12.50 &	91.33042 &	22.54782 &	2.593621 	&	1944.18 \\
KN 04E005219 &	202072961 &	9.30	 &  100.12564 &	22.91052 &	4.097024 	&	1945.3702 \\
KN 04E031147 &	202073238 &	11.80 &	90.60747 &	23.10947 &	0.791658 	&	1941.5186 \\
KN 04E019375 &	202060135 &	9.90	 &  90.36692 &	23.14098 &	2.865379 	&	1943.2567 \\
KN 04E046126 &	202060752 &	10.80 &	92.8667	 &  23.2317  &	0.770391 	&	1941.4077 \\
KN 04E016313 &	202073186 &	10.40 &	94.11483 &	23.312	 &  1.224829 	&	1941.7959 \\
KN 04E054729 &	202072978 &	12.20 &	100.26213 &	23.8078	 &  4.885995 	&	1944.6566 \\
KN 04E110687 &	202073361 &	11.10 &	90.25184 &	23.93759 &	0.361833 	&	1941.41 \\
KN 04E006288 &	202072958 &	10.50 &	94.19764 &	24.07841 &	4.024547 	&	1942.6683 \\
KN 04E069591 &	202072991 &	10.20 &	95.003	  &  26.34982 &	5.495907 	&	1945.3442 \\
KN 04E039600 &	202073144 &	11.10 &	100.21498 &	26.74048 &	1.550610 	&	1941.7757 \\
KN 04E045879 &	202073307 &	12.60 &	92.95463 &	27.00535 &	0.505216 	&	1941.7346 \\
KN 04E079988 &	202073346 &	12.50 &	99.31727 &	27.10678 &	0.384243 	&	1941.459 \\
KN 04E040654 &	202073353 &	12.00 &  92.6719	  &  27.39488 &	0.371250 	&	1941.5099 \\
KN 04E065527 &	202073266 &	12.10 &	96.92732 &	27.4025	   & 0.689607 	&	1941.9902 \\

\enddata
\tablecomments{Table showing EBs that have light curves from the KELT survey. Ephemeris from Tables~\ref{tab:knownEBs} and \ref{tab:newEBs}. See Section~\ref{sec:kelt_phot} for details. }

\end{deluxetable}

\clearpage

 \end{landscape}


\clearpage

\begin{deluxetable}{lcc}

\tabletypesize{\scriptsize}
\tablecaption{Eccentric EBs in Campaign 0 \label{tab:ecc}}
\tablewidth{0pt}
\tablehead{
\colhead{EPIC ID} & 
\colhead{$e$} &
\colhead{$\omega$}
}
\startdata


202072430 &  0.633 &   2.890 \\
202091404 &  0.563 &   2.726 \\
202062176  & 0.447 &   3.464 \\
202071945  & 0.380 &   3.733 \\
202071505  & 0.294 &   3.245 \\
202071994  & 0.247 &   2.884 \\
202068807  & 0.226 &   3.3310 \\
202072596  & 0.150 &   3.3911 

\enddata
\tablecomments{Table listing the most eccentric EBs in C0. Eccentricities calculated using the methods described in \citet{prsa15}. See Section~\ref{sec:noteworthy_ebs} for details. }

\end{deluxetable}

\clearpage


\clearpage
\begin{deluxetable}{lcc}
\tabletypesize{\scriptsize}
\tablecaption{Planet Candidates\label{tab:PCs}}
\tablewidth{0pt}
\tablehead{
\colhead{Parameter} & 
\colhead{202072704} &
\colhead{2MASS 06101557+2436535}
}
\startdata
2MASS           & 06455102+1712250              & 06101557+2436535                      \\
$K_P$ (mag)     & 11.4                          & 13.0                                  \\
RA (deg)        & 101.4626                      & 92.5648                               \\
Dec (deg)       & 17.2069                       & 24.6148                               \\
$P$ (days)      & $2.65\tablenotemark{a}$                      & $7.5559\tablenotemark{a}$                            \\
Epoch (MJD)     & $2456771.13445\pm0.00029$     & $2456776.87137^{+0.00089}_{-0.00087}$ \\
Duration (hrs)  & $3.202^{+0.055}_{-0.038}$     & $3.946^{+0.038}_{-0.036}$             \\
$b$             & $0.676^{+0.033}_{-0.055}$     & $0.00\pm0.31$                         \\
$i$ (deg)       & $81.6^{+1.0}_{-0.72}$         & $89.12^{+0.62}_{-0.87}$               \\
$R_p/R_{\ast}$  & $0.0814^{+0.0016}_{-0.0020}$  & $0.0791^{+0.0012}_{-0.00085}$         \\
$a/R_{\ast}$    & $4.66^{+0.22}_{-0.16}$        & $14.22^{+0.33}_{-0.85}$               \\

\enddata
\tablenotetext{a}{The period was fixed.}
\tablecomments{\textsc{TAP} model fit parameters for the two exoplanet candidates described in Section~\ref{sec:planet_candidates}; circular orbits have been assumed.}

\end{deluxetable}


\end{document}